\documentclass[onecolumn,preprintnumbers]{revtex4}
\usepackage{amsfonts}
\usepackage{amsmath}
\usepackage{amsthm}
\usepackage{amssymb}
\usepackage{graphicx}
\usepackage{appendix}
\usepackage{hyperref}
\usepackage{textcomp}
\usepackage{multirow}
\usepackage{color}
\usepackage{bm}
\usepackage{braket}
\usepackage{subfigure}

\begin{document}

\author{B. A. Malomed}
\affiliation{Department of Physical Electronics, School of Electrical Engineering,
Faculty of Engineering, Tel Aviv University, Tel Aviv 69978, Israel\\
Instituto de Alta Investigaci\'{o}n, Universidad de Tarapac\'{a}, Casilla
7D, Arica, Chile}
\title{Multidimensional dissipative solitons and solitary vortices}

\begin{abstract}
This article offers a review of (chiefly, theoretical) results for
self-trapped states (solitons) in two- and three-dimensional (2D and 3D)
models of nonlinear dissipative media. The existence of such solitons
requires to maintain two stable balances: between nonlinear self-focusing
and linear spreading (diffraction and/or dispersion) of the physical fields,
and between losses and gain in the medium. Due to the interplay of these
conditions, dissipative solitons exist, unlike solitons in conservative
models, not in continuous families, but as isolated solutions (\textit{%
attractors}). The main issue in the theory is stability of multidimensional
dissipative solitons, especially ones with \textit{embedded vorticity}.
First, stable 2D dissipative solitons are presented in the framework of the
complex Ginzburg-Landau equation with the cubic-quintic nonlinearity, which
combines linear loss, cubic gain, and quintic loss (the linear loss is
necessary to stabilize zero background around dissipative solitons, while
the quintic loss provided for the global stability of the setting). In
addition to fundamental (zero-vorticity) solitons, stable \textit{spiral
solitons} produced by the CGL equation are produced too, with intrinsic
vorticities $S=1$ and $2$. Stable 2D solitons were also found in a system
built of two linearly-coupled optical fields, with linear gain acting in one
and linear loss, which plays the stabilizing role, in the other. In this
case, the inclusion of the cubic loss (without quintic terms) is sufficient
for the creation of stable fundamental and vortical dissipative solitons in
the linearly-coupled system. In addition to truly localized states, weakly
localized ones are presented too, in the single-component model with
nonlinear losses, which does not include explicit gain. In that case, the
losses are compensated by the influx of power from the reservoir provided,
at the spatial infinity, by the weakly localized structure of the solution.
Other classes of 2D models which are considered in this review make use of
spatially modulated loss or gain to predict many species of robust
dissipative solitons, including localized dynamical states featuring complex
periodically recurring metamorphoses. Stable fundamental and vortical
solitons are also produced by models including a trapping or spatially
periodic potential. In the latter case, the consideration addresses 2D
\textit{gap dissipative solitons} as well. 2D two-component dissipative
models including spin-orbit coupling are considered too. They give rise to
stable states in the form of \textit{semi-vortex} solitons, with vorticity
carried by one component. In addition to the 2D solitons, the review
includes 3D fundamental and vortical dissipative solitons, stabilized by the
cubic-quintic nonlinearity and/or external potentials. Collisions between 3D
dissipative solitons are considered too.
\end{abstract}

\author{}
\maketitle
\tableofcontents

\noindent \textbf{Acronyms}

\noindent 1D -- one-dimensional

\noindent 2D -- two-dimensional

\noindent 3D -- three-dimensional

\noindent BEC -- Bose-Einstein condensate

\noindent b.c. -- boundary conditions

\noindent CGL -- complex Ginzburg-Landau (equation)

\noindent CSV -- crater-shaped vortex

\noindent CQ -- cubic-quintic (nonlinearity)

\noindent DS -- dissipative soliton

\noindent FT -- flat-top (shape of a broad soliton)

\noindent GS -- gap soliton

\noindent GVD -- group-velocity dispersion

\noindent HO -- harmonic-oscillator (trapping potential))

\noindent IS -- intersite (centered mode)

\noindent LL -- Lugiato - Lefever (equation)

\noindent NLS -- nonlinear-Schr\"{o}dinger (equation)

\noindent OS -- onsite (centered modes)

\noindent $\mathcal{PT}$ -- parity-time (symmetry)

\noindent SV -- semi-vortex

\noindent VR -- vortex ring

\section{Introduction}

Studies of pattern formation in nonlinear media combining conservative and
dissipative properties is a vast research area with many physical
realizations \cite{Cross Hohenberg,Rosanov-book-2002,Hoyle}. A broad class
of universal models for the pattern formation in many physical settings is
provided by complex Ginzburg-Landau (CGL) equations \cite%
{Aranson-Kramer,Encyclopedia}. In many cases, especially as concerns
nonlinear optics, CGL equations are consistently derived from the underlying
physical theory; in some other cases, the same equations are introduced as
useful phenomenological models.

Many patterns which were predicted theoretically and created experimentally
in nonlinear dissipative media exhibit a trend to self-trapping into
localized states, alias dissipative solitons (DSs), which are important
physical objects in many areas of physics, especially in nonlinear photonics
and optics \cite%
{PhysicaD-first,Tirapegui1,Tirapegui2,Akhmed,Liehr,Mihalache-review,added_book}%
. In most cases, DSs are adequately represented by diverse localized
solutions of CGL equations. The theory of DSs was developed in detail in
one-dimensional (1D) settings, starting from the famous, although unstable,
Pereira-Stenflo solitons \cite{Hocking,Stenflo}, which are exact solutions
of the 1D CGL equation with the cubic nonlinearity. Essentially more
challenging are studies of DSs in two- and three-dimensional (2D and 3D)
geometries. A highly promising peculiarity of multidimensional solitons is a
possibility to embed vorticity into them, thus making such solitons
self-trapped states with an intrinsic topological structure \cite%
{PhysicaD-last}. On the other hand, it is well known that multidimensional
solitons are often subject to instabilities, therefore a problem of cardinal
significance is to develop physically relevant models which admit the
creation of stable 2D and 3D solitons, especially those with internal
structure, such as localized modes with embedded vorticity \cite{review1}-%
\cite{review5}.

The objective of this article is to produce a sufficiently systematic review
of results (chiefly, theoretical ones) produced by the work in the vast area
of studies of multidimensional DSs. The review includes both relatively old
models elaborated in this area and more recent ones, along with basic
findings obtained by the studies of these models. The main topics addressed
in the article correspond to sections in the article's table of contents.
They are determined, primarily, by mechanisms providing stabilization of
DSs, both fundamental (zero-vorticity) and internally-structured (in
particular, vorticity-carrying) ones, in 2D and 3D settings. The basic
DS-stabilization mechanisms, elaborated in many theoretical and some
experimental works, which are presented in the current review, can be
summarized as follows.

(i) The use of the CGL equations with combinations of competing nonlinear
terms. In most cases, these are cubic and quintic (CQ) ones, in both
conservative and dissipative parts of the equations. In the framework of
such models, which, in particular, describe lasing optical setups with
saturable absorption, the gain is provided by the cubic term, while a lossy
quintic one must be included to secure the global stability of the system
against the blowup \cite{Fauve-Thual}. This mechanism dominates in the 2D
and 3D settings, which are considered, respectively, in Sections II and IX
of the article. A related mechanism, based on nonlinear absorption, plays
the dominant role in the stabilization of weakly localized nonlinear
dissipative modes, both fundamental and vortical ones, which are considered
in Section IV.

(ii) Stabilization of 2D DSs in a laser cavity with the linear gain and
cubic nonlinearity by means of a lossy feedback, which is linearly coupled
to the cavity. This system is considered in Section III.

(iii) Stabilization of 2D\ DSs, many of which exhibit complex intrinsic
structure and dynamics, by means of spatially modulated gain and/or loss.
This class of the models and DSs produced by them is the subject of Section
V.

(iv) Stabilization of 2D (Sections VI and VII) and 3D (Section IX) DSs with
the help of trapping potentials, which may be represented by
harmonic-oscillator (HO) or spatially-periodic (lattice) potential terms in
the respective CGL equations.

(v) Stabilization of 2D two-component DSs, built as vortex-antivortex (VAV)
bound states by means of the spin-orbit coupling (SOC), is the subject of
Section VIII.

These topics, which are included, in a sufficiently detailed form, in the
present review cover only a part of results accumulated in the work
performed in the vast area of DSs. The choice of this set of the topics is
suggested, in particular, by research interests of the author. A really
comprehensive review of the theoretical and experimental work with DSs and
related objects should be a subject of a full-size book, rather than of a
single article. Some essential topics which are not considered in the review
are mentioned in the concluding section.

\section{Spiral solitons (vortex rings) produced by the 2D complex
Ginzburg-Landau (CGL) equation with the cubic-quintic (CQ) nonlinearity}

In its simplest form, the CGL equation with the cubic nonlinearity, for 2D
complex field $U(x,y)$, may be considered as the generalization of the
corresponding cubic nonlinear Schr\"{o}dinger (NLS) equation with complex
coefficients:%
\begin{equation}
\frac{\partial U}{\partial z}=\delta U+\left( \beta +\frac{i}{2}\right)
\left( \frac{\partial ^{2}}{\partial x^{2}}+\frac{\partial ^{2}}{\partial
y^{2}}\right) U-\left( \varepsilon -ig\right) |U|^{2}U.  \tag{1}
\end{equation}%
This equation is written in terms of the light propagation in bulk media,
with propagation distance $z$ and transverse coordinates $\left( x,y\right) $
\cite{Rosanov-book-2002,KivAgr}. In Eq. (1) term $(i/2)\left(
U_{xx}+U_{yy}\right) $ represents the paraxial diffraction, $g>0$ is the
coefficient of the cubic (Kerr) self-focusing (or cubic defocusing, in the
case of $g<0$), while coefficients $\beta >0$ and $\varepsilon >0$ account
for, respectively, the linear dispersive losses (alias diffusion, or
viscosity, in terms of hydrodynamics) and the cubic loss (two-photon
absorption, in terms of optics). Finally, coefficient $\delta >0$ represents
linear gain, which is necessary for the compensation of the dispersive and
nonlinear losses.

The cubic CGL equation (1) is not an appropriate model for predicting robust
localized states, such as DSs, because the presence of the term with $\delta
>0$ destabilizes the zero solution against small perturbations, hence the
zero background around localized states is unstable, making such states
unstable as a whole too. To introduce a model which admits stable DSs, one
needs to replace the linear loss by gain in Eq. (1), i.e., reverse the sign
of $\delta $. Simultaneously, the sign of the cubic dissipative coefficient
must be reversed too, $\varepsilon \rightarrow -\varepsilon $, so that the
cubic term provides for \textit{nonlinear gain}. Finally, to prevent the
blowup of the system, it is necessary to add a quintic loss term, with some
coefficient $\mu >0$, which turns the 2D CGL equation into one of the CQ
type, as first proposed by Sergeev and Petviashvili in 1984 \cite{SergPetv}:%
\begin{equation}
\left( \frac{\partial }{\partial z}+\delta \right) U=\left( \beta +\frac{i}{2%
}\right) \left( \frac{\partial ^{2}}{\partial x^{2}}+\frac{\partial ^{2}}{%
\partial y^{2}}\right) U+\left( \varepsilon +i\sigma \right) |U|^{2}U-\left(
\mu +i\nu \right) |U|^{4}U.  \tag{2}
\end{equation}%
In this equation, necessary signs of the coefficients are%
\begin{equation}
\delta >0,\beta \geq 0,\varepsilon >0,\mu >0.  \tag{3}
\end{equation}%
The scaling freedom makes it possible to fix $|\sigma |=1$ in Eq. (2), while
signs $\sigma =+1$ and $\sigma =-1$ correspond, respectively, to the
self-focusing and defocusing cubic nonlinearity. Further, because Eq. (2)
includes the quintic dissipative term with coefficient $\mu $, it is also
natural to introduce the quintic correction to the cubic self-interaction,
with coefficient $\nu $ in Eq. (2). Usually, it is assumed that this quintic
terms represents self-defocusing, which corresponds to $\nu >0$ in Eq. (3),
as quintic self-focusing in the 2D geometry leads to the supercritical
collapse \cite{Berge,Sulem,Fibich}. Nevertheless, in the presence of the
quintic loss, the quintic self-interaction may have the focusing sign too ($%
\nu <0$), as the quintic loss arrests the collapse in that case \cite%
{Mihalache-et-al-2006,Mihalache-et-al-2007a,Mihalache-et-al-2007b}.

While the CQ combination of the nonlinear dissipation and gain was
originally postulated phenomenologically \cite{SergPetv}, this pair of terms
provides an appropriate model of nonlinear amplifiers in optics, which are
built as a juxtaposition of the usual saturable amplifier (featuring the
linear gain in a combination with cubic and quintic losses) and saturable
absorber \cite{Haus-2000,Keller-2003,Bao-et-al-2009,Elsass-et-al-2010}. Such
setups are very efficient for the generation of stable temporal solitons in
fiber lasers \cite{Ferrmann-and-Hartl-2009}. The two-dimensional CGL
equation with the CQ nonlinearity was derived as a dynamical model for
coupled arrays of lasers by \cite{Zykin-Skryabin-Kartashov-2021}.

Stationary solutions to Eq. (2), which carry a real propagation constant $k$
and integer vorticity $S$, are looked for in the usual form, written in
polar coordinates $\left( r,\theta \right) $ (instead of the Cartesian
coordinates $\left( x,y\right) $),%
\begin{equation}
U(r,\theta ,z)=A(r)\exp (ikz+iS\theta ),  \tag{4}
\end{equation}%
with amplitude function $A(r)$ determined by the radial equation, obtained
from the substitution of ansatz (4) in CGL equation (2):%
\begin{equation}
\left( ik+\delta \right) A=\left( \beta +\frac{i}{2}\right) \left( \frac{%
d^{2}}{dr^{2}}+\frac{1}{r}\frac{d}{dr}-\frac{S^{2}}{r^{2}}\right) A+\left(
\varepsilon +ig\right) |A|^{2}A-\left( \mu +i\nu \right) |A|^{4}.  \tag{5}
\end{equation}%
DS solution $A(r)$ of Eq. (5) is a complex function, while in the
conservative model, with $\delta =\beta =g=\nu =0$, relevant solutions,
including vortex-ring (VR) ones, take a real form. A principal difference
from the conservative counterpart is that DS solutions exist at isolated
values of $k$, while solitons in conservative models (as well as in
parity-time ($\mathcal{PT}$)-symmetric ones, which include spatially
separated mutually-balanced linear-gain and loss terms, lying at the border
between conservative and dissipative systems \cite%
{Konotop-Yang-Zezyulin-2016,Suchkov-et-al-2016}), form continuous families,
parameterized by propagation constant $k$, that takes values in certain
intervals populated by the solitons (most typically, it is the semi-infinite
bandgap). This difference is explained by the fact that, in addition to the
balance between the nonlinear self-focusing and diffractive expansion, which
is necessary for the existence of any soliton in conservative models, DSs
must satisfy the additional constraint, \textit{viz}., the balance between
the gain and losses. It is also relevant to mention 2D localized states in
the form of ``chiral bubbles", which are produced by the CQ CGL equation
with real coefficients, and additional chiral terms. Such states were
experimentally observed in nematic liquid crystals \cite{Clerc}.

\begin{figure}[tbp]
\subfigure[]{\includegraphics[scale=0.7]{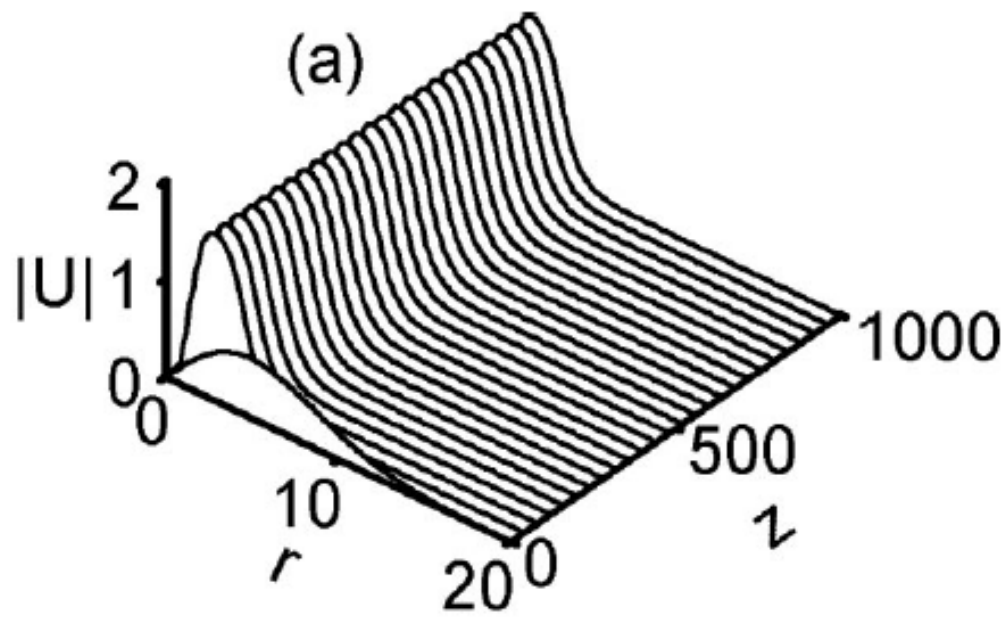}}\subfigure[]{%
\includegraphics[scale=0.7]{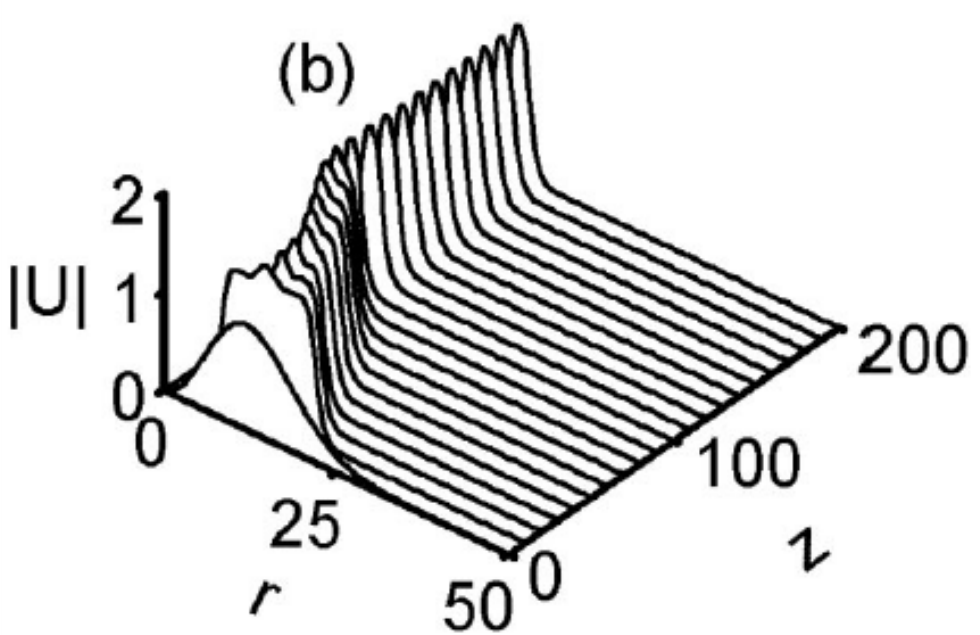}}
\caption{Formation of spiral solitons with winding numbers $S=1$ (a) and $%
S=2 $ (b), as produced by simulations of Eq. (2) with parameters $\protect%
\delta =0.5$, $\protect\beta =0.5$, $\protect\varepsilon =2.5$, $\protect\nu %
=0.1$, $\protect\mu =1$, and $\protect\sigma =+1$. The input is taken as per
Eq. (3) with the initial amplitude function $A_{0}(r)=0.2r\exp \left(
-(r/7)^{2}\right) $ in (a), and $A_{0}(r)=0.02r^{2}\exp \left(
-(r/12)^{2}\right) $ in (b) (source: Ref. \protect\cite%
{Crasovan-Malomed-Mihalache-2001}).}
\label{fig14.1=fig167}
\end{figure}

Basic result for the existence and stability of localized solutions of Eqs.
(2) and (5) with $S=0,1,$ and $2$ were reported in Ref. \cite%
{Crasovan-Malomed-Mihalache-2001}. In most cases, direct simulations of Eq.
(2) demonstrate that an input with winding number $S$ (introduced as in Eq.
(4)) and a radial shape which decays at $r\rightarrow \infty $ and, in the
case of $S\geq 1$, vanishes $\sim r^{S}$ at $r\rightarrow 0$ (or remains
finite at $r=0$ in the case of $S=0$), quickly transforms into a stable VR,
as shown in Fig. \ref{fig14.1=fig167}.

\begin{figure}[tbp]
\includegraphics[scale=0.5]{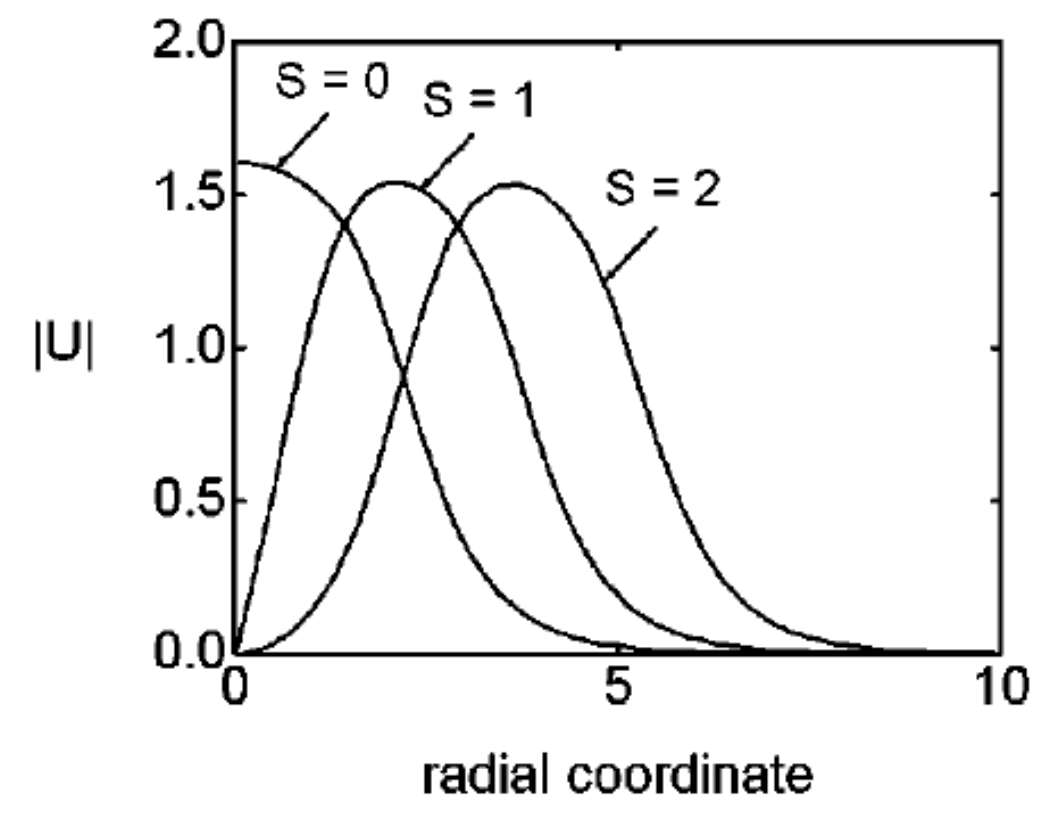}
\caption{Radial profiles of the stable spiral dissipative solitons with $S=1$
and $2$, whose formation is demonstrated in Fig. \protect\ref{fig14.1=fig167}%
, and the same for the fundamental soliton, with $S=0$ (source: Ref.
\protect\cite{Crasovan-Malomed-Mihalache-2001}).}
\label{fig14.2=fig168}
\end{figure}

\begin{figure}[tbp]
\subfigure[]{\includegraphics[scale=0.5]{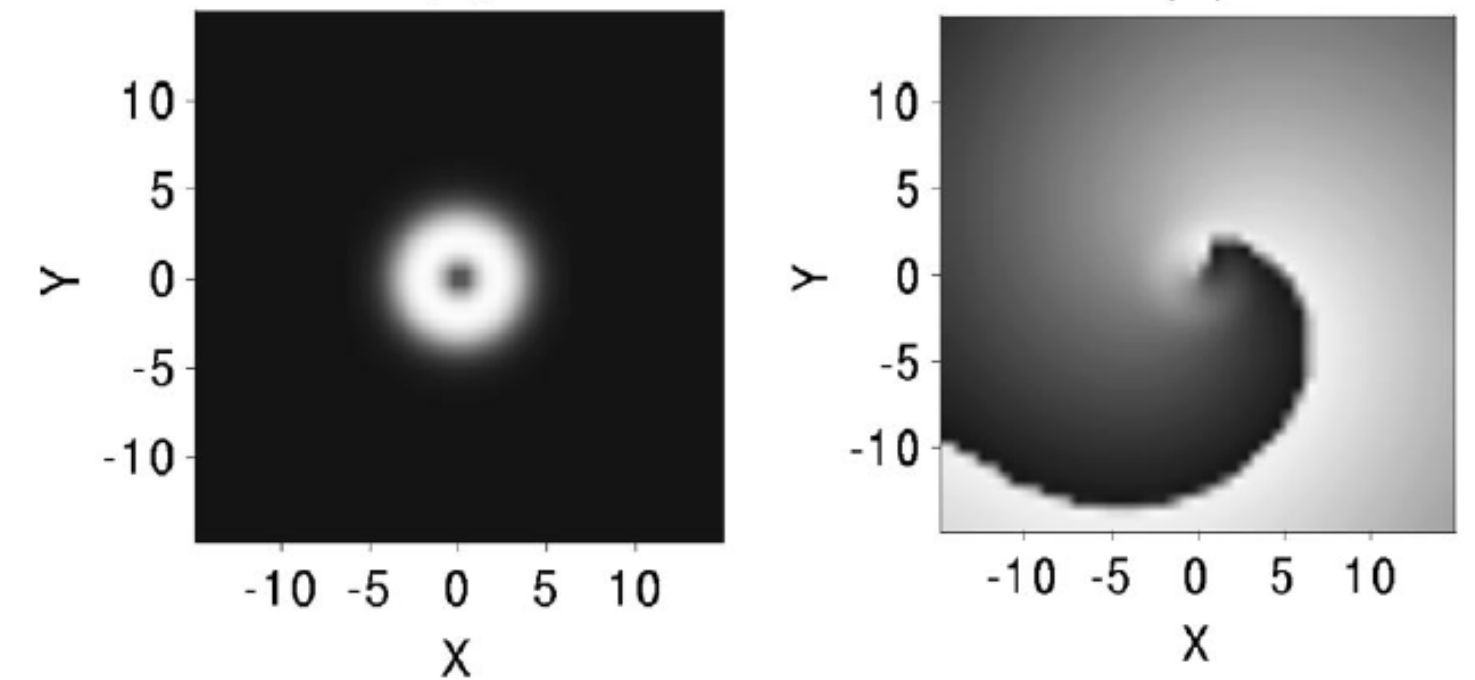}}\subfigure[]{%
\includegraphics[scale=0.5]{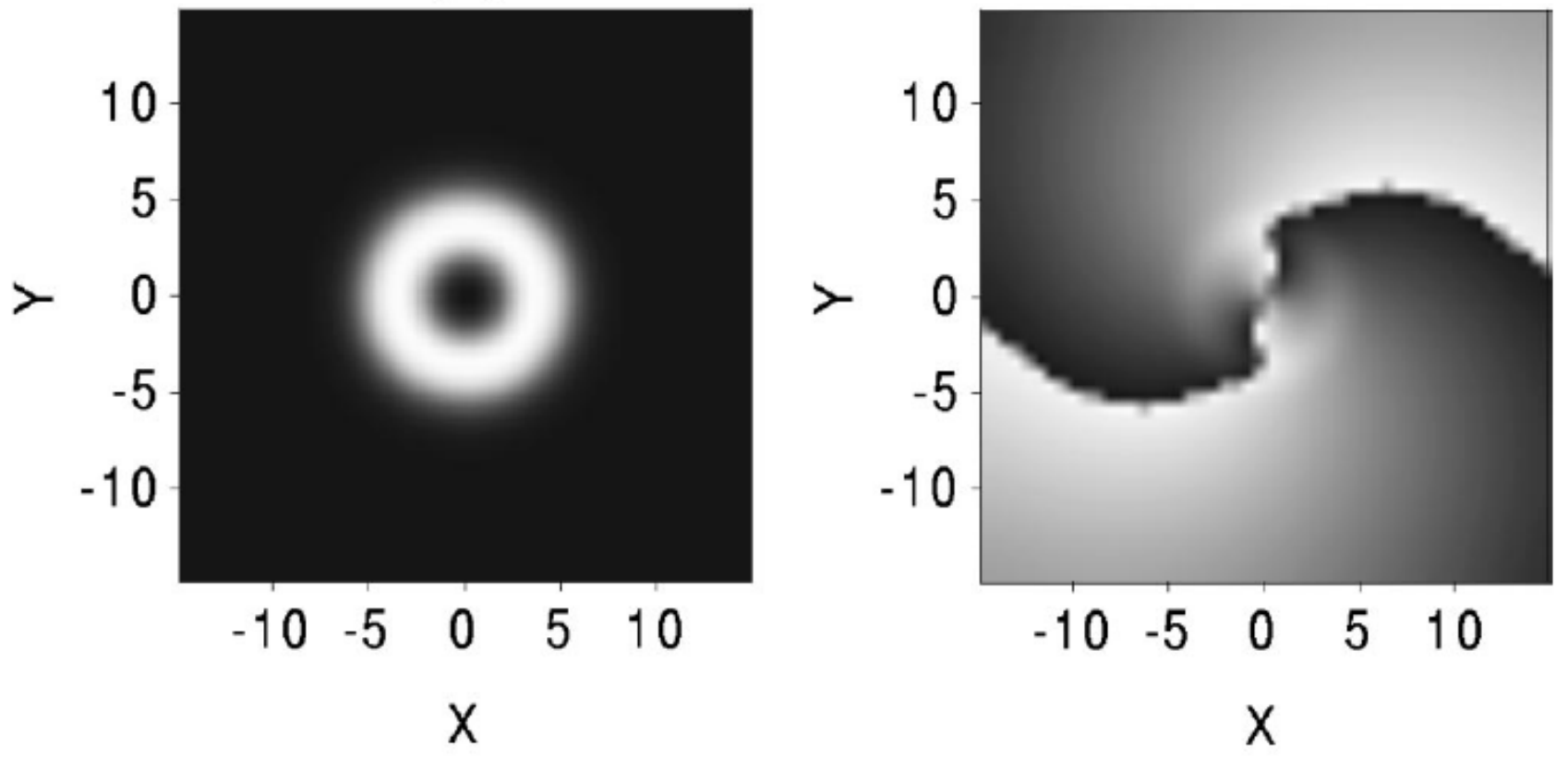}}
\caption{Top views of the amplitude, $\left\vert A(r)\right\vert $, and
phase, defined as per Eq. (6), of the same stable spiral solitons with $S=1$
and $2$ (left and right pairs of panels, respectively) which are presented
in Figs. \protect\ref{fig14.1=fig167} and \protect\ref{fig14.2=fig168}
(source: Ref. \protect\cite{Crasovan-Malomed-Mihalache-2001}).}
\label{fig14.3=fig169}
\end{figure}

Radial profiles of the stable solitons produced in Fig. \ref{fig14.1=fig167}%
, along with their fundamental-vortex counterpart with $S=0$, are displayed
in Fig. \ref{fig14.2=fig168}. Further, for the same VRs with $S=1$ and $2$
top views of the spatial distribution of the amplitude, $\left\vert
A(r)\right\vert $, and phase, which is defined, according to Eq. (4), as%
\begin{equation}
\phi (r,\theta )=S\theta +\arg \left( A(r)\right) ,  \tag{6}
\end{equation}%
are plotted in Fig. \ref{fig14.3=fig169}. Note that, due to\ the presence of
the $r$-dependent term in Eq. (6), constant-phase lines, $\phi (r,\theta )=%
\mathrm{const}$, in Fig. \ref{fig14.3=fig169} are not circumferences, which
they are in VR solutions of the NLS equation with real coefficients, but
\textit{spirals}, therefore the respective VR solutions are categorized as
\textit{spiral solitons}.

\begin{figure}[tbp]
\includegraphics[scale=0.75]{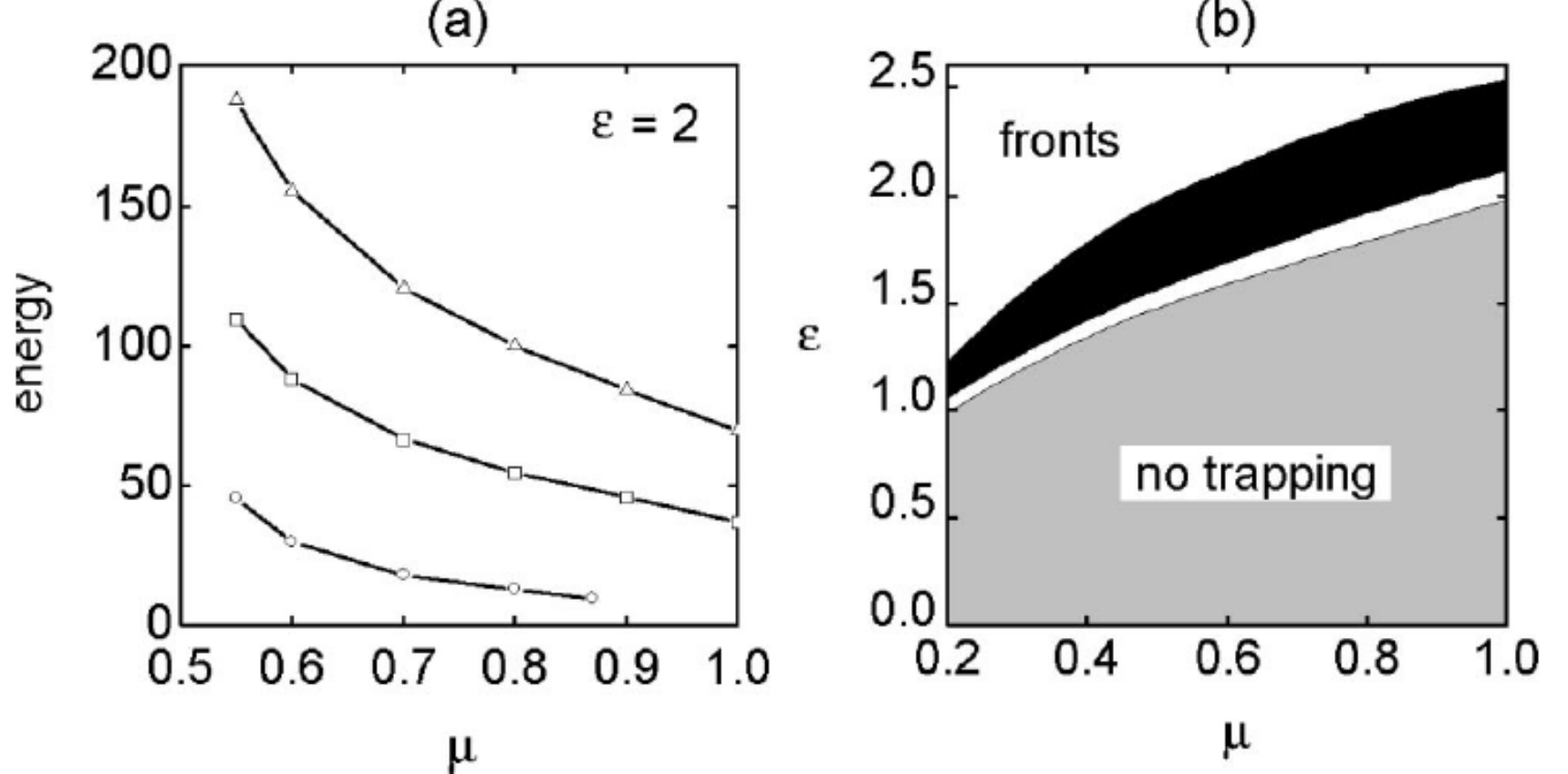}
\caption{(a) The total power of the stable two-dimensional DSs, defined as
per Eq. (7) (here called \textquotedblleft energy") vs. the quintic loss
coefficient $\protect\mu $ for a fixed value of the cubic-gain coefficient
in Eq. (2), $\protect\varepsilon =2.5$. Other parameters are $\protect\delta %
=\protect\beta =0.5$, $\protect\nu =0.1$, and $\protect\sigma =+1$. Chains
of circles, squares, and triangles represent, respectively, the DSs with
winding numbers $S=0$, $1$, and $2$. (b) The existence domain in the
parameter plane ($\protect\mu ,\protect\varepsilon $) for the same stable
solitons, and the same values of $\protect\delta $, $\protect\beta $, and $%
\protect\nu $ as in panel (a). In the black region of full multistability,
the stable solitons with $S=1$, $2$, and $3$ coexist. The narrow lower white
strip is a bistability area, in which their exist stable spiral solitons
with $S=1$ and $2$, while the fundamental ones, with $S=0$, are absent. No
stable solitons exist in the gray region (\textquotedblleft no trapping").
In the upper white region, localized inputs do not self-trap into solitons,
but rather expand in the radial direction in the form of \textquotedblleft
fronts" (source: \protect\cite{Crasovan-Malomed-Mihalache-2001}).}
\label{fig14.4=fig170}
\end{figure}

The results produced by the systematic numerical analysis if Eq. (5) are
summarized by means of the plots displayed in Fig. \ref{fig14.4=fig170}.
Because most important control parameters in this model are the coefficients
of the cubic gain and quintic loss, i.e., $\varepsilon $ and $\mu $,
respectively, and the main characteristic of the 2D solitons is, as usual,
their total norm (or power, in terms of the realization of the model in
optics),%
\begin{equation}
P=2\pi \int_{0}^{\infty }\left\vert A(r)\right\vert ^{2}rdr,  \tag{7}
\end{equation}%
three families of stable DSs, with winding numbers $S=0$, $1$, and $2$, are
presented by means of curves $P(\mu )$ in Fig. \ref{fig14.4=fig170}(a) for
fixed typical values of other parameters. Most essential results are
displayed in Fig. \ref{fig14.4=fig170}(b) by means of the stability chart in
parameter plane $\left( \mu ,\varepsilon \right) $. The stability was
verified by means of direct simulations of the perturbed evolution of the
solitons, as well as through computation of stability eigenvalues for small
perturbations around the stationary soliton solutions \cite%
{Crasovan-Malomed-Mihalache-2001}. As seen in Fig. \ref{fig14.4=fig170}(b),
the stability area expands with the increase of $\mu $ and $\varepsilon $,
and multistability, i.e., coexistence of the stable solitons with different
values of $S$ is a noteworthy feature of the model.

An essential peculiarity of spiral solitons produced by the two-dimensional
CGL equation (2), with $S\geq 1$, is that they are \emph{completely unstable}
against spontaneous splitting in the absence of the dispersive losses
(diffusion), $\beta =0$, while the fundamental DSs, with $S=0$, may be
stable in this case. Indeed, all the stable spiral solitons presented in
Figs. \ref{fig14.1=fig167} -- \ref{fig14.4=fig170} were obtained as
solutions of Eq. (2) with $\beta >0$. The instability of the spiral solitons
in the case of $\beta =0$ is an essential problem, because natural models of
laser cavities, which are an essential source of CGL equations, do not
include the diffusion term, for the reason that light propagates by rays,
but not by diffusion (nevertheless, a diffusion term may appear in an
effective CGL model derived from the system of Maxwell-Bloch equations, as
shown by \cite{Coullet-Gil-Rocca-1989}). It is shown below, in Section V,
that, in the absence of the diffusion term, VR modes may be stabilized by
the HO potential added to the CGL equation \cite{Mihalache-et-al-2010b}, or,
as shown in Section VI, by a spatially periodic potential \cite%
{Leblond-Malomed-Mihalache-2009}. Another possibility to stabilize localized
vortex modes, outlined below in Section IV, is provided by spatial
modulation of the local loss or gain \cite%
{Skarka-et-al-2010,Lobanov-et-al-2011,Kartashov-et-al-2013}.

A well-known feature of conservative NLS equation with the CQ nonlinearity
is that the competition of the cubic self-focusing and quintic defocusing
gives rise to the characteristic flat-top (FT) shape of stable 1D and 2D
solitons. The fundamental and spiral solitons produced by the CGL equation
(2) share this feature: FT profiles are generated by Eq. (2) close to the
boundary between localized and delocalized states, as shown in Fig. \ref%
{fig14.5=fig171}.
\begin{figure}[tbp]
\includegraphics[scale=0.5]{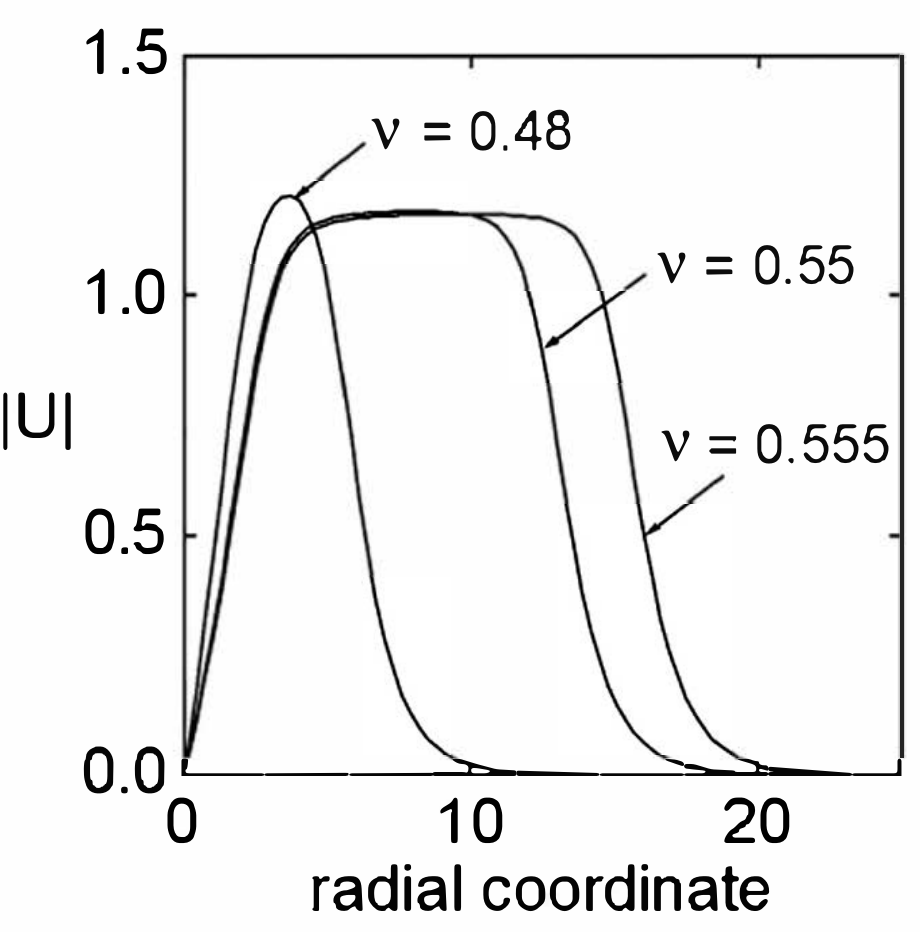}
\caption{The transition from the narrow stable VR radial profile (for $S=1$)
to a broad FT one with the variation of the quintic self-defocusing
coefficient $\protect\nu $ in Eq. (2), while other coefficients are fixed as
$\protect\varepsilon =0.3$, $\protect\delta =0.05$, $\protect\beta =0.03$, $%
\protect\mu =0.2$, and $\protect\sigma =+1$ (source: Ref. \protect\cite%
{Crasovan-Malomed-Mihalache-2001b}).}
\label{fig14.5=fig171}
\end{figure}

\begin{figure}[tbp]
\includegraphics[scale=0.5]{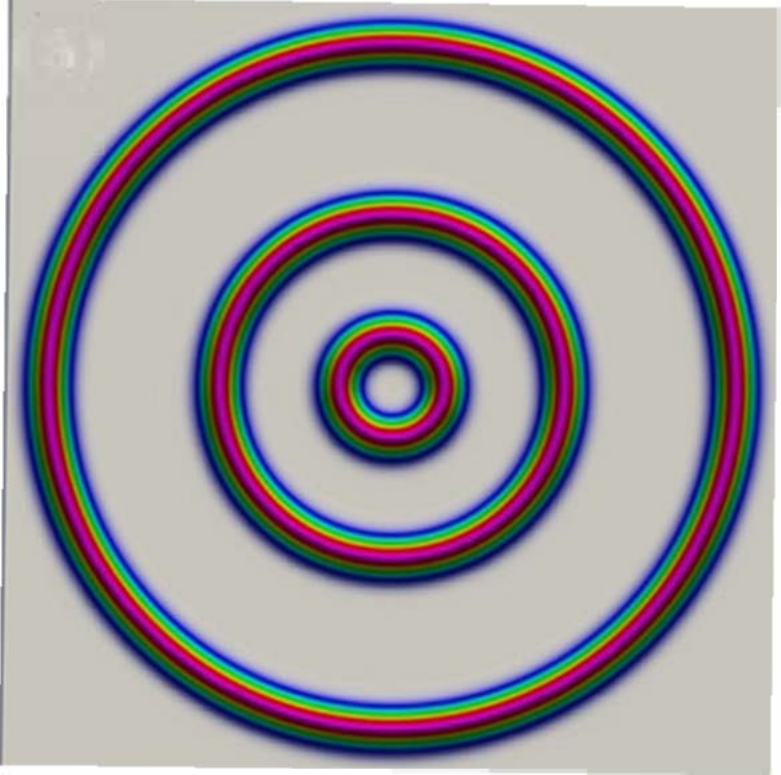}
\caption{An example of a stable nested multi-ring state produced by the
numerical solution of Eq. (2): an inner VR with $S=3$ embedded into the
middle one one with $S=10$, which is embedded into the outer VR with $S=20$
(source: Ref. \protect\cite{Aleksic-et-al-2015}).}
\label{fig14.6=fig172}
\end{figure}

Stable VR solutions with higher values of the winding number, up to $S=20$,
were found by means of numerical solutions of Eq. (2), in combination with
some results produced by the variational approximation based on the complex
Lagrangian formally corresponding to Eq. (2) \cite{Aleksic-et-al-2015}. The
same work reported the existence of stable concentric multi-ring states. In
particular, there are solutions in which inner rings carry VRs with smaller
values of $S$, which are surrounded by very broad rings carrying much larger
$S$, see an example in Fig. \ref{fig14.6=fig172}. Dynamics of complex
multi-vortex patterns was addressed by means of simulations of Eq. (2) in
Ref. \cite{Wu-Wang-2018}.

\begin{figure}[tbp]
\includegraphics[scale=0.5]{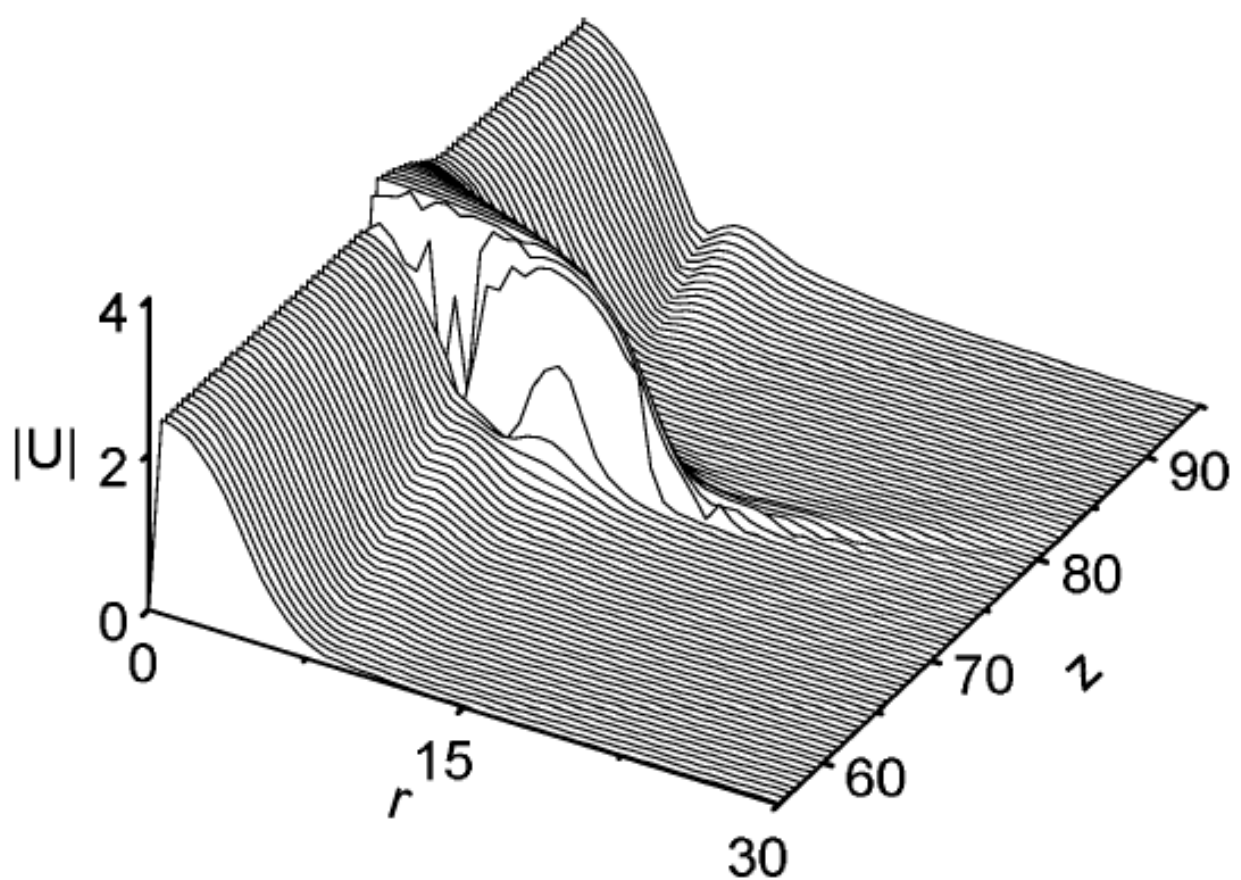}
\caption{An example of a sudden burst in the VR with $S=1$, revealed by
simulations of Eq. (2) with $\protect\delta =\protect\beta =0.05$, $\protect%
\varepsilon =1.75$, $\protect\mu =0.2$, $\protect\nu =1.4$, and $\protect%
\sigma =+1$. The long-time evolution exhibits a chain of bursts at random
moments of time (source: \protect\cite{Crasovan-Malomed-Mihalache-2001b}).}
\label{fig14.7=fig173}
\end{figure}

At parameter values where stationary spiral-vortex solutions of Eq. (2) are
unstable, simulations demonstrate a possibility of robust nonstationary
states. Most typical among them are periodically oscillating breathers,
which were found, e.g., at $\delta =\beta =0.05$, $\varepsilon =0.3$, $\mu
=0.2$, $\nu =2$, and $\sigma =+1$ \cite{Crasovan-Malomed-Mihalache-2001}. A
remarkable dynamical state is an \textit{erupting} one, in the form of a
randomly recurring sudden bursts of the quasi-stationary soliton, as shown
in Fig. \ref{fig14.7=fig173} (such solutions are similar to erupting states
found in the 1D CGL equation \cite{Soto-Crespo-Akhmed-Ank-2000}). An example
of the burst is displayed in Fig. \ref{fig14.6=fig172}.

A special case of the CGL equation is one without dispersive losses, i.e.,
with $\beta =0$ in Eq. (2) (recall that, in this case, only fundamental DSs
may be stable, while all VRs are subject to splitting instability). This
form of the CGL equation keeps the property of the Galilean invariance,
hence a stable fundamental DS can be set in motion with an arbitrary speed
by means of the Galilean boost. In turn, this possibility makes it possible
to explore collisions between moving DSs. Simulations of the collisions
between them demonstrate that, depending on parameters, the collisions may
be quasi-elastic (the solitons separate after the collision, although with
essentially different speeds), inelastic, with the two solitons merging into
a single one, and destructive, leading to complete decay of the solitons
triggered by the collision \cite{Sakaguchi-2005}.

\section{Dissipative vortex solitons in a laser cavity stabilized by the
linearly coupled lossy feedback}

As said above, the necessity to maintain stable zero background around the
DS suggests to keep linear loss, $\delta >0$, in the CGL\ equation (2).
Then, the cubic gain, $\varepsilon >0$, is necessary to compensate the loss,
and the quintic loss, $\mu >0$, must be added to prevent the blowup.
Nevertheless, a system of linearly coupled equations of the CGL type, which
admits the existence of stable DSs in the presence of the usual cubic
dissipation (without \textquotedblleft exotic" quintic terms) was introduced
(originally, in the 1D form) in Ref. \cite{Malomed-Winful-1996}. The
original 1D system governs the copropagation of amplitudes $u(z,\tau )$ and $%
v(z,\tau )$ of optical waves in a dual-core fiber, with inter-core
linear-coupling coefficient $\kappa $. In this setup, one (\textquotedblleft
active") core carries linear gain, accounted for by coefficient $\gamma
_{0}>0$, while the other (\textquotedblleft passive") core is dissipative,
with loss coefficient $\Gamma _{0}>0$. The corresponding CGL system is
\begin{equation}
iu_{z}+\left( \frac{1}{2}-i\gamma _{1}\right) u_{\tau \tau }+(1+i\gamma
_{2})|u|^{2}u-i\gamma _{0}u+\kappa v=0,  \tag{8}
\end{equation}%
\begin{equation}
iv_{z}+\left( \frac{1}{2}-i\gamma _{1}\right) v_{\tau \tau
}+|v|^{2}v+i\Gamma _{0}v+\kappa u=0.  \tag{9}
\end{equation}%
Here, $z$ and $\tau \equiv t-z/V_{\mathrm{gr}}$ are, as usual, the
propagation distance and reduced time, with group velocity $V_{\mathrm{gr}}$
of the carrier wave \cite{KivAgr}, the anomalous-group-velocity-dispersion
(GVD) and Kerr coefficients are scaled to be $1$, while $\gamma _{1}\geq 0$
and $\gamma _{2}>0$ account for the dispersive and nonlinear losses,
respectively. Possible phase- and group-velocity mismatch between the two
cores is not included in Eqs. (8) and (9), but its effect was studied too,
see a review of the DS dynamics in 1D systems of this type in Ref. \cite%
{Malomed-2007}. The competition of the linear gain and loss in the two cores
of the system keeps stability of the zero solution if the linear-loss
coefficient takes values in interval
\begin{equation}
\gamma _{0}<\Gamma _{0}<\kappa ^{2}/\gamma _{0}  \tag{10}
\end{equation}%
(note that the dispersive-loss coefficient $\gamma _{1}$ does not appear in
Eq. (10)). It was demonstrated by means of an analytical perturbation
theory, that may be applied in the case of small coefficients $\gamma
_{0,1,2}$ and $\Gamma _{0}$ \cite{Malomed-Winful-1996}, and in a general
form by means of a numerical solution \cite{Atai-Malomed-1996}, that
condition (10) is compatible with the existence of a pair of DSs produced by
Eqs. (8) and (9), one of which (a higher-amplitude narrow one) is stable,
while a lower-amplitude broad DS is unstable (it plays the role of a
separatrix between attraction basins of the stable zero solution and stable
DS).

The system of the linearly-coupled CGL equations with the cubic-only
nonlinearity was extended into a 2D form and applied to construct stable
two-dimensional DSs, including vortical ones \cite{Paulau-et-al-2011}. The
so defined 2D system is%
\begin{equation}
\frac{\partial E}{\partial t}=(d+iD)\left( \frac{\partial ^{2}}{\partial
x^{2}}+\frac{\partial ^{2}}{\partial y^{2}}\right) E+g_{0}E+g_{2}|E|^{2}E+F,
\tag{11}
\end{equation}%
\begin{equation}
\frac{\partial F}{\partial t}=-\lambda F+\sigma E.  \tag{12}
\end{equation}%
This model governs the time evolution of the optical-field amplitude $E$ in
a laser cavity of the VCSEL type with paraxial diffraction in the plane $%
\left( x,y\right) $, $F$ being the feedback field generated by an external
reflector. Real coefficients $g_{0}$, $\lambda $, $\sigma $, $d$ and the
real part of the complex one $g_{2}$ are counterparts of constants $\gamma
_{0}$, $\Gamma _{0}$, $\kappa ^{2}$, $\gamma _{1}$ and $\gamma _{2}$ in Eqs.
(8) and (9). Unlike Eq. (9), the passive component of the present system,
given by Eq. (12), does not include the diffraction and cubic nonlinearity,
as these are not essential ingredients of the setup.

\begin{figure}[tbp]
\includegraphics[scale=0.83]{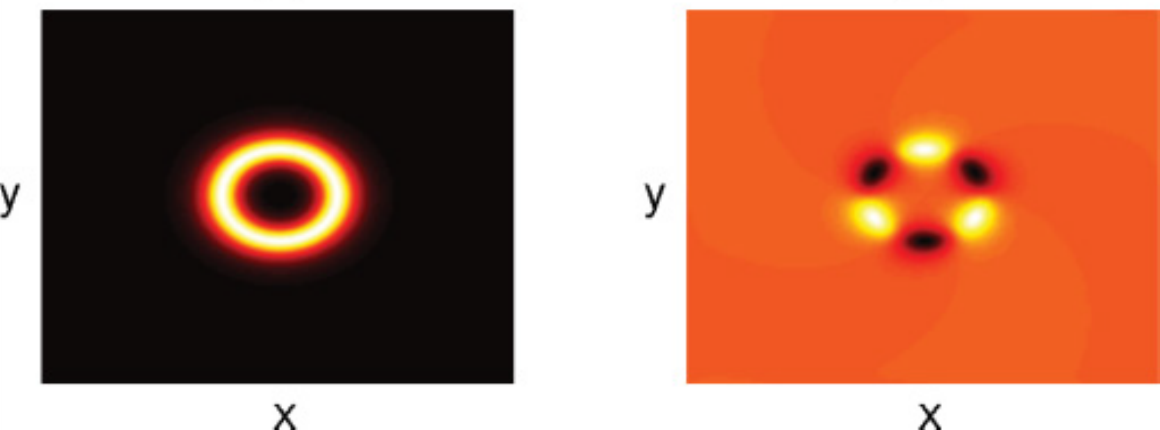}
\caption{Spatial profiles of the absolute value (a) and snapshot of the real
part (b) of field $E(x,y)$ for a stable DS with embedded vorticity $S=3$,
which is produced by numerical solution of the system of coupled CGL
equations (11) and (12). See further details in Ref. \protect\cite%
{Paulau-et-al-2011}.}
\label{fig14.8=fig174}
\end{figure}

Numerical solution of Eqs. (11) and (12) readily create stable
two-dimensional DSs with embedded vorticity $S=0,1,2,3,...$ . An example is
presented in Fig. \ref{fig14.8=fig174}. Additional results demonstrate that,
for fixed parameters of Eqs. (11) and (12), the radius of stable VRs
produced by these equations grows linearly with $S$ \cite{Paulau-et-al-2011}%
.

A similar 2D system, with saturable nonlinearity instead of the cubic term,
was considered in Ref. \cite{Paulau-et-al-2010}. In that case too, stable
vortex DSs were produced by a numerical solution of the system.

\section{Weakly localized modes in media with nonlinear losses (multi-photon
absorption)}

Specific quasi-soliton states are generated by the NLS equation which
includes higher-order nonlinear loss:%
\begin{equation}
U_{z}=i\left( U_{xx}+U_{yy}\right) +i\alpha |U|^{2}U-|U|^{2M-2}U.  \tag{13}
\end{equation}%
Here $\alpha >0$ is an effective Kerr coefficient, and the last term with $%
M=2,3,4,...$ accounts for the $M$-photon absorption in the optical medium.\
It is commonly known that, in terms of the polar coordinates $\left(
r,\theta \right) $, solutions of the linearized equation (13) carrying
vorticity $S$ are given by Bessel functions:%
\begin{equation}
U=b\exp \left( ikz+iS\theta \right) J_{S}\left( \sqrt{-k}r\right) ,  \tag{14}
\end{equation}%
where $k<0$ is the propagation constant, and $b$ is an arbitrary amplitude.
Full equation (13) gives rise to solutions which may be considered as the
Bessel modes distorted by the Kerr self-focusing and nonlinear losses. The
asymptotic form of the solutions at $r\rightarrow 0$ and $r\rightarrow
\infty $ keeps the same form as produced by solution (14) of the linearized
equation:%
\begin{equation}
U\approx b\exp \left( ikz+iS\theta \right) \frac{\left( \sqrt{-k}r\right)
^{S}}{2^{S}S!},  \tag{(15a)}
\end{equation}%
\begin{equation}
U\approx A\exp \left( ikz+iS\theta \right) r^{-1/2}\cos \left( \sqrt{-k}%
r+\delta \right) ,  \tag{15b}
\end{equation}%
where $A$ and $\delta $ are amplitude and phase constants.

The weak localization exhibited by Eq. (15b) cannot correspond to a true
soliton, because it gives rise to divergence of the respective norm
(total-power) integral at $r\rightarrow \infty $:%
\begin{equation}
N=2\pi \int_{0}^{R}\left\vert U(r)\right\vert ^{2}rdr\simeq \pi A^{2}R.
\tag{16}
\end{equation}%
On the other hand, the divergence makes it possible to construct stationary
solutions to Eq. (13), in spite of the \textquotedblleft naive expectation"
that the absence of gain terms in the equation does not allow one to
compensate the loss. The compensation of the loss in the stationary states
generated by Eq. (13) is provided by the influx of power from the infinite
reservoir which the Bessel solution keeps at $r\rightarrow \infty $ \cite%
{Porras-et-al-2004,Porras-and-Ruiz-2014,Porras-et-al-2016}. The local influx
rate is determined by the current vector, which is proportional to the
squared amplitude and gradient of the local phase:%
\begin{equation}
\mathbf{j}=\left\vert U\right\vert ^{2}\nabla \left( \arg (U)\right) .
\tag{17}
\end{equation}%
The corresponding compensation condition is tantamount to a local property
which all stationary solution share:%
\begin{equation}
\nabla \cdot \mathbf{j}+|U|^{2M}=0.  \tag{18}
\end{equation}

It is relevant to stress the difference of this class of weakly localized
solutions and well-localized DSs (such as those displayed, e.g., in Figs. %
\ref{fig14.1=fig167}-\ref{fig14.3=fig169}, \ref{fig14.5=fig171}, \ref%
{fig14.6=fig172}, and \ref{fig14.8=fig174}). As explained above, DSs exist
as isolated states, with uniquely selected values of parameters, including,
in particular, the propagation constant. On the contrary to that, the
solutions to Eq. (13), specified by b.c. (15a) and (15b), exist for all
values of $k<0$.

\begin{figure}[tbp]
\includegraphics[scale=0.8]{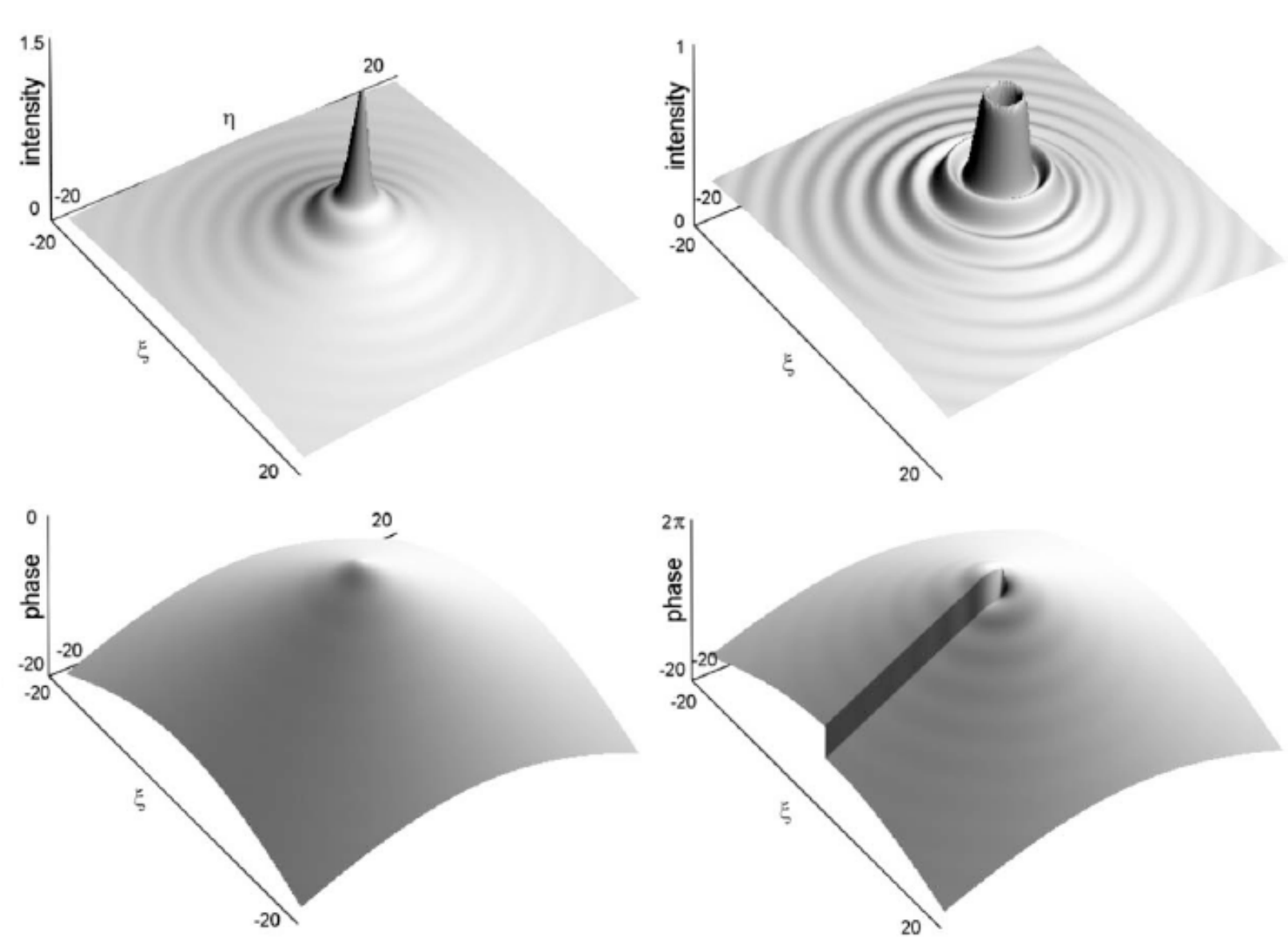}
\caption{The left and right columns show the distribution of the local
intensity, $|U|^{2}$ (top panels), and phase, $\arg (U)$ (bottom panels), in
stable weakly localized (Bessel-like) modes with $S=0$ and $S=1$,
respectively. The modes are produced by numerical solution of Eq. (13) with $%
M=4$ and $\protect\alpha =0$ (i.e., the single nonlinear terms represents
the higher-order loss). The solutions with $S=0$ and $S=1$ correspond,
respectively, to $b=1.174$ and $1.666$ in Eq. (15a). The Cartesian
coordinates are denoted here as $\left( \protect\xi ,\protect\eta \right) $
(source: Ref. \protect\cite{Porras-and-Ruiz-2014}).}
\label{fig5.9=fig175}
\end{figure}

\begin{figure}[tbp]
\includegraphics[scale=0.8]{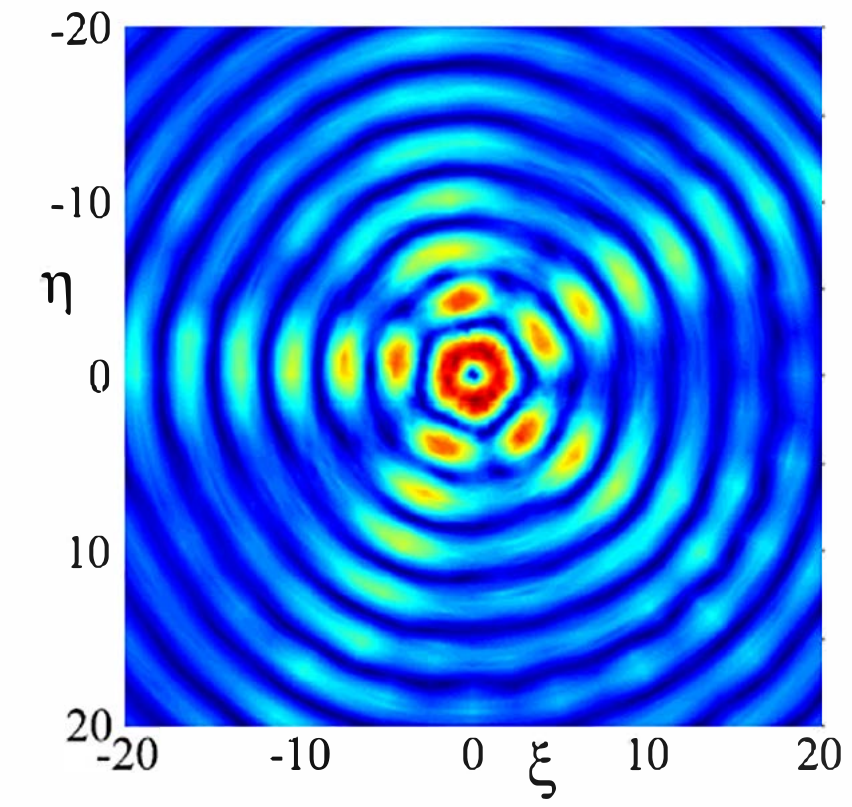}
\caption{A rotating fragmented pattern produced by the development of the
azimuthal instability of a weakly localized state with vorticity $S=1$. The
result is produced by simulations of Eq. (13) with $M=4$ and $\protect\alpha %
=1.6$. The unstable state is defined by Eq. (15a) with $b=1.6$ (source: Ref.
\protect\cite{Porras-et-al-2016}).}
\label{fig5.10=fig176}
\end{figure}

These weakly localized solutions with the divergent total power may be
stable or unstable, depending on parameters. Examples of stable states with $%
S=0$ and $1$ are presented in Fig. \ref{fig5.9=fig175}. Stable stationary
states with $S=0$ were observed in an experiment which dealt with light
propagation in water \cite{Porras-et-al-2004}.
Those weakly localized vortex states which are unstable are split by
azimuthal perturbations into steadily rotating fragmented modes (which are
similar to \textit{azimuthons}, which were originally introduced in
conservative systems \cite{Des-Sukh-Kiv-2005}, and then extended to
dissipative ones \cite{azim-dissip-1, azim-dissip-2}). An example is
presented in Fig. \ref{fig5.10=fig176}.

\begin{figure}[tbp]
\includegraphics[scale=0.6]{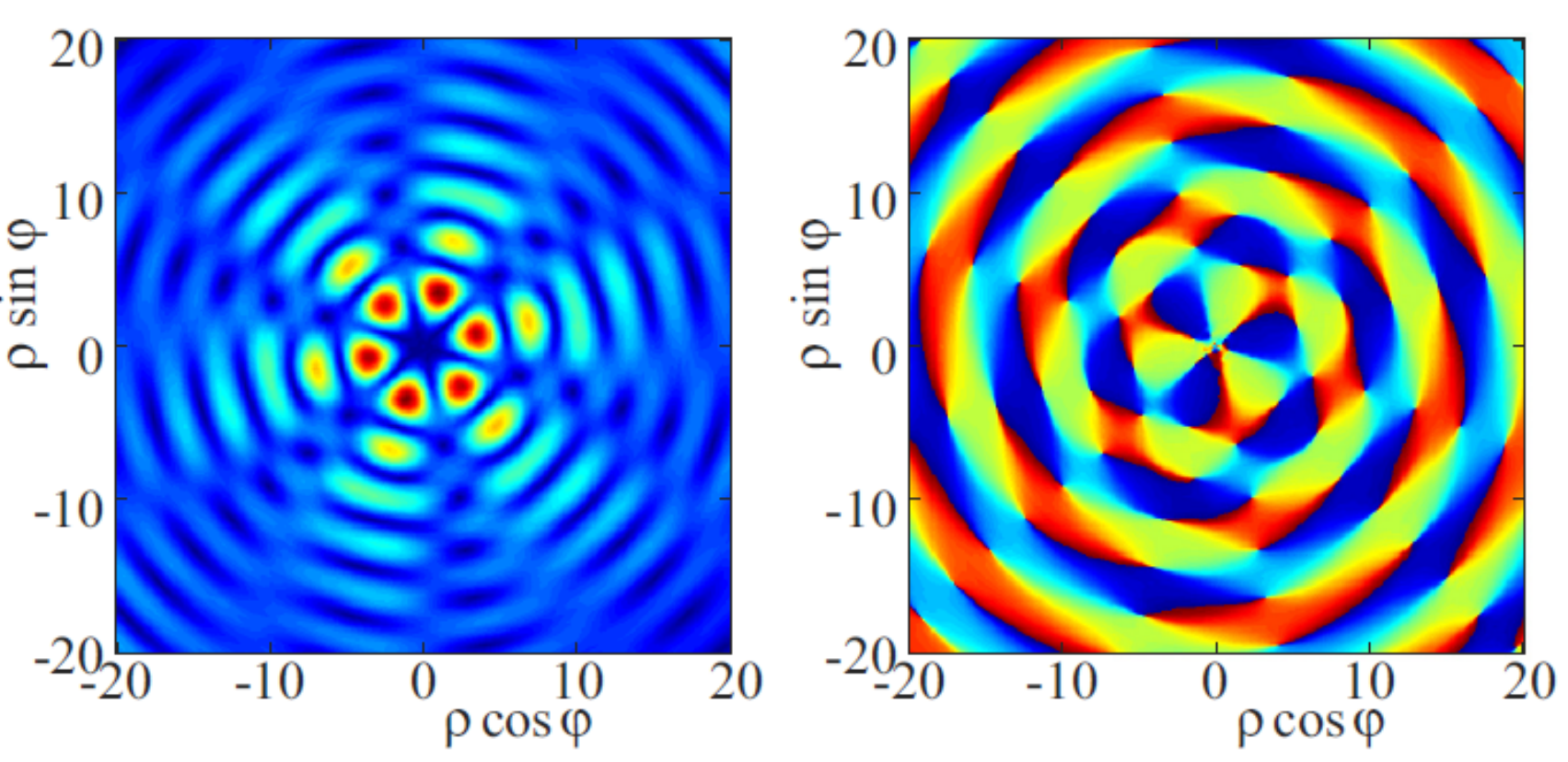}
\caption{The intensity and phase distribution in a steadily rotating
azimuthon produced by the numerical solution of Eq. (13) with $M=4$ and $%
\protect\alpha =1.6$. The input was taken as a pair of Bessel beams
corresponding to $S=\pm 1$, with $b=1.6$ in Eq. (15a) (source: Ref.
\protect\cite{Ruiz-et-al-2020}). }
\label{fig14.11=fig177}
\end{figure}

Numerical solution of Eq. (13) produces stable steadily rotating azimuthons
of a different type, which are generated by an input in the form of two
slightly mismatched Bessel beams with opposite vorticities, $\pm S$ \cite%
{Ruiz-et-al-2020}. In their established form, the azimuthons also represent
a species of stable solutions of Eq. (13). An example, fragmented into six
segments, is displayed in Fig. \ref{fig14.11=fig177}.

\section{Localized states in 2D nonlinear dissipative media with spatially
modulated gain or loss}

The use of spatially nonuniform gain or loss offers possibilities for the
creation and stabilization of various nonlinear states in the 2D geometry
which do not exist or are unstable in uniform dissipative media. This
section aims to report some essential findings produced by analysis of 2D
CGL models of this type.

\subsection{Vortex solitons supported by the 2D cubic-quintic CGL\ equations
with spatially modulated linear losses}

As mentioned above, the CGL equation (2) with constant coefficients and
without the diffusion term ($\beta =0$) cannot support stable VRs. It was
shown \cite{Skarka-et-al-2010} that vortex DSs may be stabilized by
introducing radial modulation of the linear loss in Eq. (2):
\begin{equation}
\frac{\partial E}{\partial z}=-\delta (r)E+\frac{i}{2}\left( \frac{\partial
^{2}}{\partial x^{2}}+\frac{\partial ^{2}}{\partial y^{2}}\right) E+\left(
\varepsilon +i\right) |E|^{2}E-\left( \mu +i\nu \right) |E|^{4}E,  \tag{19}
\end{equation}%
where the modulation format is chosen so as to have the minimum loss at $r=0$%
,%
\begin{equation}
\delta (r)=\gamma +Vr^{2},~\mathrm{with}~\ \gamma >0,V>0.~  \tag{20}
\end{equation}

\begin{figure}[tbp]
\includegraphics[scale=0.6]{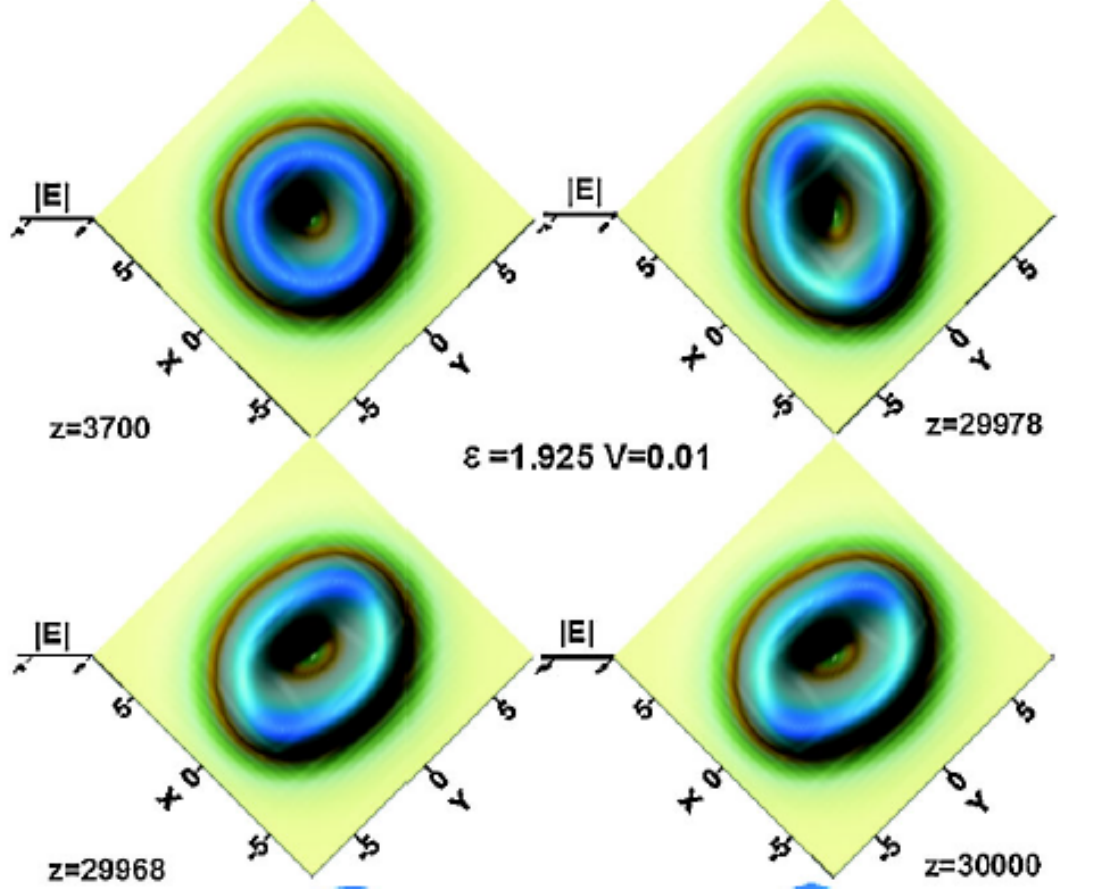}
\caption{A stably rotating elliptic vortex, produced by simulations of Eq.
(19) with $\protect\gamma =0.29$, $V=0.01$, $\protect\varepsilon =1.925$, $%
\protect\mu =1.4$, $\protect\nu =0.4$ The configurations displayed at $%
z=29968$ and $z=30000$ correspond to the rotation of the ellipse by $180^{0}$%
, hence the rotation period is $\Delta z=64$ (source: Ref. \protect\cite%
{Skarka-et-al-2010}). }
\label{fig14.12=fig178}
\end{figure}

Systematic numerical analysis of the model based on Eqs. (19) and (20)
(which was combined with the variational approximation based on the use of
the formal complex Lagrangian of Eq. (19)) gives rise to several species of
\emph{stable} localized states with embedded vorticity, depending on values
of parameters in Eqs. (19) and (20). First, these may be usual vortex
solitons of the crater-shaped-vortex (CSV) type. More interesting are
self-trapped vortex modes which spontaneously develop an elliptic shape,
that performs rotation at a constant angular velocity around $r=0$, see an
example in Fig. \ref{fig14.12=fig178}.

\begin{figure}[tbp]
\includegraphics[scale=0.6]{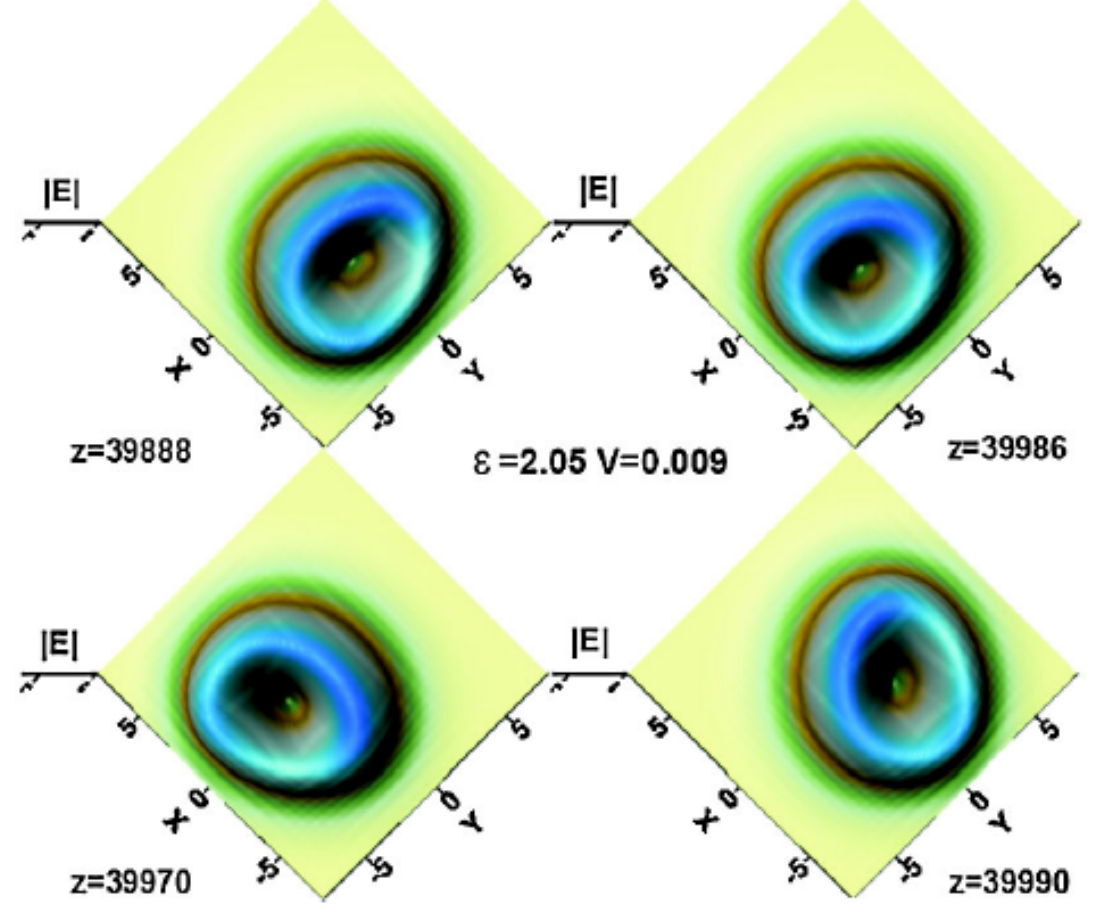}
\caption{A stable vortex with the eccentric shape, which combines inner
spinning and the orbital motion around the center. The solution is produced
by simulations of Eq. (19) with $\protect\gamma =0.29$, $V=0.09$, $\protect%
\varepsilon =2.05$, $\protect\mu =1.4$, $\protect\nu =0.4$ The
configurations displayed at $z=29968$ and $z=30000$ correspond to the
rotation of the ellipse by $180^{0}$ (source: Ref. \protect\cite%
{Skarka-et-al-2010}).}
\label{fig14.13=fig179}
\end{figure}

Another stable dynamical mode produced by simulations of Eq. (19) is a still
stronger deformed elliptic vortex, which develops an eccentric shape
(shifted off the pivot, $r=0$), as shown in Fig. \ref{fig14.13=fig179}. This
vortex features superposition of two modes of rotational motion: spinning
around its own center, and precession of the center around the point $r=0$,
with the respective periods locked by ratio $1:4$.

\begin{figure}[tbp]
\includegraphics[scale=0.6]{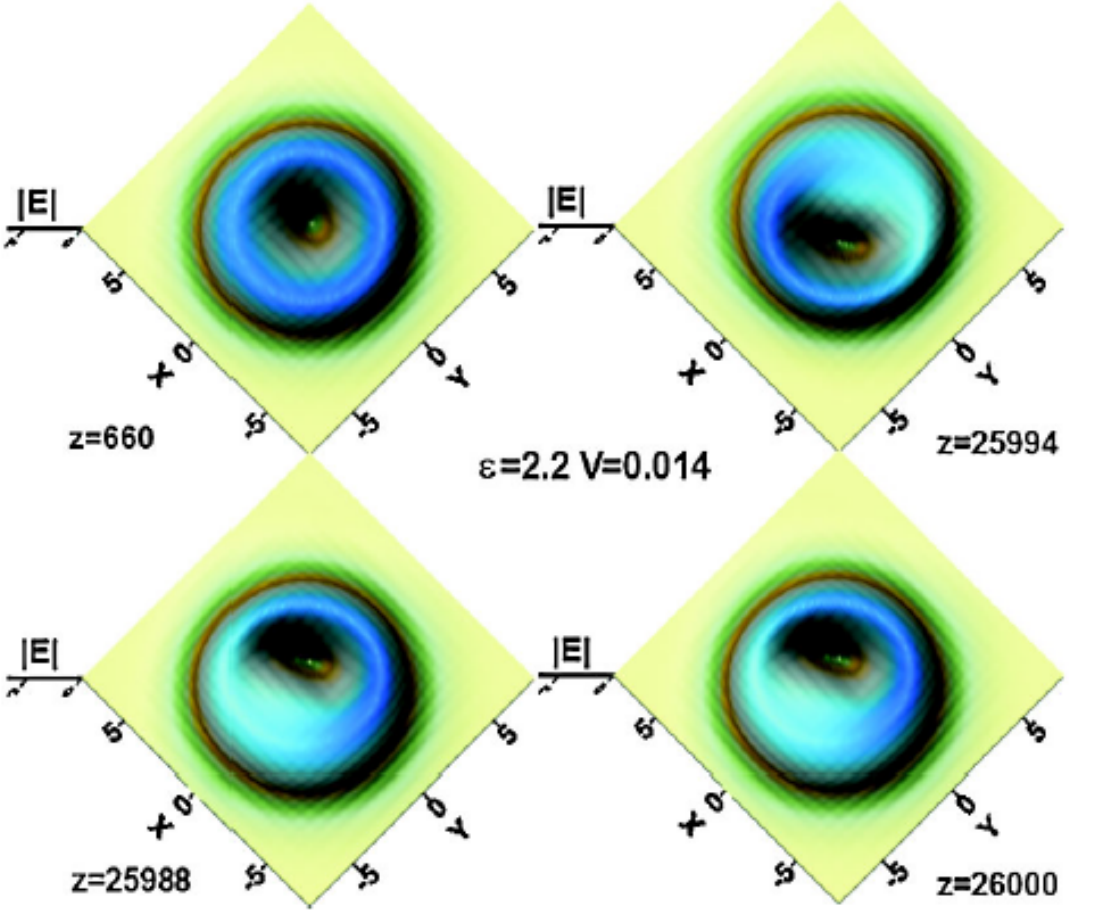}
\caption{A stable crescent-shaped spinning vortex, which performs orbital
motion around the center. The solution is produced by simulations of Eq.
(19) with $\protect\gamma =0.29$, $V=0.014$, $\protect\varepsilon =2.2$, $%
\protect\mu =1.4$, $\protect\nu =0.4$ (source: \protect\cite%
{Skarka-et-al-2010}).}
\label{fig14.14=fig180}
\end{figure}

A still stronger deformation transforms the original isotropic vortex into
an internally spinning crescent-shaped one, which also performs precession
at a constant angular velocity around the central point. An example of this
stable mode is displayed in Fig. \ref{fig14.14=fig180},

\subsection{Complex modes supported by the 2D cubic-quintic CGL equation
with localized linear gain\newline
}

A natural extension of the above analysis is based on Eq. (19) with
modulation format%
\begin{equation}
\delta (r)=\gamma -\Gamma r^{2}.  \tag{21}
\end{equation}%
Unlike the format defined above by Eq. (20), expression (21) implies the
presence of the linear gain in the circle,
\begin{equation}
r^{2}<\gamma /\Gamma .  \tag{22}
\end{equation}%
Results of simulations of this model, based on Eqs. (19) and (21), were
reported in Ref. \cite{Skarka-et-al-2014}, fixing, in particular,%
\begin{equation}
\gamma =0.08,\mu =1.4,\nu =0.4,  \tag{23}
\end{equation}%
and varying parameters $\varepsilon $ (the strength of the cubic gain) and $%
\Gamma $ (thus changing the size of the gain area, see Eq. (22)). Actually, $%
\varepsilon $ and $\Gamma $ are the most essential parameters which
determine the selection of various stable states.

\begin{figure}[tbp]
\includegraphics[scale=0.6]{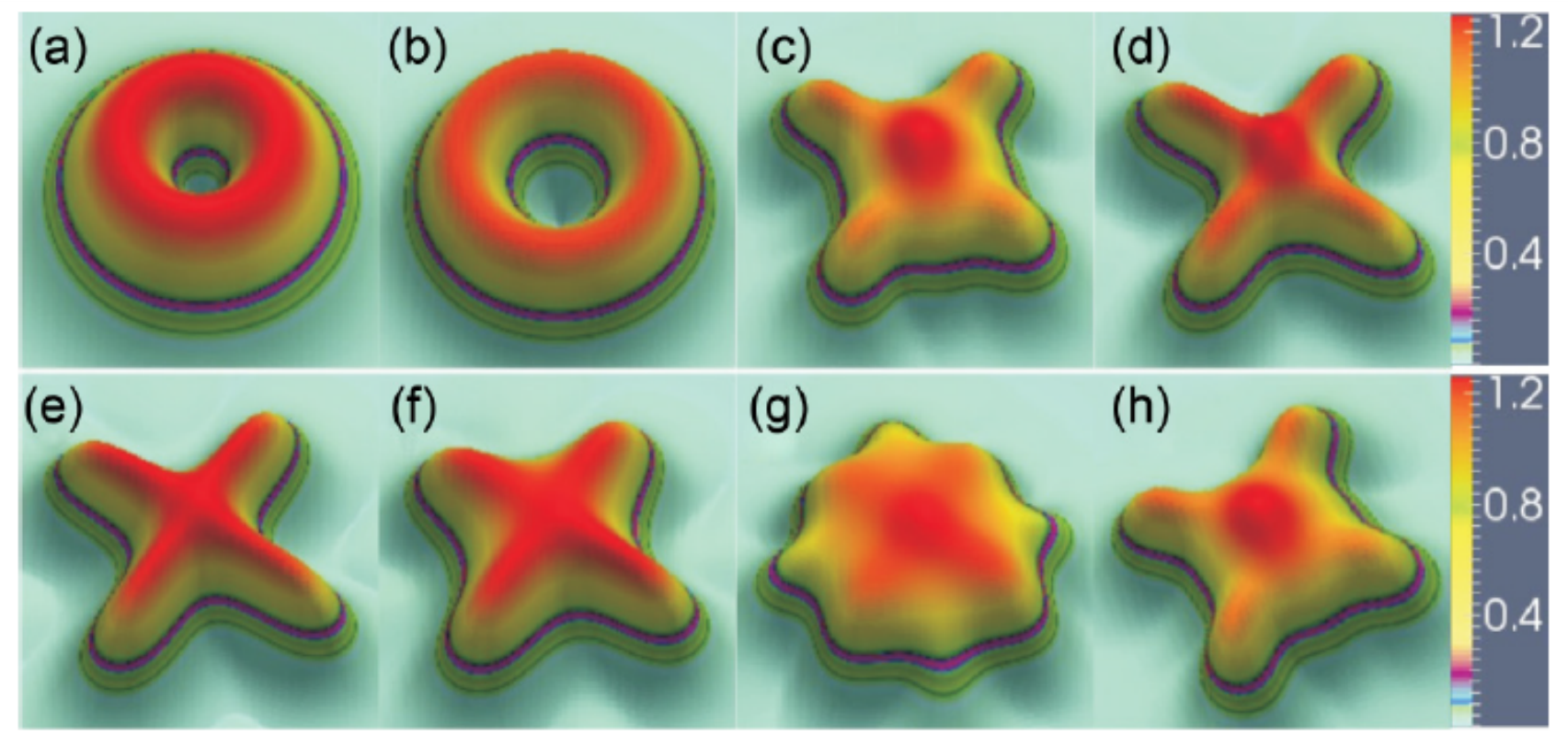}
\caption{The evolution of the VR input (panel (a)) governed by Eqs. (19) and
(21) with parameters given by Eq. (23) and $\protect\varepsilon =1.7,\Gamma
=0.018$. First, a vortex soliton is formed, in panel (b). Then, the
azimuthal instability breaks the isotropy of the vortex, expels the
vorticity, and produces a \textquotedblleft Celtic cross" in (c). This is
followed by a periodic chain of metamorphoses. Having passed half a period, $%
\Delta z=200$, the pattern recovers the Celtic-cross shape (source:
\protect\cite{Skarka-et-al-2014}).}
\label{fig14.15=fig181}
\end{figure}

The analysis has revealed a great variety of robust localized patterns.
First, in a large part of the parameter plane $\left( \varepsilon ,\Gamma
\right) $, the model produces elliptically deformed rotating vortices
similar to those found in the case of the modulation format (20), such as
crescent-shaped ones (cf. Fig. \ref{fig14.14=fig180}). More interesting are
star-shaped modes into which the vortex input is transformed by the
azimuthal modulational instability. In particular, at $\varepsilon =1.7$, $%
\Gamma =0.018$ the instability development expels the vortex' pivot (phase
singularity), and thus converts the pattern into a cross-shaped one
(four-armed star), as shown in Fig. \ref{fig14.15=fig181}. The star performs
a cycle of metamorphoses, with half period $\Delta z\approx 200$, as shown
in the figure. The computation of the the angular momentum of the pattern,
defined by the standard expression, $M=-i\int \int dxdyE^{\ast }\left(
x\partial E/\partial y-y\partial E/\partial x\right) $ (the asterisk stands
for the complex conjugate), demonstrates that, in the course of the cyclic
evolution, it varies symmetrically between positive and negative values, in
an interval of $-0.18<M<+0.18$ (i.e., the memory of the initial sign of the
angular momentum, carried by the VR in Fig. \ref{fig14.15=fig181}(b), is
completely lost).

\begin{figure}[tbp]
\includegraphics[scale=0.6]{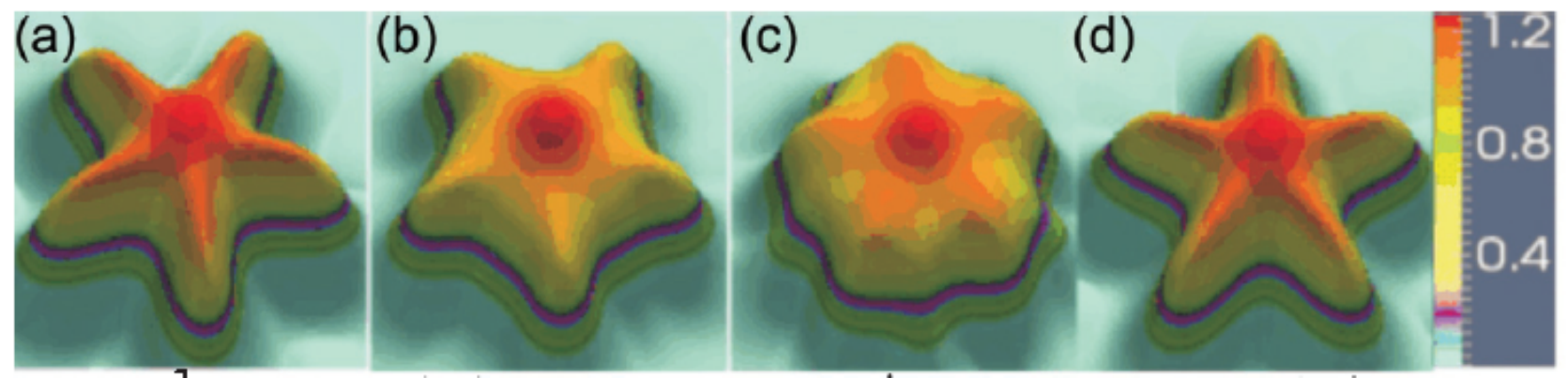}
\caption{The periodic chain of metamorphoses of the five-armed star,
produced by simulations of Eq. (19) with parameters given by Eqs. (21) and $%
\protect\varepsilon =1.7,\Gamma =0.014$. The period is $\Delta z=40$. In the
course of the evolution the start rotates, so that its angular momentum
takes values in interval (24) (source: \protect\cite{Skarka-et-al-2014}).}
\label{fig14.16=fig182}
\end{figure}

Decreasing $\Gamma $ to $0.014$, the simulations produce a periodically
transforming pattern in the form of a five-armed star, which is displayed in
Fig. \ref{fig14.16=fig182}. In this case, the periodic metamorphoses occur
simultaneously with steady rotation of the star (the cross-shaped pattern
observed in Fig. \ref{fig14.15=fig181} does not rotate). Accordingly, this
dynamical state has a definite sign of the angular momentum, which varies in
interval%
\begin{equation}
-9.7<M<-8.3.  \tag{24}
\end{equation}%

\begin{figure}[tbp]
\includegraphics[scale=0.6]{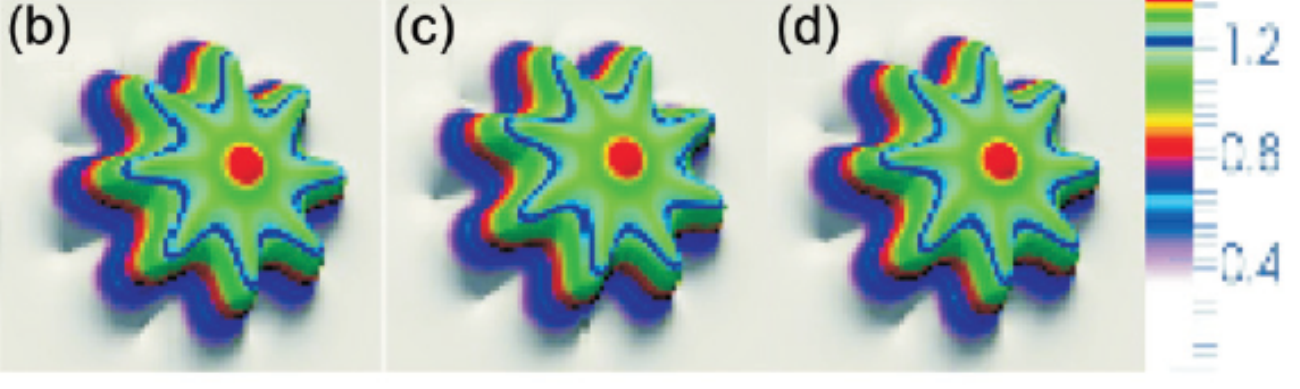}
\caption{Steady rotation of the stable pattern shaped as the eight-arned
star (without intrinsic metamorphosis, unlike the situations displayed in
Figs. \protect\ref{fig14.15=fig181} and \protect\ref{fig14.16=fig182}),
produced by simulations of Eq. (19) with parameters given by Eqs. (21) and $%
\protect\varepsilon =1.75,\Gamma =0.007$. The rotation period is $\Delta
z=192$ (source: \protect\cite{Skarka-et-al-2014}).}
\label{fig14.17=fig183}
\end{figure}

Further decrease of $\Gamma $ gives rise to star-shaped patterns with the
number of arms from six to ten, see an example of the eight-armed star in
Fig. \ref{fig14.17=fig183}. All these higher-order modes with the azimuthal
structure feature only steady rotation, without a change in their shapes,
unlike the metamorphoses exhibited by the four- and five-armed stars.

\subsection{Stable vortices supported by the cubic CGL equation with
localized gain applied along a ring}

\begin{figure}[tbp]
\includegraphics[scale=0.6]{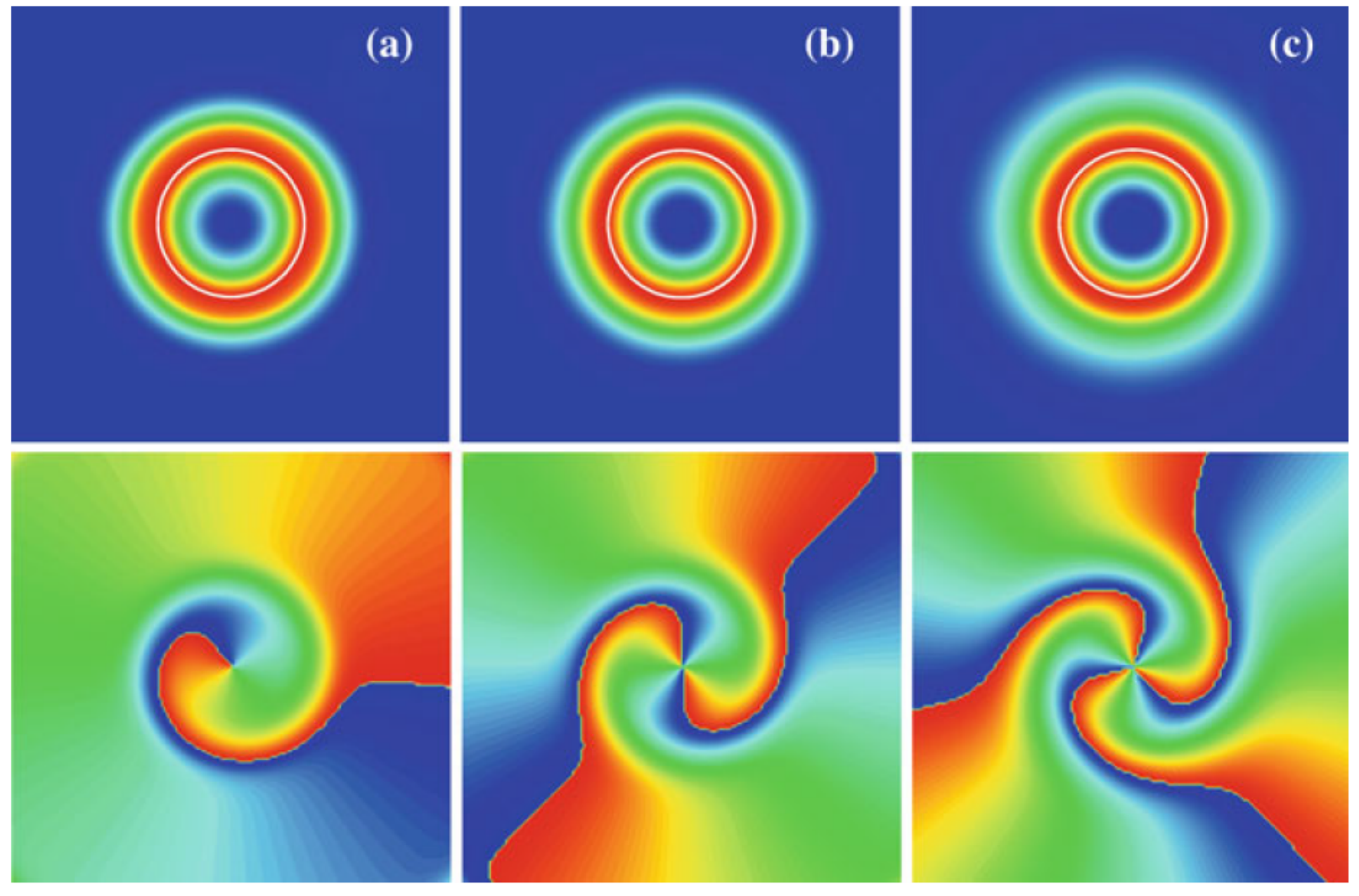}
\caption{Distributions of $|U(x,y)|$ and $\arg (U(x,y))$ (the phase of the
complex field) in stable vortex solitons with winding numbers $S=1$ (a), $%
S=2 $ (b), and $S=3$ (c), as produced by numerical solution of Eq. (25).
White rings designate the location of the local gain maximum, $r=r_{0}$. The
parameters are $\protect\alpha =2$, $\protect\gamma =3,$ $r_{0}=5.25,$ and $%
d=1.75$ (source: \protect\cite{Lobanov-et-al-2011}).}
\label{fig14.18=fig184}
\end{figure}

It is also natural to consider a possibility to maintain stable localized
vortex states by means of localized gain applied along a ring with a finite
radius, rather than at the center. This possibility was considered \cite%
{Lobanov-et-al-2011} in the framework of the cubic (rather than
cubic-quintic, cf. Eq. (19) CGL equation:%
\begin{equation}
iU_{z}+\frac{1}{2}\left( U_{xx}+U_{yy}\right) +|U|^{2}U=i\gamma \exp \left( -%
\frac{(r-r_{0})^{2}}{d^{2}}\right) U-i\alpha |U|^{2}U.  \tag{25}
\end{equation}%
The results is that vortices attached to the gain-carrying ring may be
stable if $\gamma $ in Eq. (25) exceeds a certain threshold value, which, in
turn, strongly grows with the increase of vorticity $S$. Examples of stable
vortical modes with winding numbers $S=1,2,$ and $3$ are displayed in Fig. %
\ref{fig14.18=fig184}.

Stable VRs supported by a similar ring-shaped local gain configuration, but
in the system of two dissipative equations for the fundamental-frequency and
second-harmonic waves coupled by the $\chi ^{(2)}$ (quadratic) nonlinearity
were recently reported too \cite{Lobanov-et-al-2022}.

\section{Two-dimensional DSs (dissipative solitons) in the HO
(harmonic-oscillator) trapping potential}

\subsection{The CGL equation with the CQ nonlinearity and linear loss}

As mentioned above, the CGL equation (2), which is relevant as a model in
optics, does not include the diffusion term, i.e., it has $\beta =0$. Then,
the problem is that vortex-soliton solutions of such an equation with
constant coefficients are completely unstable. In addition to what is
considered above (spatial modulation of the linear loss/gain),
dissipative-VR solutions can be stabilized by means of a trapping potential
\cite{Mihalache-et-al-2010b,Kalashnikov-and-Wabnitz-2021}, the most natural
form of which is the HO potential,%
\begin{equation}
U(r)=\left( \Omega ^{2}/2\right) r^{2}.  \tag{26}
\end{equation}%
Thus, the corresponding form of Eq. (19), which includes potential (26), is
\begin{equation}
\frac{\partial E}{\partial z}=-\delta \cdot E+\frac{i}{2}\left( \frac{%
\partial ^{2}}{\partial x^{2}}+\frac{\partial ^{2}}{\partial y^{2}}\right)
E+\left( \varepsilon +i\right) |E|^{2}E-\left( \mu +i\nu \right) |E|^{4}E-%
\frac{i}{2}\Omega ^{2}r^{2}E.  \tag{27}
\end{equation}%
As shown in Ref. \cite{Mihalache-et-al-2010b}, adequate results can be
obtained, in particular, by fixing%
\begin{equation}
\delta =0.5,\mu =1,\nu =0.1,  \tag{28}
\end{equation}%
while the most important parameters, \textit{viz}., the cubic gain $%
\varepsilon $ and HO-trap strength $\Omega ^{2}$, are subject to variation.

Stationary axisymmetric solutions of Eq. (27) with a real propagation
constant $k$ and integer vorticity $S$ can be looked for in the usual form,
using the polar coordinates $\left( r,\theta \right) $:%
\begin{equation}
E(x,y,z)=\exp \left( ikz+iS\theta \right) U(r),  \tag{29}
\end{equation}%
where complex function $U(r)$ satisfies the radial equation,%
\begin{equation}
\left( \delta +ik+\frac{i}{2}\Omega ^{2}r^{2}\right) U=\frac{i}{2}\left(
\frac{d^{2}}{dr^{2}}+\frac{1}{r}\frac{d}{dr}-\frac{S^{2}}{r^{2}}\right)
U+\left( \varepsilon +i\right) |U|^{2}U-\left( \mu +i\nu \right) |U|^{4}U.
\tag{30}
\end{equation}%
Appropriate numerical solutions of both equations (27) and (30) could be
found starting with a straightforward form of the input,%
\begin{equation}
U_{0}(r)=A_{0}r^{|S|}\exp \left( -r^{2}/w_{0}^{2}\right) ,  \tag{31}
\end{equation}%
with amplitude $A_{0}$ and $w_{0}$. The stationary solutions are
characterized, as usual, by values of the integral power (norm),%
\begin{equation}
P=2\pi \int_{0}^{\infty }\left\vert U(r)\right\vert ^{2}rdr.  \tag{32}
\end{equation}%
Values of $k$ were found as eigenvalues at which Eq. (30) produces
appropriate solutions for $U(r)$.

\begin{figure}[tbp]
\includegraphics[scale=0.5]{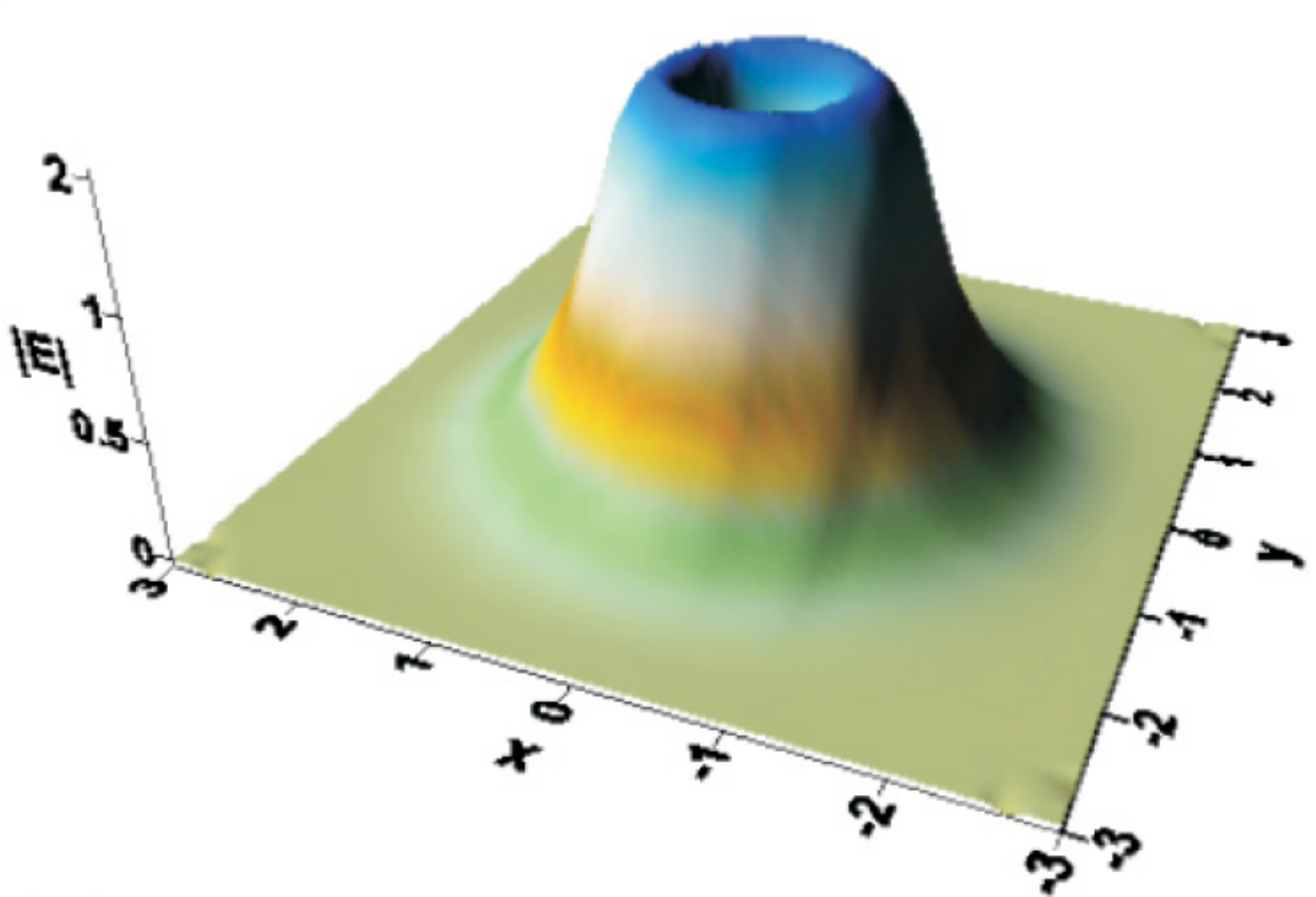}
\caption{The profile of $\left\vert E(x,y)\right\vert $ for the stable
stationary solution of Eq. (27) in the form of the CSV (crater-shaped
vortex), with winding number $S=1$, see Eq. (29)). The parameters are taken
as per Eq. (28), along with $\protect\varepsilon =2.22$ and $\Omega
=\allowbreak 1.7$ (source: \protect\cite{Mihalache-et-al-2010b}).}
\label{fig14.19=fig185}
\end{figure}

Numerical solutions with $S=1$ readily produce stable solutions in the form
of CSVs, see a typical example in Fig. \ref{fig14.19=fig185}.
The amplitude and phase structures of the stable CSV with $S=1$ are
displayed, at slightly different values of parameters, in Fig. \ref%
{fig14.20=fig186}. As expected, its shape shows a spiral structure.
\begin{figure}[tbp]
\includegraphics[scale=0.6]{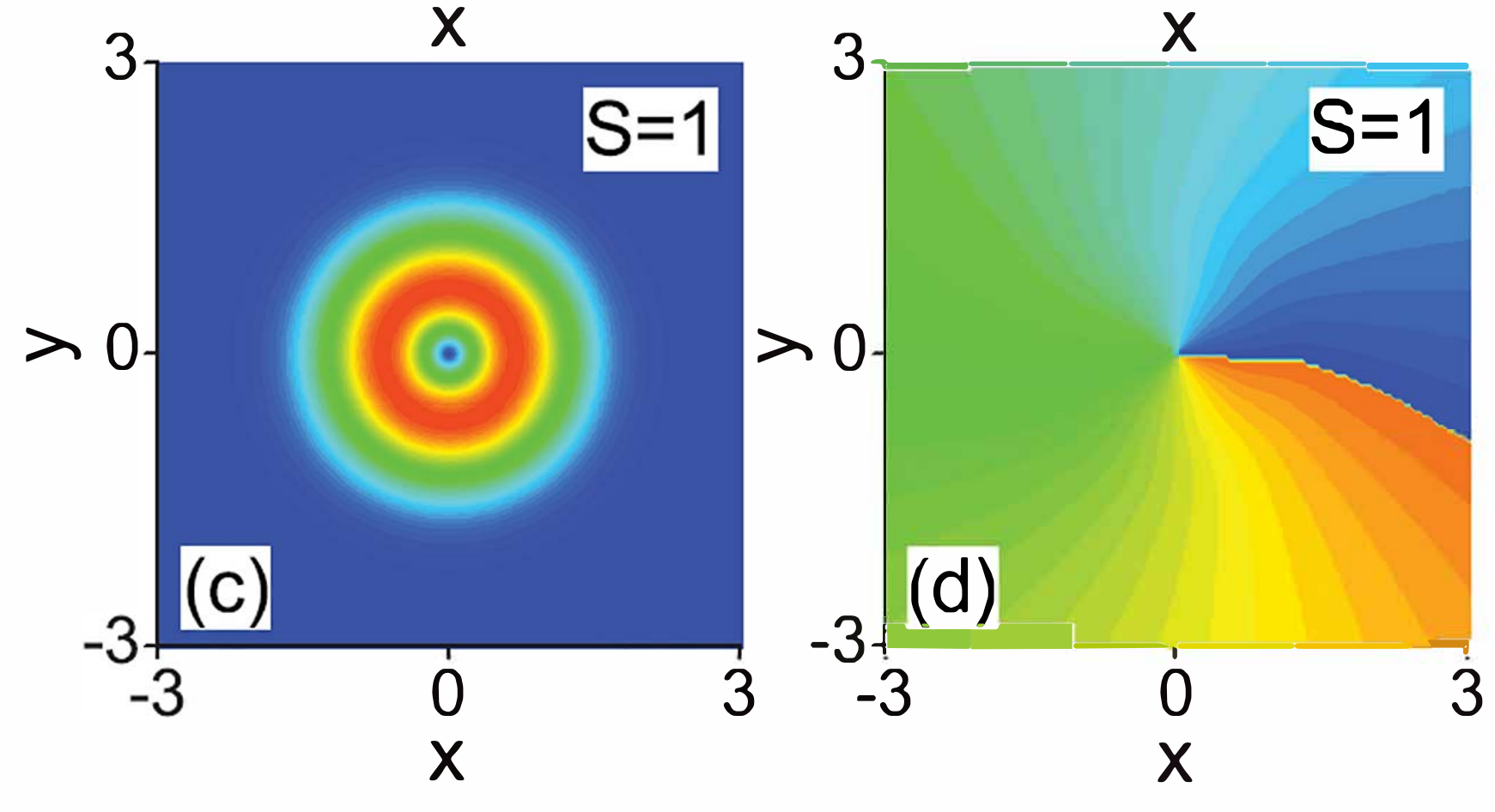}
\caption{The left and right panels display, respectively, the top view of
amplitude $\left\vert E(x,y)\right\vert $ and a snapshot of the phase, $\arg
\left( E(x,y)\right) $, of the stable stationary solution of Eq. (27) in the
form of the CSV with $S=1$. The parameters are taken as per Eq. (28), along
with $\protect\varepsilon =1.8$ and $\Omega =2$ (source: \protect\cite%
{Mihalache-et-al-2010b}).}
\label{fig14.20=fig186}
\end{figure}

\begin{figure}[tbp]
\includegraphics[scale=0.6]{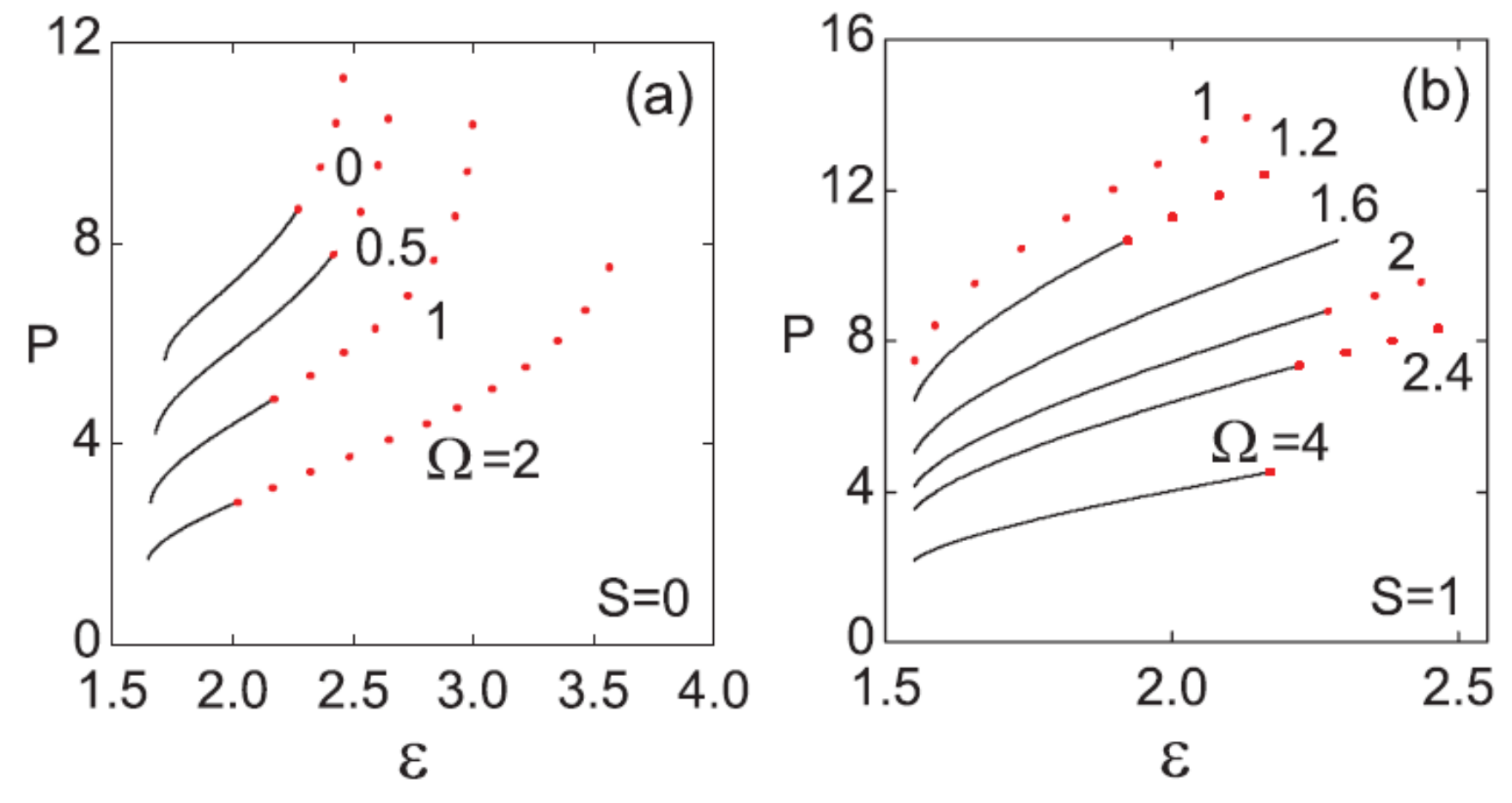}
\caption{Panels (a) and (b) show dependences of the integral power (32) on
the cubic gain, $\protect\varepsilon $, in Eq. (27), at different fixed
values of strength $\Omega $ of the HO trap, for families of fundamental ($%
S=0$) and VR ($S=1$) states, respectively. Other parameters are fixed as in
Eq. (28) (source: \protect\cite{Mihalache-et-al-2010b}).}
\label{fig14.21=fig187}
\end{figure}

The results are summarized in Fig. \ref{fig14.21=fig187}(b) by means of
plots which show curves $P(\varepsilon )$ for power (32) vs. the cubic gain,
at different fixed values of the trapping strength, $\Omega $, for families
of the CSV solutions, distinguishing stable and unstable solutions. In
accordance with what is stated above, the vortices are completely unstable
if the trapping potential is absent or too weak, $\Omega \leq 1$. The
stability region attains its maximum close to $\Omega =1.6$. This is a value
at which the characteristic radial size, determined by the HO potential,%
\begin{equation}
r_{\mathrm{HO}}=\Omega ^{-1/2},  \tag{33}
\end{equation}%
is roughly equal to the natural width of the 2D\ solitons created by the
CQ-CGL equation. When the further increase of $\Omega $ makes the trapping
potential too tight, the stability region gradually shrinks. For the sake of
comparison, similar curves $P(\varepsilon )$ for fundamental DSs, with $S=0$%
, are presented in Fig. \ref{fig14.21=fig187}(a). It is seen that the
stability region of the fundamental DSs monotonously shrink with the
increase of $\Omega $.

\begin{figure}[tbp]
\includegraphics[scale=0.5]{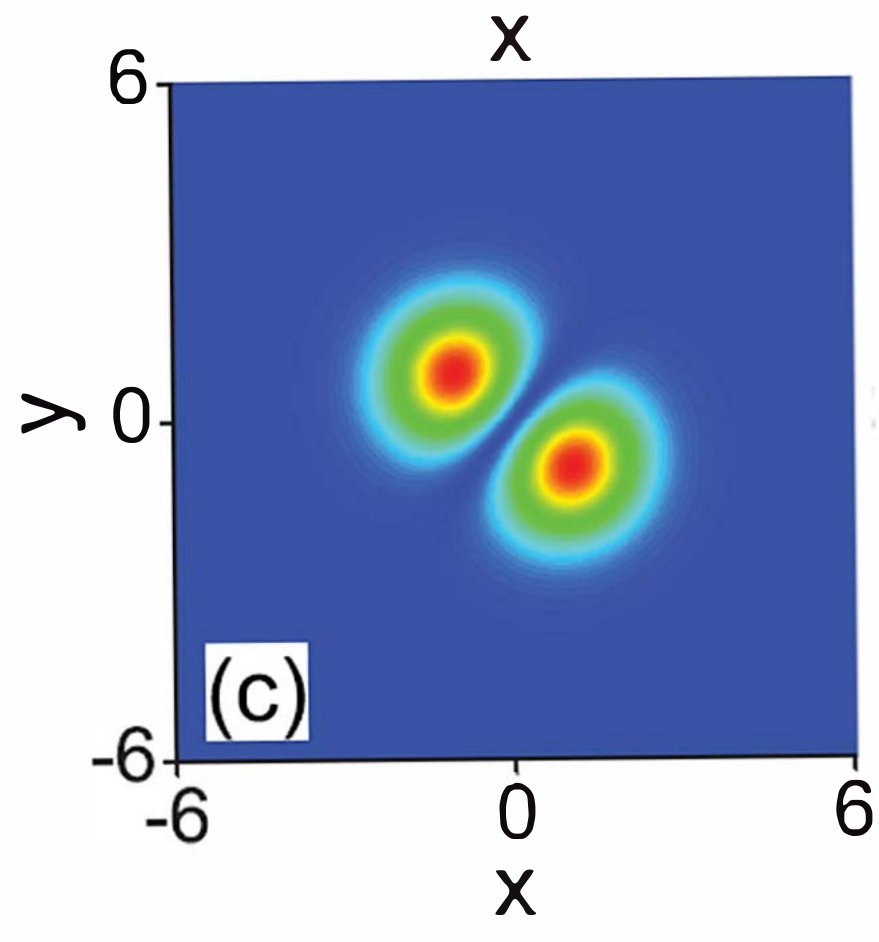} %
\includegraphics[scale=0.5]{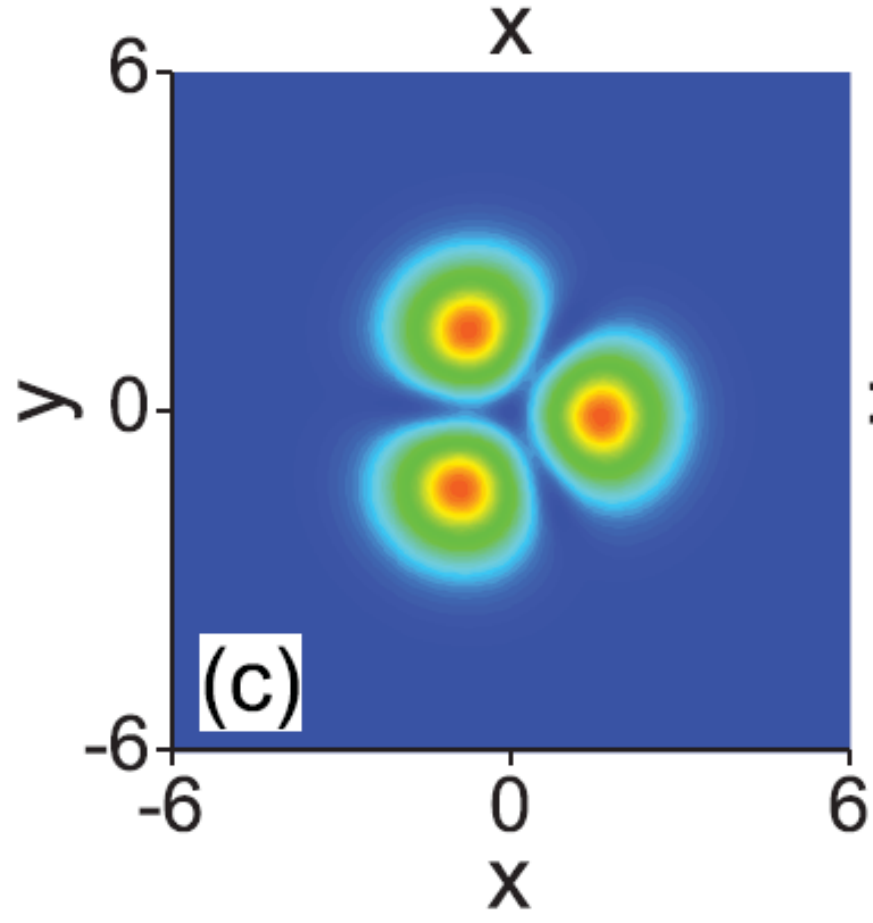}
\caption{The left and right panels show examples of dipoles and tripoles,
produced by the spitting instability of the CSV states with $S=1$ and $2$,
respectively. The parameters are $\protect\varepsilon =1.8$, $\Omega =0.5$
in (a), and $\protect\varepsilon =1.7$, $\Omega =0.5$ in (b). Other
parameters are fixed as in Eq. (28) (source: \protect\cite%
{Mihalache-et-al-2010b}).}
\label{fig14.22=fig188}
\end{figure}

Those CSV\ states with $S=1$ which are unstable spontaneously split into
stable dipole modes, as shown in the left panel of Fig. \ref{fig14.22=fig188}%
. As concerns the CSVs with $S=2$, they can be easily obtained as stationary
solutions (using Eq. (30)), but simulations of Eq. demonstrate that they all
are unstable, spontaneously splitting into \textit{tripoles} which, by
themselves, are stable states, see an example in the right panel Fig. \ref%
{fig14.22=fig188}. Further simulations demonstrate that the emerging dipoles
are static modes, while the tripoles feature rotation at a constant angular
speed. In fact, both dipoles and tripoles can be directly constructed as
solutions of Eq. (30) (for the tripoles, it should be rewritten as the
stationary equation in the rotating reference frame).

\subsection{The cubic CGL equation with linear gain}

As mentioned above, the CGL equation normally generates stable localized
solutions if it includes the linear loss, which secures the stability of the
zero background around the localized state. Nevertheless, the equation which
combines the \emph{linear gain}, diffusion term, and the HO trapping
potential also admits stable solutions, because small perturbations fueled
by the linear gain far from the soliton roll down the HO potential profile
towards the location of the soliton, where the cubic loss helps to suppress
them. In this case, the presence of the linear gain requires the inclusion
of the cubic loss, while quintic terms are not necessary, which makes the
CGL equation essentially simpler \cite{Mayteevarunyoo-Malomed-Skryabin-2018a}%
:%
\begin{equation}
i\frac{\partial \psi }{\partial t}=-\frac{1}{2}(1-i\eta )\left( \frac{%
\partial ^{2}}{\partial x^{2}}+\frac{\partial ^{2}}{\partial y^{2}}\right)
\psi -\sigma |\psi |^{2}\psi +i\gamma (1-|\psi |^{2})\psi +\frac{1}{2}\Omega
^{2}r^{2}\psi ,  \tag{34}
\end{equation}%
cf. Eq. (27). Equation (34) is written as the dissipative Gross-Pitaevskii
equation for the exciton field in a 2D semiconductor microcavity. Here $%
\gamma $ is the linear gain (the coefficient of the cubic loss is made equal
to $\gamma $ by means of scaling), $\eta $ is the diffusion coefficient (the
same as $\beta $ in Eq. (2)), and $\sigma =+1$ or $=1$ for the self-focusing
or defocusing cubic term, respectively.

At the threshold of emergence of nontrivial solutions with an infinitesimal
amplitude $\varepsilon $, real chemical potential $\mu $ and integer
vorticity $S\geq 0$, in the form of

\begin{equation}
\psi \left( x,y,t\right) =\varepsilon u_{\mathrm{thr}}(r)\exp \left( -i\mu
t+iS\theta \right) ,  \tag{35}
\end{equation}%
complex function $u(r)$ satisfies the linearized equation which follows from
the substitution of ansatz (35) in Eq. (34):
\begin{equation}
\mu u_{\mathrm{thr}}=-\frac{1}{2}(1-i\eta )\left( \frac{d^{2}}{dr^{2}}+\frac{%
1}{r}\frac{d}{dr}-\frac{S^{2}}{r^{2}}\right) u_{\mathrm{thr}}(r)+i\gamma u_{%
\mathrm{thr}}(r)+\frac{1}{2}\Omega ^{2}r^{2}u_{\mathrm{thr}}(r).  \tag{36}
\end{equation}%
This equation admits the \emph{exact solution},%
\begin{equation}
u_{\mathrm{thr}}(r)=r^{S}\exp \left( -\frac{\Omega }{2\sqrt{1-i\eta }}%
r^{2}\right) ,  \tag{37}
\end{equation}%
with chemical potential%
\begin{equation}
\mu _{\mathrm{thr}}=\frac{(1+S)\Omega \eta }{\sqrt{2\left( \sqrt{1+\eta ^{2}}%
-1\right) }},  \tag{38}
\end{equation}%
at the threshold value of the linear gain:
\begin{equation}
\gamma _{\mathrm{thr}}=\left( 1+S\right) \Omega \sqrt{\frac{1}{2}\left(
\sqrt{1+\eta ^{2}}-1\right) }.  \tag{39}
\end{equation}%
This result means that, for given $\eta $, nontrivial states with vorticity $%
S$ exist at $\gamma >\gamma _{\mathrm{thr}}$.

\begin{figure}[tbp]
\caption{A sequence of stably rotating states (patterns of the local
intensity, $\left\vert u\left( x,y,t\right) \right\vert ^{2}$, and local
phase, $\arg \left( u\left( x,y,t\right) \right) $, are shown), which were
produced by simulations of Eq. (34) with $\protect\sigma =+1$, $\protect%
\gamma =2.5$, and $\Omega ^{2}=2$, at decreasing values of diffusivity $%
\protect\eta $, which are indicated in the panels. The sequence includes a
crescent vortex ($S=1$), and multi-vortex complexes with $S=2,3,4,6,$ and $7$%
. The rotation angular velocities are $\protect\omega (S=1)\approx 3.9$, $%
\protect\omega (S=2)\approx 1.7$, $\protect\omega (S=3)\approx 1.6$, and $%
\protect\omega (S=4,6,7)\approx 1.5$. At $\protect\eta >0.5$, the model
supports a a stable axisymmetric vortex with $S=1$, while at $\protect\eta %
<0.22$ a transition to vortex turbulence occurs. (source: Ref. \protect\cite%
{Mayteevarunyoo-Malomed-Skryabin-2018a}).}
\label{fig14.add1=fig_extra5}\includegraphics[scale=0.75]{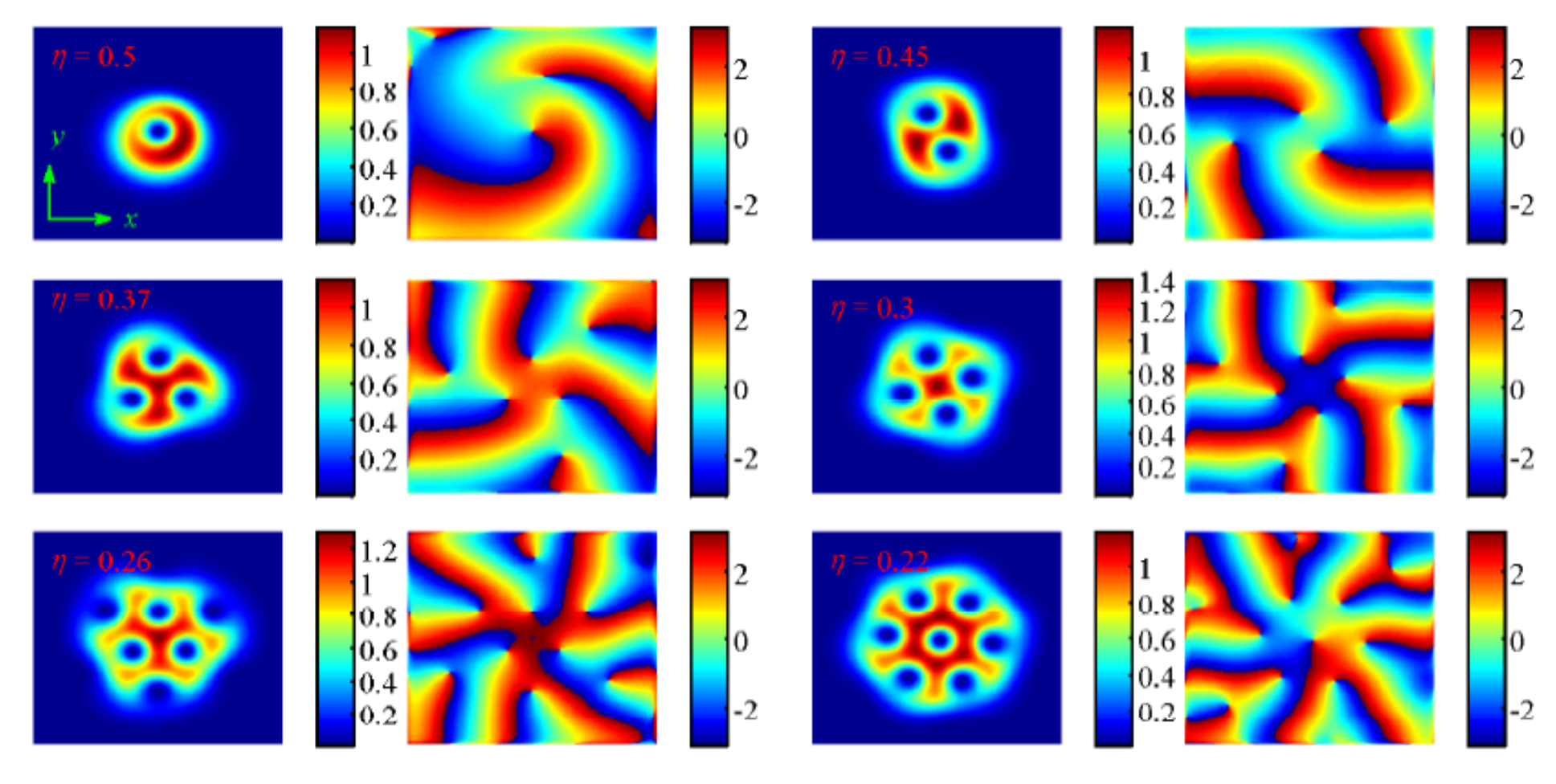}
\end{figure}

Fixing the value of the linear gain, e.g., $\gamma =2.5$, numerical solution
of Eq. (34) with $\Omega ^{2}=2$ and $\sigma =+1$ gives rise to a stable
solution in the form of an axisymmetric vortex with $S=1$ at $\eta >0.5$.
Making the system \textquotedblleft more agile" by decreasing $\eta $, it
was observed that, as shown in Fig. \ref{fig14.add1=fig_extra5}, at $\eta
=0.5$, the axisymmetric vortex is replaced by a rotating deformed one, with
a crescent-like shape of the intensity pattern. At still smaller values of $%
\eta $, the single vortex passes a cascade of splittings. As a result, the
role of the stable state is played, consecutively, by rotating clusters of $%
2 $, $3$, $4$, $6$, and $7$ vortices. Finally, the system falls into a state
of vortex chaos (\textquotedblleft turbulence") at $\eta <0.22$.

Note that the sequence of the rotating patterns displayed in Fig. \ref%
{fig14.add1=fig_extra5} does not include a cluster of five vortices. Such a
stable state is produced too by the simulations of Eq. (34) with $\sigma =-1$
(the self-defocusing cubic term).

\begin{figure}[tbp]
\includegraphics[scale=0.75]{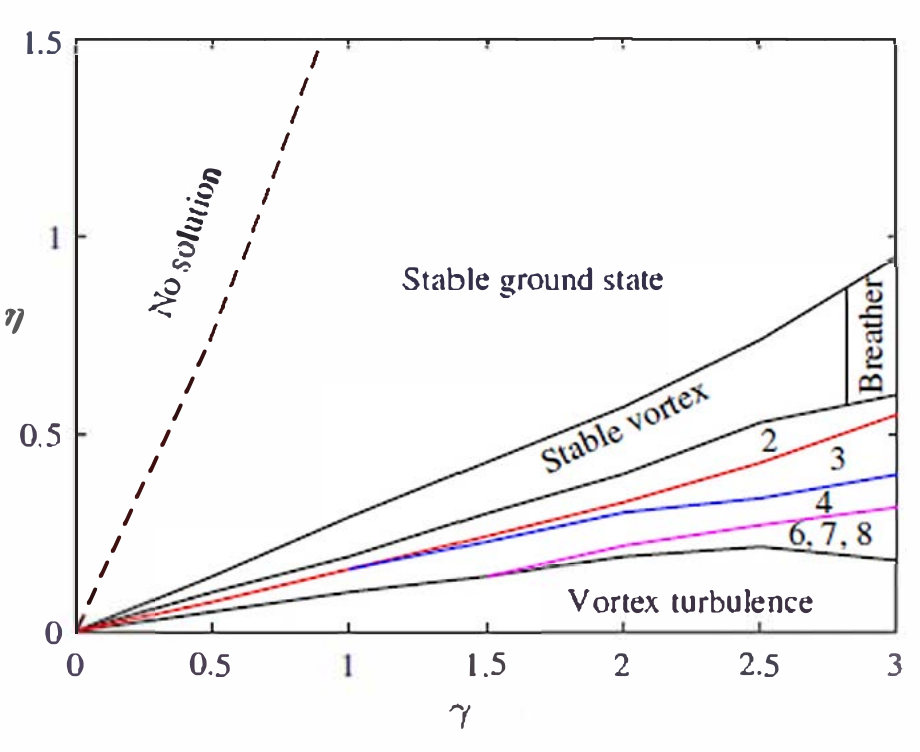}
\caption{Stability borders of different patterns in the plane of $\left(
\protect\gamma ,\protect\eta \right) $ for $\protect\sigma =0$ and $\Omega
^{2}=1$, produced by the numerical solution of Eq. (34). The analytical
existence boundary, predicted by Eq. (39), and its numerically identified
counterpart completely overlap, in the form of the dashed line. Digits in
subregions represent the number of vortices in rotating clusters populating
these subregions. In particular, $1$ implies the presence of the single ($%
S=1 $) rotating crescent-shaped vortex, different from the axisymmetric one
in the area of \textquotedblleft Stable vortex". Label "Breather" identifies
a small region in which the dominant state is one with $S=0$ which performes
regular internal oscillations (source: \protect\cite%
{Mayteevarunyoo-Malomed-Skryabin-2018a}).}
\label{fig14.add2=fig_extra6}
\end{figure}

For the case of $\sigma =0$, when the nonlinearity is represented, in Eq.
(34), solely by the cubic loss, the results are summarized in the stability
chart plotted in Fig. \ref{fig14.add2=fig_extra6} in the plane of parameters
$\left( \gamma ,\eta \right) $. Note, in particular, that the exact
analytical result (39) correctly predicts the boundary of the existence of
nontrivial states.

\section{Dissipative vortex solitons stabilized by spatially periodic
(lattice) potentials}

\subsection{Crater-shaped (tightly confined) vortex solitons}

It is known that the conservative 2D NLS equation with the CQ nonlinearity
and a spatially periodic (lattice) potential cannot produce stable vortical
modes (\textquotedblleft eddies") of the CSV type, that would be essentially
\textquotedblleft squeezed" in a single cell of the lattice \cite{Zyss}.
Nevertheless, the CQ-CGL equation with the same spatially-periodic
potential, which does not include the diffusion term ($\beta =0$ in the
corresponding equation (2)), can produce stable eddies of the CSV type \cite%
{Mihalache-et-al-2010b}. The respectively modified equation (27) is%
\begin{equation}
\frac{\partial E}{\partial z}=-\delta \cdot E+\frac{i}{2}\left( \frac{%
\partial ^{2}}{\partial x^{2}}+\frac{\partial ^{2}}{\partial y^{2}}\right)
E+\left( \varepsilon +i\right) |E|^{2}E-\left( \mu +i\nu \right) |E|^{4}E-ip%
\left[ \cos (2x)+\cos (2y)\right] E.  \tag{40}
\end{equation}%
Here coordinates $\left( x,y\right) $ are scaled so that the period of the
lattice potential with amplitude $p>0$ is $\pi $. The form of the potential
adopted in Eq. (40) implies that point $x=y=0$ corresponds to a local
maximum of the potential, rather than a minimum. This definition is natural
if the intention is to produce CSV solutions, with the zero amplitude
(vortex' pivot) which should be placed at a local potential maximum.

\begin{figure}[tbp]
\includegraphics[scale=0.6]{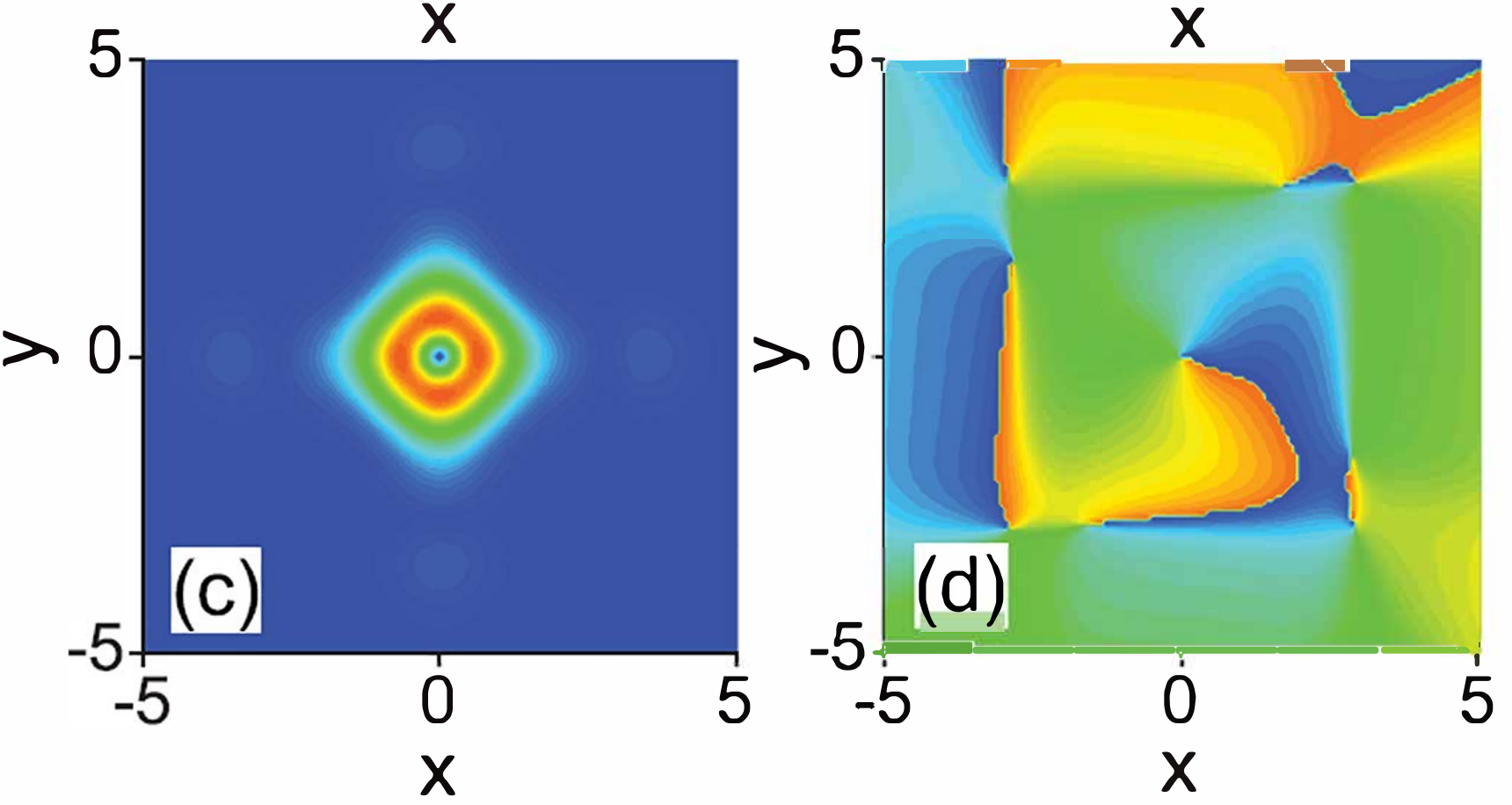}
\caption{The left and right panels display the amplitude and phase
distribution of a stable \textquotedblleft crater-shaped" vortex (CSV) with
winding number $S=1$, as produced by the numerical solution of Eq. (40).
The CSV is essentially squeezed in a single cell of the
potential lattice with strength $p=2$ and cubic-gain coefficient $\protect%
\varepsilon =2$. Other parameters of Eq. (40) are fixed as in Eq. (28) (source: Ref.
\protect\cite{Mihalache-et-al-2010b}).}
\label{fig14.23=fig189}
\end{figure}

\begin{figure}[tbp]
\includegraphics[scale=0.6]{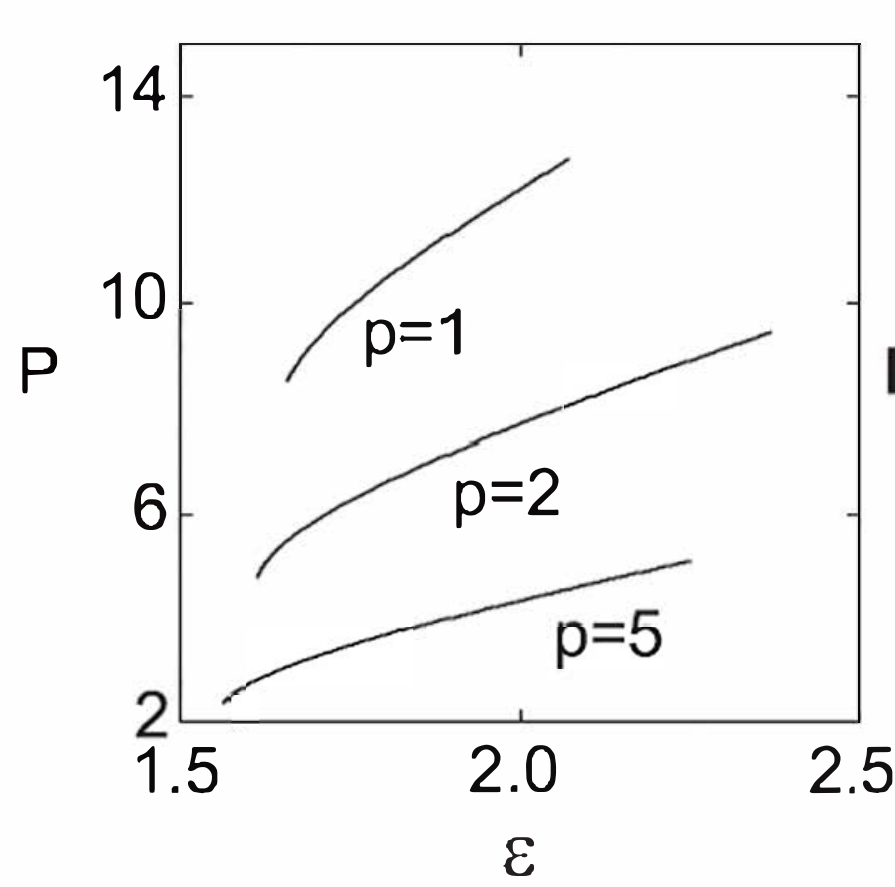}
\caption{Dependence of the integral power (41) of the CSVs (crater-shaped
vortices) with $S=1$ on the cubic-gain coefficient, $\protect\varepsilon $,
produced by the numerical solution of Eq. (40) at indicated fixed values of
strength $p$ of the lattice potential. Other parameters are fixed as in Eq.
(28) (source: Ref. \protect\cite{Mihalache-et-al-2010b}).}
\label{fig14.24=fig190}
\end{figure}

The numerical solution of Eq. (40) readily produces stationary states of the
CSV type, a part of which are stable. A typical example of a stable CSV
solution is displayed in Fig. \ref{fig14.23=fig189}. The results are
summarized in Fig. \ref{fig14.24=fig190} which presents families of stable
CSV solutions by dint of the dependence of its integral power,%
\begin{equation}
P=\int \int \left\vert E(x,y\right\vert ^{2}dxdy  \tag{41}
\end{equation}%
(cf. Eq. (32)), on the cubic-drive's strength, $\varepsilon $.

\begin{figure}[tbp]
\includegraphics[scale=0.5]{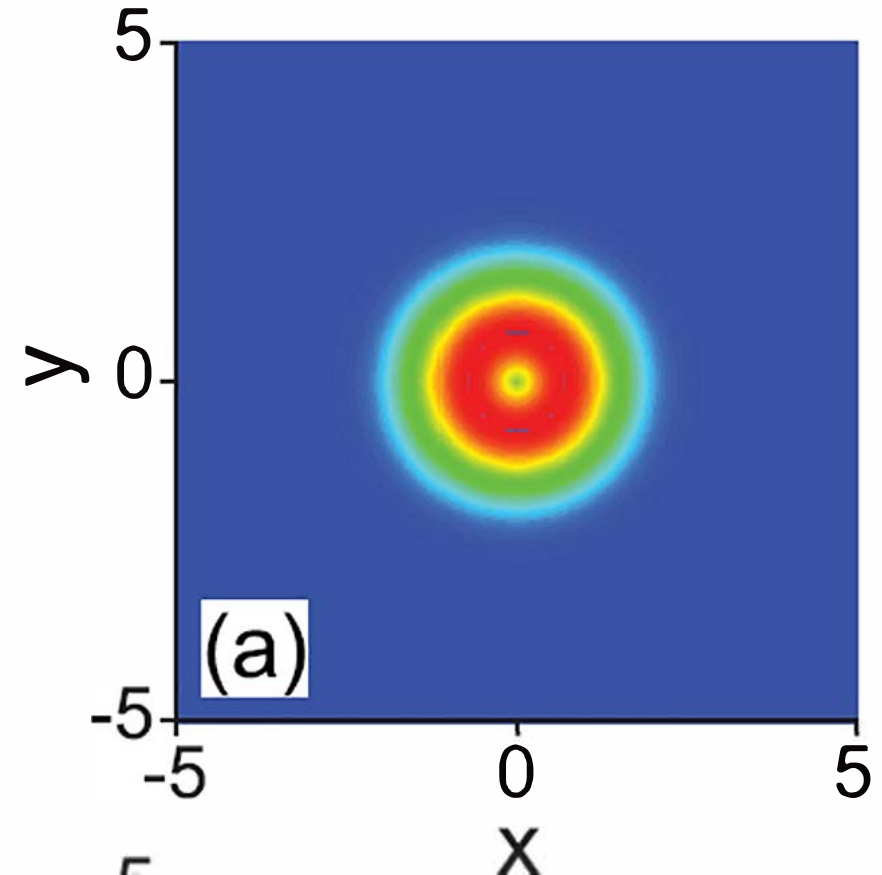} %
\includegraphics[scale=0.5]{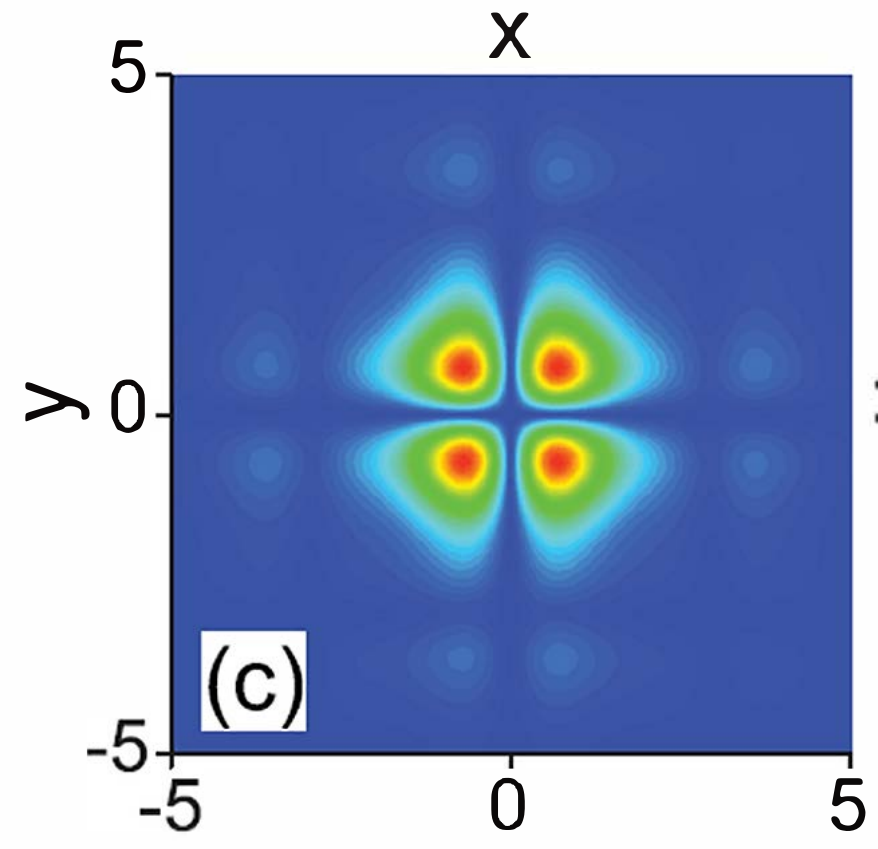}
\caption{(a) The amplitude profile of an unstable CVS model with winding
number $S=2$, obtained as a numerical solution of Eq. (40) with $\protect%
\varepsilon =1.8$ and $p=2$. Other parameters are fixed as in Eq. (28). (c)
The amplitude profile of a stable quadrupole into which the unstable vortex
from (a) is transformed by the development of the instability (source: Ref.
\protect\cite{Mihalache-et-al-2010b}).}
\label{fig14.25=fig191}
\end{figure}

CSV modes with winding number $S=2$ can be found too, as stationary
solutions of Eq. (40), but they all are unstable. Spontaneous development of
the instability transforms them into stable quadrupoles, as shown in Fig. %
\ref{fig14.25=fig191}. Such localized quadrupoles also constitute families
of stable solutions.

A relevant extension of the analysis of the model based on Eq. (40) is to
study a possibility of setting DSs in motion by kicking them. A critical
strength of the kick leading to depinning of a soliton originally pinned to
the underlying potential lattice, and various scenarios of complex pattern
formation, following the depinning and also collisions between traveling
DSs, were studied by means of systematic simulations by in Ref. \cite%
{Besse-et-al-2013}.

\subsection{Multi-peak vortex solitons}

When the CSV solutions of the conservative NLS equation with the 2D lattice
potential are unstable, the same equation can easily give rise to stable
rhombus- and square-shaped solutions, built of four local peaks, which carry
the phase pattern corresponding to vorticity $S=\pm 1$. Such onsite
(OS)-centered \cite{Baizakov-Malomed-Salerno-2003} and intersite (IS)-centered
\cite{Yang-Musslimani-2003} multi-peak vortex complexes are well known as
stable solutions of the 2D conservative NLS equation with the lattice
potential, see also the review . The CGL equation (40) produces similar
multi-peak vortex states, as shown by Leblond, Malomed, and Mihalache
(2009). As a typical set of parameter values which makes it possible to find
stable compound-vortex modes, one can take%
\begin{equation}
\delta =0.4,\varepsilon =1.85,\mu =1,\nu =0.1,V_{0}\equiv -p=1.  \tag{42}
\end{equation}%
Note that the negative sign of $p\equiv -V_{0}$ in Eq. (42) implies that,
unlike the situation illustrated above by Figs. \ref{fig14.23=fig189}-\ref%
{fig14.25=fig191}, the central point, $x=y=0$, is a local minimum of the
cellular potential (rather than a maximum).

\begin{figure}[tbp]
\includegraphics[scale=0.4]{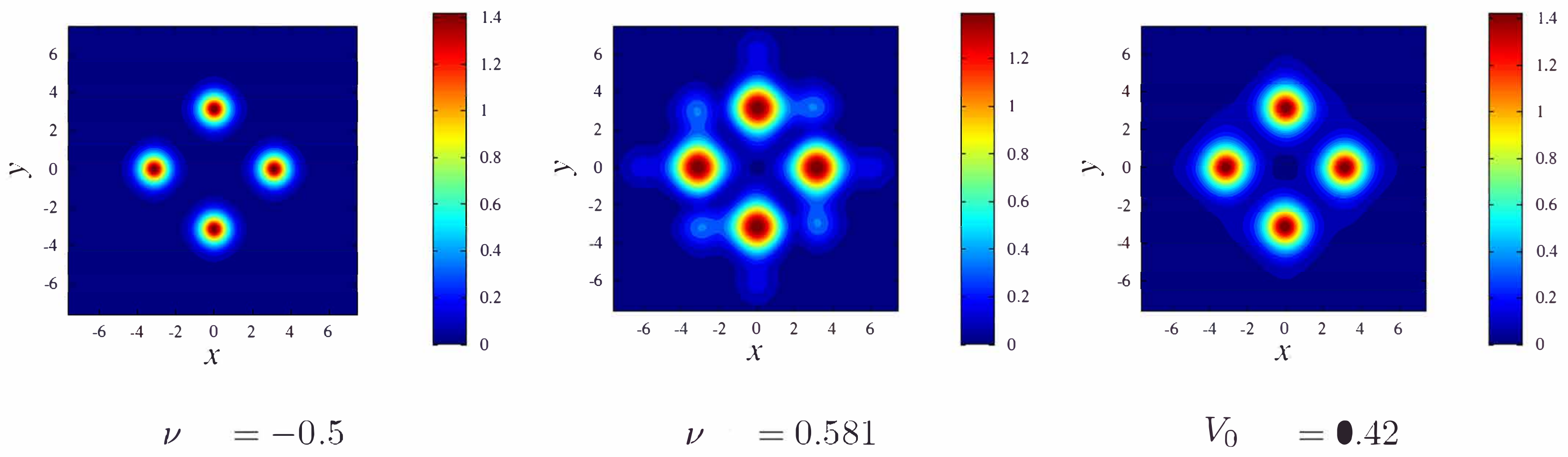}
\caption{Patterns of $\left\vert E(x,y)\right\vert $ in stable OS-centered
(rhombus-shaped) four-peak vortex complexes with winding number $S=1$,
obtained as numerical solutions of Eq. (40) with parameters taken as per Eq.
(42), except for $\protect\nu $ (in the left and central panels) or $V_{0}$
(in the right one), whose values are indicated in the figure (source: Ref.
\protect\cite{Leblond-Malomed-Mihalache-2009}).}
\label{fig14.26=fig192}
\end{figure}

A set of top-view plots of $\left\vert E(x.y)\right\vert $ for the stable
OS-centered vortex rhombuses are displayed in Fig. \ref{fig14.26=fig192}. In
particular, the difference between the left and central panels is that the
quintic self-interaction, represented by coefficient $\nu $, is,
respectively, focusing ($\nu <0$) and defocusing ($\nu >0$) in them.
Accordingly, the pattern in the left panel is essentially more
self-compressed than in the central one.

Families of the OS-centered vortex complexes are characterized in Fig. \ref%
{fig14.27=fig193} by dependences of their integral power, calculated
according to Eq. (41), and propagation constant (found as the eigenvalue of
Eq. (40)) on the cubic-gain coefficient $\varepsilon $. It is seen that the
rhombic vortices are unstable if $\varepsilon $ is too small or two large,
being stable in the broad intermediate region.
\begin{figure}[tbp]
\includegraphics[scale=0.6]{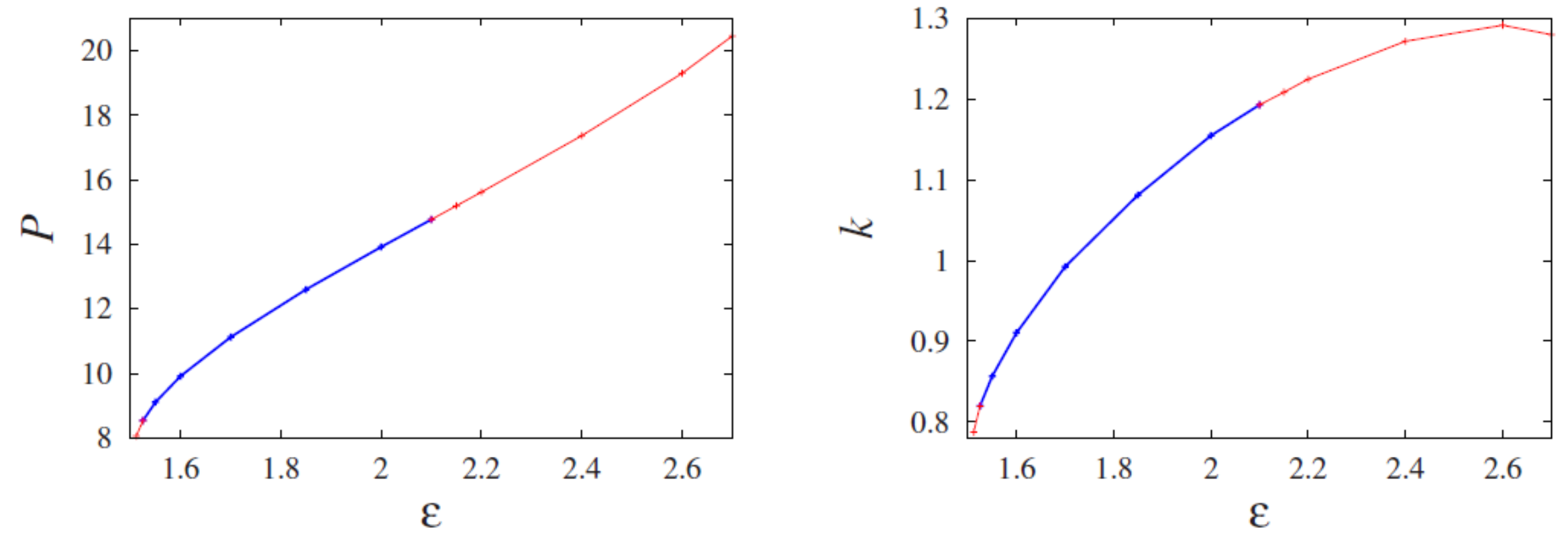}
\caption{Dependence of the integral power (41) and propagation constant $k $
(the eigenvalue of Eq. (40)) for the family of OS-centered vortex complexes,
built of four peaks (see Fig. \protect\ref{fig14.26=fig192}), on strength $%
\protect\varepsilon $ of the cubic gain. Blue (lower) and red (upper)
segments denote, respectively, stable and unstable subfamilies (source: Ref.
\protect\cite{Leblond-Malomed-Mihalache-2009}).}
\label{fig14.27=fig193}
\end{figure}

\begin{figure}[tbp]
\includegraphics[scale=0.4]{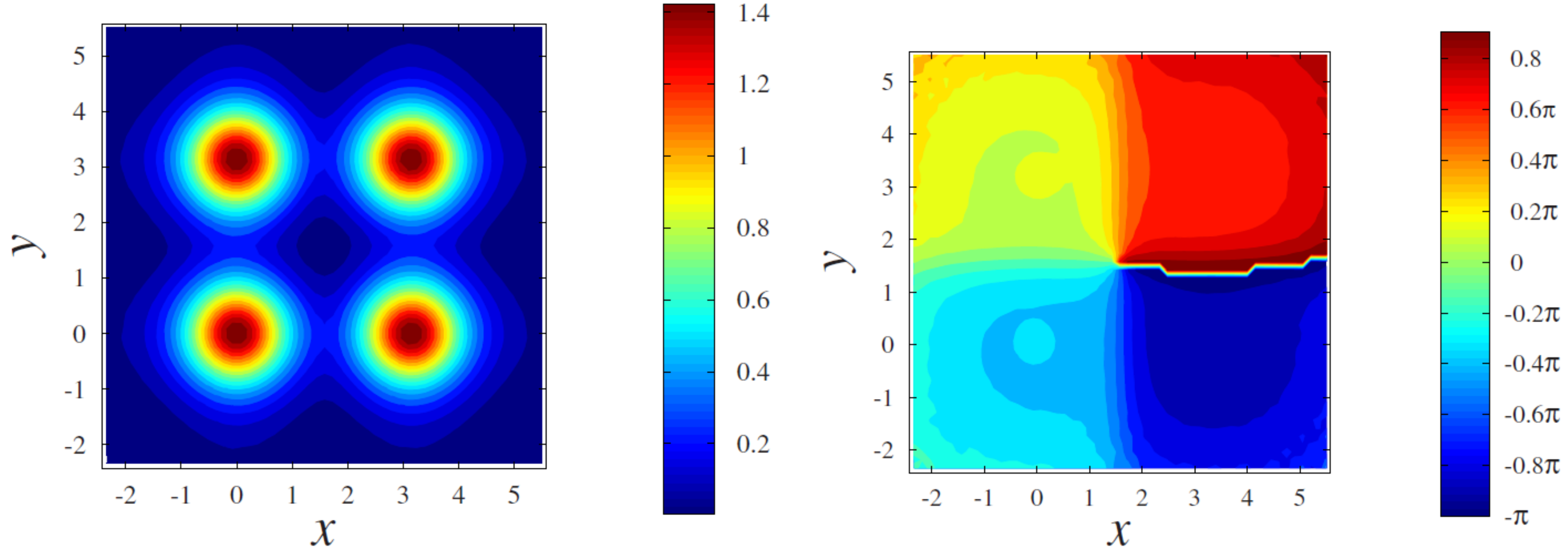}
\caption{Amplitude and phase patterns of a stable IS-centered
(square-shaped) vortex complex with winding number $S=1$, obtained as a
numerical solution of Eq. (40) with parameters taken as per Eq. (42)
(source: Ref. \protect\cite{Leblond-Malomed-Mihalache-2009}).}
\label{fig14.28=fig194}
\end{figure}

An example of the amplitude and phase structure in a stable IS-centered
(square-shaped) vortex complex is displayed in Fig. \ref{fig14.28=fig194}.
In particular, the phase pattern clearly corroborates the presence of
vorticity $S=1$ in this complex.

\begin{figure}[tbp]
\includegraphics[scale=0.4]{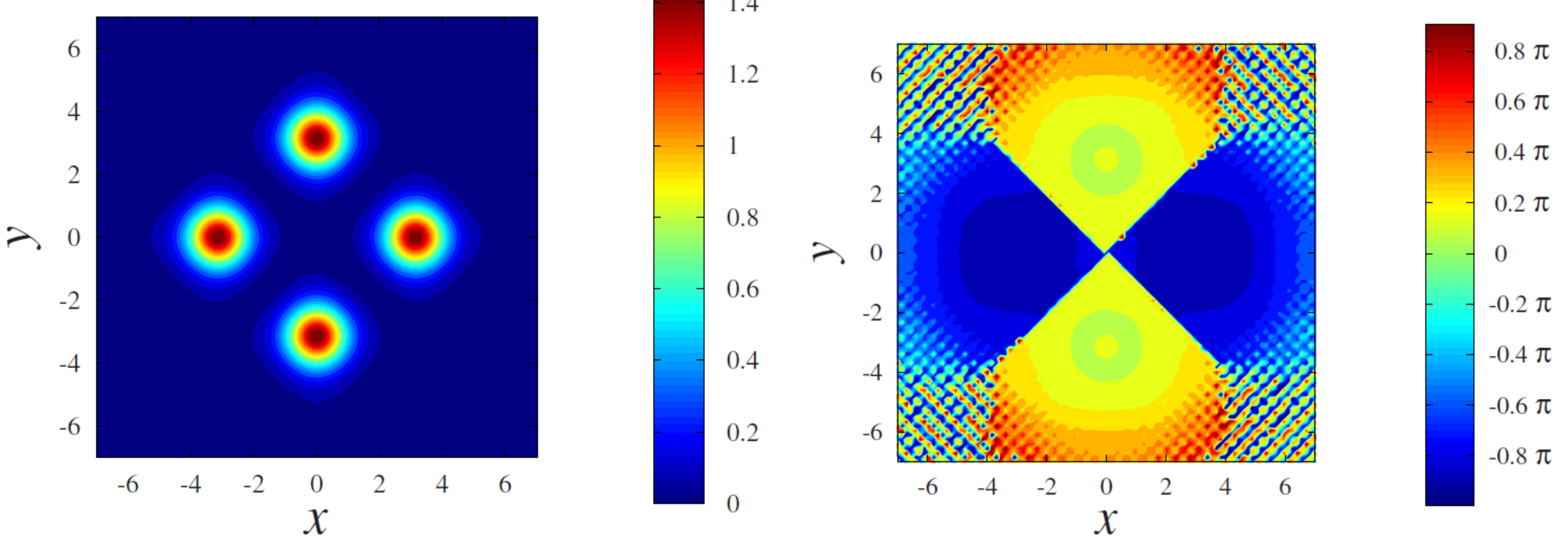}
\caption{Amplitude and phase patterns of a stable OS-centered
(square-shaped) quadrupole, produced by the numerical solution of Eq. (40)
with parameters taken as per Eq. (42) (source: Ref. \protect\cite%
{Leblond-Malomed-Mihalache-2009}).}
\label{fig14.29=fig195}
\end{figure}

In addition to the OS- and IS-centered vortex complexes, Eq. (40) generates
stable compound states in the form of \textit{quadrupoles}, which do not
carry vorticity. An example of an OS-centered one (a rhombic quadrupole) is
displayed in Fig. \ref{fig14.29=fig195}. These solutions also form families
parameterized by values of the cubic gain $\varepsilon $, cf. Fig. \ref%
{fig14.27=fig193}.

\subsection{Dissipative gap solitons (GSs) generated by the CGL equation
with the 2D lattice potential}

The results presented above are based on the CGL equation (40) with the
self-focusing cubic nonlinearity. It is also interesting to consider the
model with the self-defocusing cubic term. It is well known that the
respective 2D conservative NLS equation with the lattice potential gives
rise to families of gap solitons (GSs) \cite{KivOst,HS-GS}.

GSs as solutions of the 1D CGL equation with the lattice potential were
introduced in Ref. \cite{Sak-Mal-2008}. The 2D setting was then developed in
Ref. \cite{Sak-Mal-2009}, where the CGL\ equation with the 2D lattice
potential was adopted in the following form, which does not include the
diffusion term (recall that term is irrelevant for realizations in optics):%
\begin{equation}
\left( \frac{\partial }{\partial z}+\Gamma _{1}\right) \psi =\frac{i}{2}%
\left( \frac{\partial ^{2}}{\partial x^{2}}+\frac{\partial ^{2}}{\partial
y^{2}}\right) \psi -i|\psi |^{2}\psi +iA\left\{ \left[ \cos \left(
2q_{0}x\right) +\cos \left( 2q_{0}y\right) \right] \right\} \psi +(\Gamma
_{2}|\psi |^{2}-\Gamma _{3}|\psi |^{4})\psi .  \tag{43}
\end{equation}%
In comparison to Eq. (40), this CGL equation does not include the quintic
conservative term, which is not essential in the present context, and, with $%
A>0$, the point $x=y=0$ is a minimum of the potential.

Assuming that the loss and gain coefficients $\gamma _{1,2,3}$ in Eq. (43)
are small parameters, an analytical approximation for constructing DS
solutions was developed in Ref. \cite{Sak-Mal-2009}. To this end, Eq. (43)
was, at first, considered in the conservative limit, with $\gamma _{1,2,3}=0$%
, in which case Eq. (43) has the usual Lagrangian. Using this fact, the
variational approximation was elaborated for the fundamental solitons,
adopting the following ansatz:%
\begin{equation}
\psi _{\mathrm{fund}}=B\exp \left( ikz-\frac{x^{2}+y^{2}}{2W^{2}}\right)
\cos (qx)\cos (qy),  \tag{44}
\end{equation}%
where amplitude $A$ and width $W$ are variational parameters, $q$ being an
adjustment constant. In the zero-order approximation with respect to the
loss and gain terms, $\gamma _{1,2,3}=0$, the integral power (norm) of
ansatz (44),%
\begin{equation}
P=\int \int \left\vert \psi \left( x,y\right) \right\vert ^{2}dxdy=(\pi
/4)B^{2}W^{2}\left[ 1+\exp \left( -q^{2}W^{2}\right) \right] ^{2},  \tag{45}
\end{equation}%
is an arbitrary parameter of the soliton family. Then, the loss and gain are
taken into regard by means of the balance equation for the power, which, by
itself, is an exact corollary of Eq. (43):

\begin{equation}
\frac{dP}{dz}=2\left( -\Gamma _{1}P+\Gamma _{2}\int \int |\psi
|^{4}dxdy-\Gamma _{3}\int \int |\psi |^{6}dxdy\right) .  \tag{46}
\end{equation}%
Finally, the stationary solution of the CGL equation is analytically
predicted as one corresponding to $dP/dz=0$ in Eq. (45).

\begin{figure}[tbp]
\includegraphics[scale=0.72]{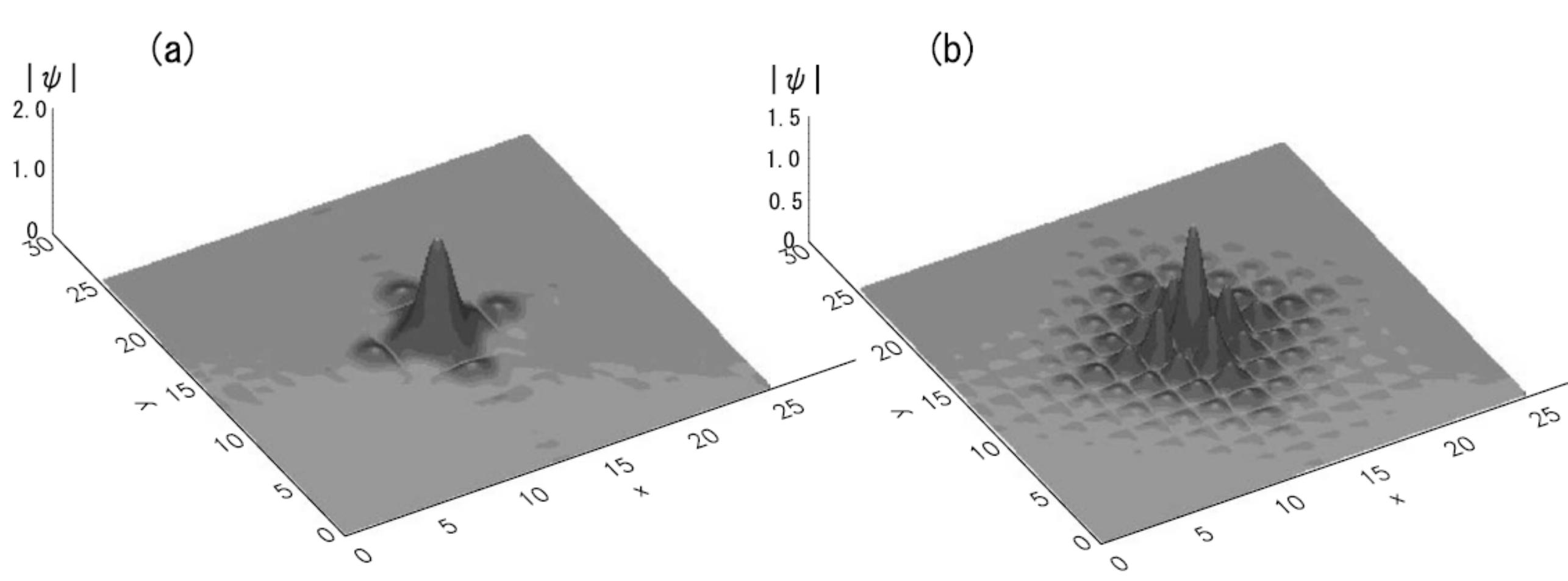}
\caption{Shapes of $\left\vert \protect\psi (x,y)\right\vert $ in stable
fundamental solitons, produced as numerical solutions of Eq. (43) with
parameters $A=2$, $\Gamma _{1}=0.025$, $\Gamma _{2}=0.1$, $\Gamma _{3}=0.05$
and $q_{0}=1$ in (a) (a tightly bound soliton), and $q_{0}=1.75$ in (b) (a
loosely bound soliton). Source: Ref. \protect\cite{Sak-Mal-2009}.}
\label{fig14.30=fig196}
\end{figure}

Similar to their counterparts produced by the conservative NLS equation \cite%
{HS-GS}, GS solutions of the CGL equation (43) also feature loosely and
tightly bound shapes. Examples of stable GSs of these types are shown in
Fig. \ref{fig14.30=fig196}. Their shapes are additionally illustrated, and
compared to the analytical approximation outlined above, in Fig. \ref%
{fig14.31=fig197}.
\begin{figure}[tbp]
\includegraphics[scale=0.70]{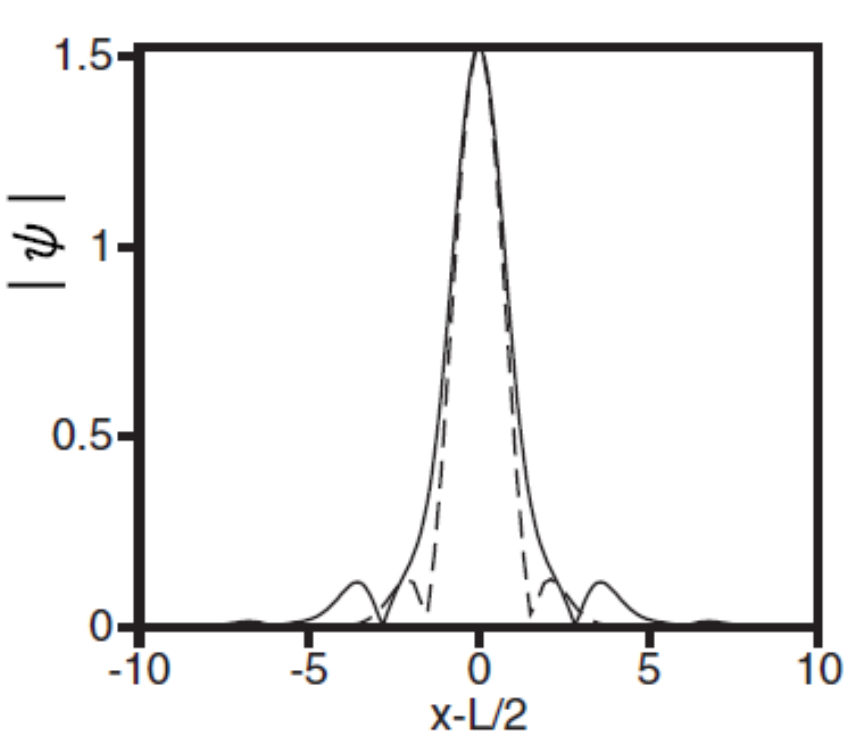} %
\includegraphics[scale=0.70]{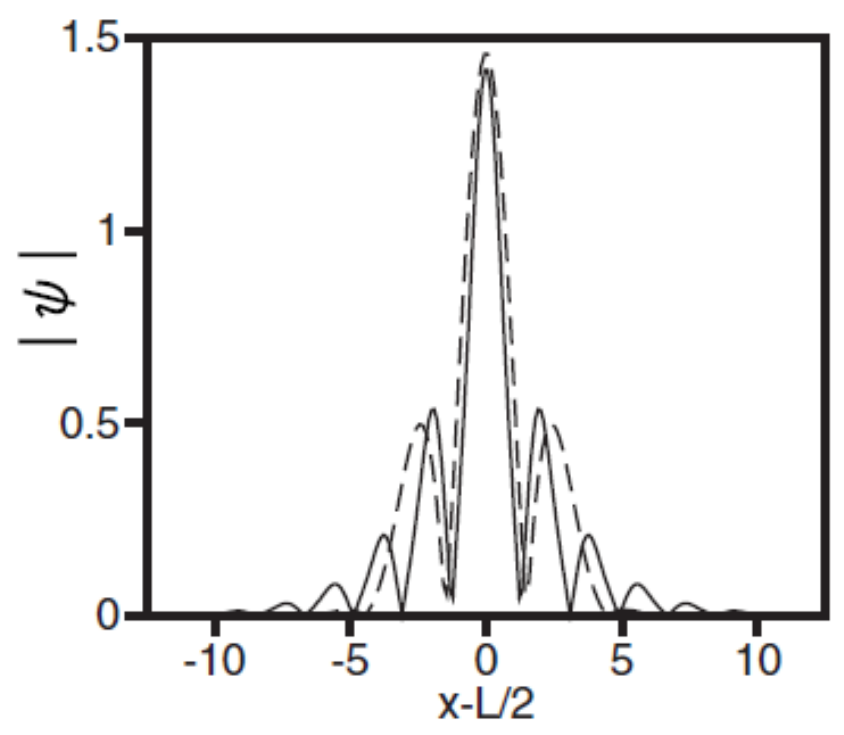}
\caption{Solid lines in the left and right panels show cross sections of the
2D profiles of the same fundamental dissipative GSs (tightly and loosely
bound ones, respectively) which are displayed in Figs. \protect\ref%
{fig14.30=fig196}(a) and (b). Dashed lines depict the analytical prediction
of the VA based on ansatz (44), with $q=1$ and $1.05$, in the left and right
panels, respectively, and the balance condition $dP/dz=0$, as it follows
from Eq. (46) (source: Sakaguchi and Malomed (2009)).}
\label{fig14.31=fig197}
\end{figure}

Numerical results for the fundamental dissipative GSs are summarized in the
stability chart in the parameter plane of $\gamma _{1}\equiv 20\Gamma _{1}$
(the linear-loss coefficient and $A$ (it determines depth $4A$ of the
lattice potential in Eq. (43)), which is presented in Fig. \ref%
{fig14.32=fig198}. Naturally, stable GSs exist if the underlying potential
is strong enough.
\begin{figure}[tbp]
\includegraphics[scale=0.5]{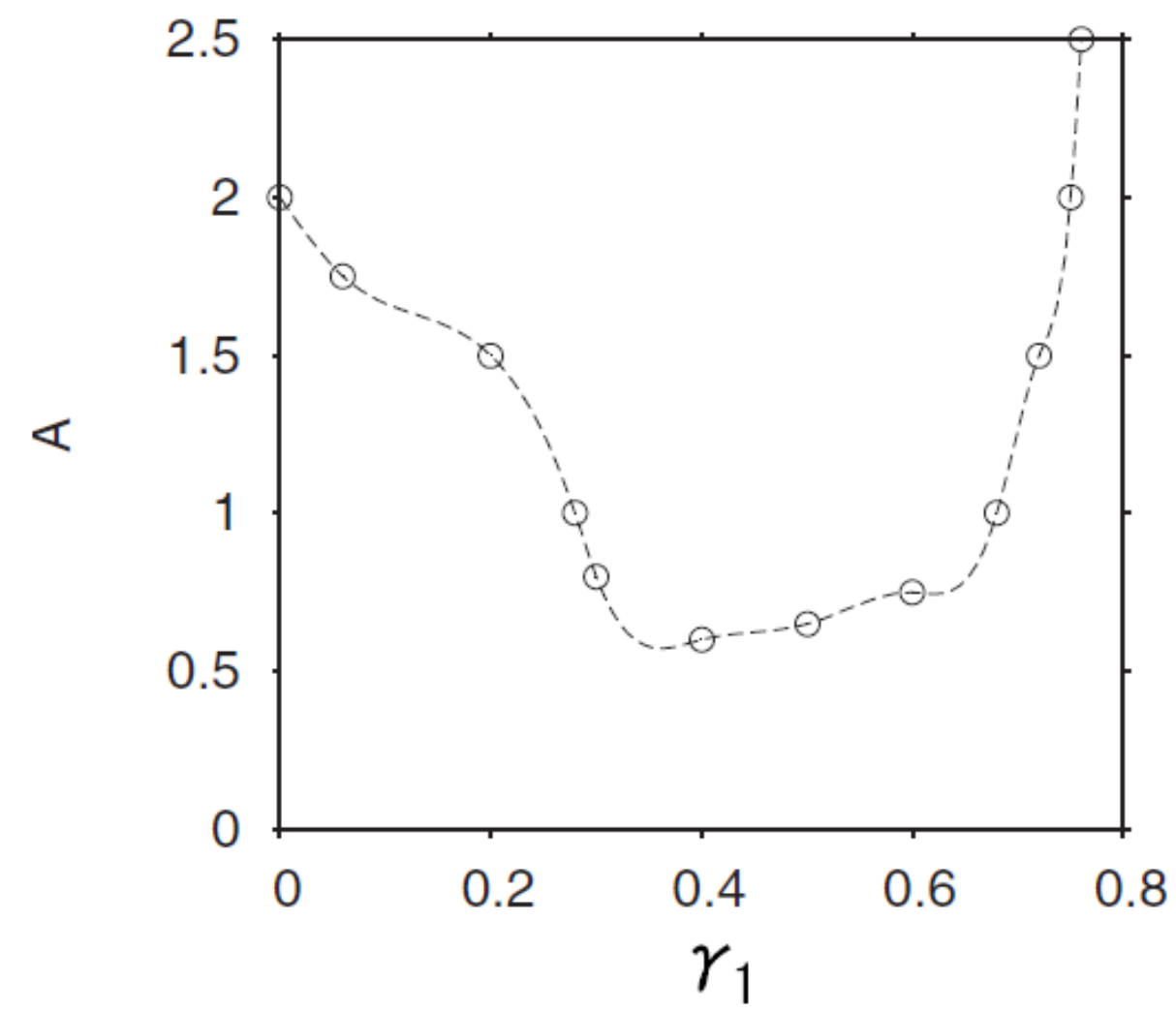}
\caption{Stable fundamental dissipative GSs exist, as solutions to Eq. (43),
above the plotted boundary in the parameter plane $\left( 20\Gamma
_{1},A\right) $. Other parameters are $\Gamma _{2}=0.1$, $\Gamma _{3}=0.05$,
and $q_{0}=1$ (source: Sakaguchi and Malomed (2009)).}
\label{fig14.32=fig198}
\end{figure}

Vortex-soliton solutions with winding number $S=1$ were produced by the
numerical solution of Eq. (43) with input

\begin{equation}
\left( \psi _{\mathrm{vort}}\right) _{0}=B(x+iy)\exp \left( -\frac{%
x^{2}+y^{2}}{2W^{2}}\right) \cos (qx)\cos (qy),  \tag{47}
\end{equation}%
varying parameters $B$, $W$, and $q$. The results are stable vortex solitons
of the same two types as identified in the case of the fundamental solitons,
i.e., loosely and tightly bound ones. Examples are presented in Fig. \ref%
{fig14.33=fig199}.
\begin{figure}[tbp]
\includegraphics[scale=0.71]{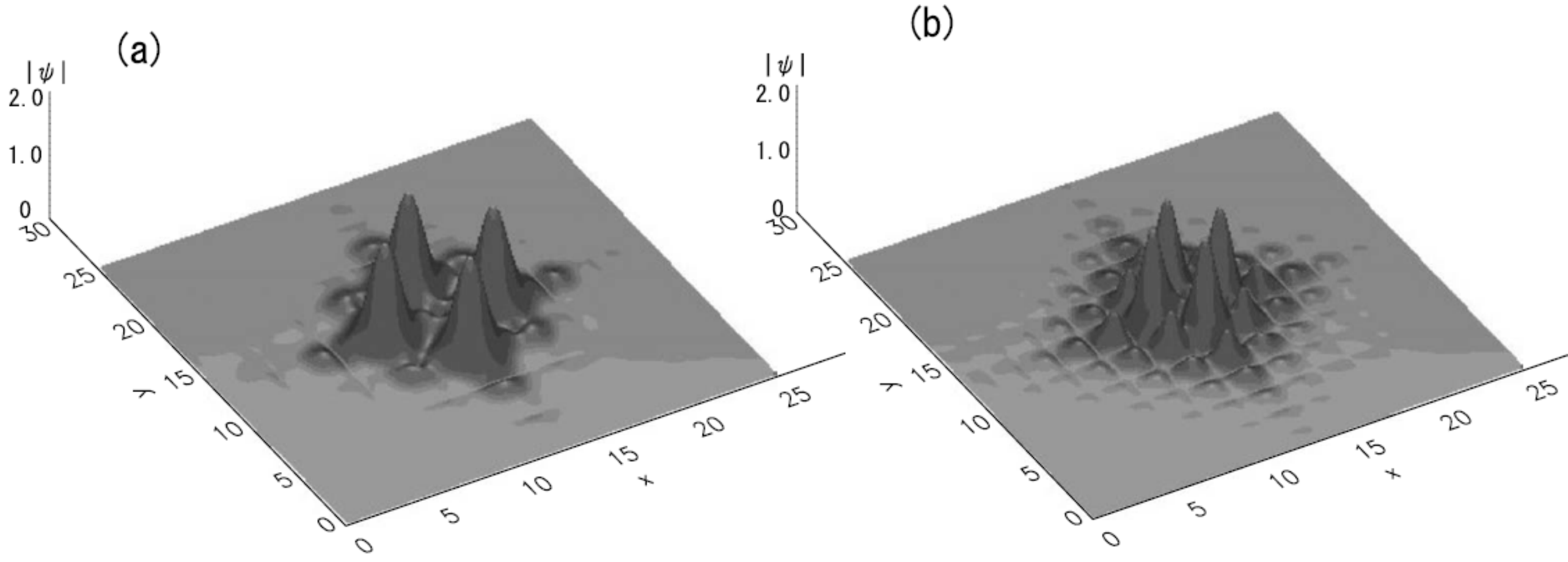}
\caption{Shapes of $\left\vert \protect\psi (x,y)\right\vert $ in stable
vortex solitons with $S=1$, produced as numerical solutions of Eq. (43) with
parameters $A=2$, $\Gamma _{1}=0.025$, $\Gamma _{2}=0.1$, $\Gamma _{3}=0.05$
and $q_{0}=1$ in (a) (a tightly bound vortex soliton), and $q_{0}=1.6$ in
(b) (a loosely bound one). Source: Ref. \protect\cite{Sak-Mal-2009}.}
\label{fig14.33=fig199}
\end{figure}

\section{Two-component dissipative solitons of the vortex-antivortex (VAV)
type stabilized by SOC (spin-orbit coupling)}

\subsection{The model}

Parallel to the well-known work aimed at the emulation of SOC in binary BEC
mixtures \cite{GalSpiel}, SOC effects were realized in exciton-polariton
fields populating semiconductor microcavities \cite%
{Schulz,Shelykh,Sala,Duffer,Lafont}. Actually, SOC in polaritonic
microcavities may have two different physical origins, \textit{viz}., the
direct SOC acting on excitons, or SOC mixing two photonic modes. The latter
option makes it possible to realize SOC in microcavities operating in the
regime of weak coupling between light and matter, when polaritons are not
formed, while SOC originates from the splitting of cavity modes, $\psi _{\pm
}$, whose electric-field components are perpendicular (TE) and parallel (TM)
to the in-plane component of the carrier wave vector.

Here, this option is addressed, following Ref. \cite%
{Mayteevarunyoo-Malomed-Skryabin-2018b}. To this end, a planar waveguide is
considered with saturable gain and saturable absorption, which are provided
by laser setups \cite%
{Rosanov-book-2002,Elsass-et-al-2010,Genevet,Turconi,Gustave,Veretenov2019}.
The corresponding system of coupled CGL equations, including scaled time $t$
and in-cavity coordinates $x$ and $y$, is

\begin{equation}
i\partial _{t}\psi _{+}=-(\partial _{x}^{2}+\partial _{y}^{2})\psi
_{+}+if\psi _{+}+\beta (\partial _{x}-i\partial _{y})^{2}\psi _{-},  \tag{48}
\end{equation}%
\begin{equation}
i\partial _{t}\psi _{-}=-(\partial _{x}^{2}+\partial _{y}^{2})\psi
_{-}+if\psi _{-}+\beta (\partial _{x}+i\partial _{y})^{2}\psi _{+},  \tag{49}
\end{equation}%
\begin{equation}
f=-1+\frac{g}{1+\varepsilon (|\psi _{+}|^{2}+|\psi _{-}|^{2})}-\frac{a}{%
1+(|\psi _{+}|^{2}+|\psi _{-}|^{2})}.  \tag{50}
\end{equation}%
Here the coefficient of the linear loss is scaled to be $1$, the saturable
gain and absorption are represented, respectively, by coefficients $g>0$ and
$a>0$, and SOC is represented by terms with coefficient $\beta $, which
perform linear mixing between the modes via the second spatial derivatives
\cite{Flayac}, unlike the first derivatives which represent SOC in the BEC\
models \cite{GalSpiel}. Further, positive $\varepsilon <1$ defines relative
saturation strength of the gain and absorption in Eq. (50). The generic
situation may be adequately represented by parameters
\begin{equation}
\varepsilon =0.1,~a=2,  \tag{51}
\end{equation}%
which implies weak saturation of the gain in comparison with absorption,
while the gain and SOC\ strengths, $g$ and $\beta $, may be varied as
physically relevant parameters controlling modes generated by the system.

Note that the nonlinearity in Eqs. (48) and (49) is entirely dissipative, as
the equations do not include SPM or XPM terms. The domination of the
dissipative terms over the conservative nonlinearity is the situation which
is possible in laser cavities.

The objective is to constructs solutions of Eqs. (48) and (49) in the form
of 2D solitons morphed as a bound state of two components with vorticities $%
m-1$ and $m+1$, to comply with the condition that, due to the SOC acting
through the second spatial derivatives in these equations, the difference
between the components' vorticities must be $\Delta m=2$. In polar
coordinates $\left( r,\theta \right) $, the relevant solutions with real
chemical potential $\mu $ can be defined as

\begin{equation}
\psi _{+}=\phi _{+}(r)\exp \left[ -i\mu t+i(m-1)\theta \right] ,~\psi
_{-}=\phi _{-}(r)\exp \left[ -i\mu t+i(m+1)\theta \right] ,  \tag{52}
\end{equation}%
with complex amplitude functions $\phi _{\pm }(r)$ satisfying the radial
equations:%
\begin{equation}
\mu \phi _{+}=-\left[ \frac{d^{2}}{dr^{2}}+\frac{1}{r}\frac{d}{dr}-\frac{1}{%
r^{2}}\left( m-1\right) ^{2}\right] \phi _{+}+if\phi _{-}+\beta \left( \frac{%
d^{2}}{dr^{2}}+\frac{2m+1}{r}\frac{d}{dr}+\frac{m^{2}-1}{r^{2}}\right) \phi
_{-},  \tag{53}
\end{equation}

\begin{equation}
\mu \phi _{-}=-\left[ \frac{d^{2}}{dr^{2}}+\frac{1}{r}\frac{d}{dr}-\frac{1}{%
r^{2}}\left( m+1\right) ^{2}\right] \phi _{-}+if\phi _{-}+\beta \left( \frac{%
d^{2}}{dr^{2}}-\frac{2m-1}{r}\frac{d}{dr}+\frac{m^{2}-1}{r^{2}}\right) \phi
_{+},  \tag{54}
\end{equation}%
where $f$ is given by Eq. (50) with $\left\vert \psi _{\pm }\right\vert $
replaced by $\left\vert \phi _{\pm }\right\vert $.

The necessary condition of the stability of the zero solution of Eqs. (48)
and (49) obviously amounts to condition
\begin{equation}
g<1+a.  \tag{55}
\end{equation}%
On the other hand, a necessary condition for the ability of the saturable
gain to maintain nontrivial modes is that the largest value of $f(n\equiv
|\psi _{+}|^{2}+|\psi _{-}|^{2})$ in Eq. (50), which is attained at density
\begin{equation}
n_{0}=\frac{\sqrt{a}-\sqrt{\varepsilon g}}{\sqrt{\varepsilon }\left( \sqrt{g}%
-\sqrt{\varepsilon a}\right) },  \tag{56}
\end{equation}%
must be positive. The substitution of $n_{0}$ in Eq. (50) yields a lower
bound for $g$, which, if combined with Eq. (55), defines the interval in
which the gain coefficient may take its values:
\begin{equation}
\sqrt{\varepsilon a}+\sqrt{1-\varepsilon }<g<1+a.  \tag{57}
\end{equation}%
In particular, for values adopted in Eq. (51) interval (57) amounts to
\begin{equation}
1.396<g<3.  \tag{58}
\end{equation}%
The b.c. for Eqs. (53) and (54) at $r\rightarrow 0$ is that $\phi _{\pm }$
must be vanishing as $r^{\left\vert m\mp 1\right\vert }$ at $r\rightarrow 0$%
, except for the case of $m\mp 1=0$, when the b.c. is $d\phi _{\pm
}/dr|_{r=0}=0$. At $r\rightarrow \infty $, soliton solutions must feature
the exponential decay,
\begin{equation}
\phi _{\pm }(r)\sim r^{-1/2}\exp \left( -\left( \lambda _{\mathrm{r}%
}+i\lambda _{\mathrm{i}}\right) r\right) ,  \tag{59}
\end{equation}%
with $\lambda _{\mathrm{r}}>0$.

\subsection{Results}

There are two most essential species of soliton complexes produced by
numerical solution of Eqs. (53) and (54) \cite%
{Mayteevarunyoo-Malomed-Skryabin-2018b}. First, this is the
vortex-antivortex (VAV) state, corresponding to $m=0$ in Eq. (52). The name
is due to the obvious fact that the components carry opposite vorticities, $%
-1$ and $+1$. The VAVs are similar to the HD (hidden-vorticity) states
considered in two-component atomic BEC \cite{Brtka}, cf. Eq. (5.26). The
difference is that the incoherently coupled GP equations for the binary
atomic BEC admit solutions with any combination of vorticities in the two
components \cite{Carmon}, while the vortex and antivortex components in VAV
are inherently coupled by the SOC terms, which admits solely the difference $%
\Delta m=2$ between the two components, as mentioned above

The second species is the SV (semi-vortex), corresponding to $m=1$ in Eq.
(5.52), with component vorticities $0$ and $2$. In SV solutions of the
coupled Gross-Pitaevskii equations for the binary atomic BEC under the
action of the SOC, SVs have two components with vorticities $\left(
0,1\right) $ or $\left( -1,0\right) $, due to the fact that the SOC is
represented in those equations by the first-order spatial derivatives mixing
the components, hence the corresponding SVs must have $\Delta m=1$ \cite%
{GalSpiel}.

\begin{figure}[tbp]
\includegraphics[scale=0.9]{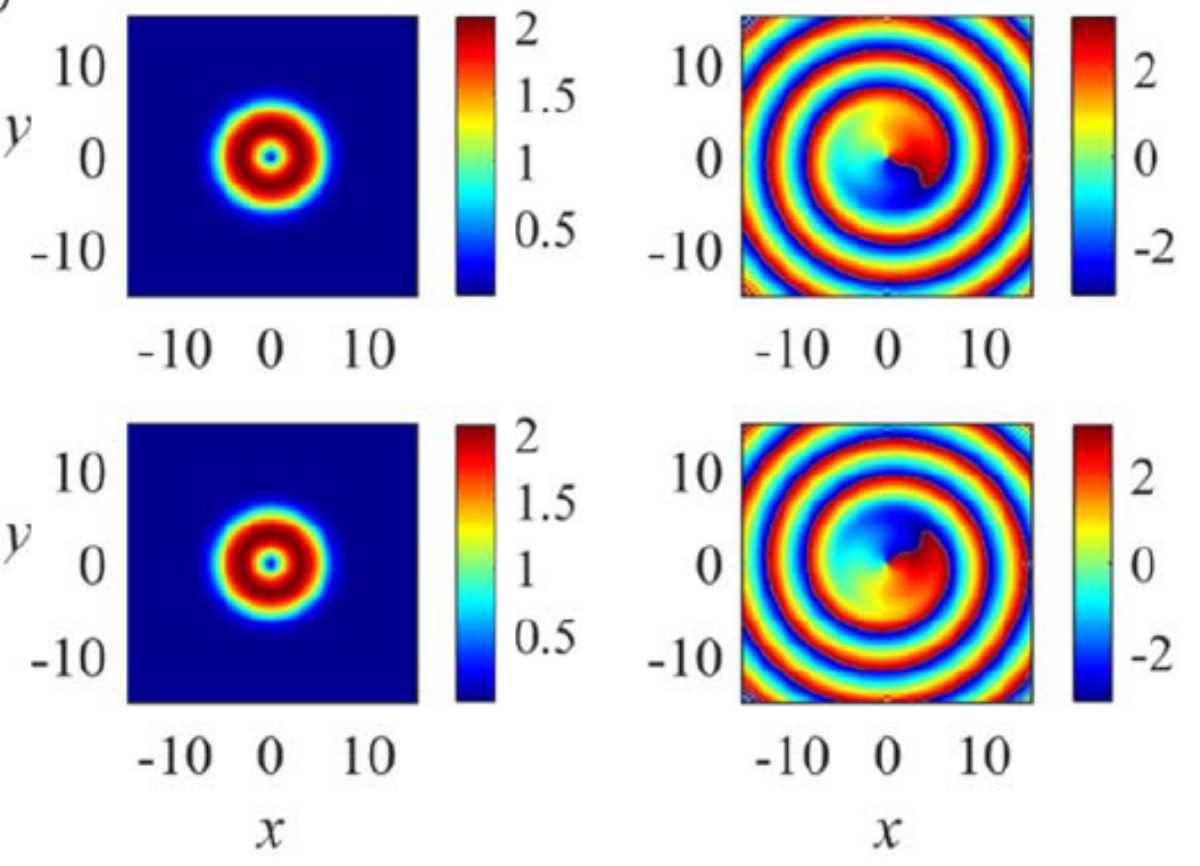}
\caption{Top views of amplitudes $\left\vert \protect\psi _{\pm
}(x,y)\right\vert $ and phases $\arg \left( \protect\psi _{\pm }(x,y\right)
) $ of the two components of a stable VAV (vortex-antivortex) solution
produced by Eqs. (53) and (54) with $\protect\beta =0.8$ and $g=2.12$. Other
parameters are fixed as per Eq. (51). The chemical potential of this VAV is $%
\protect\mu =0.072$ (source: Ref. \protect\cite%
{Mayteevarunyoo-Malomed-Skryabin-2018b}).}
\label{fig14.add3=fig_extra7}
\end{figure}

Examples of numerically found stable VAV and stable SV are displayed,
respectively, in Figs. \ref{fig14.add3=fig_extra7} and \ref%
{fig14.add4=fig_extra8}. Their stability was identified by means of
calculation of eigenvalues for small perturbations added to the stationary
solutions, and verified by direct simulations of the perturbed evolution.
\begin{figure}[tbp]
\includegraphics[scale=0.9]{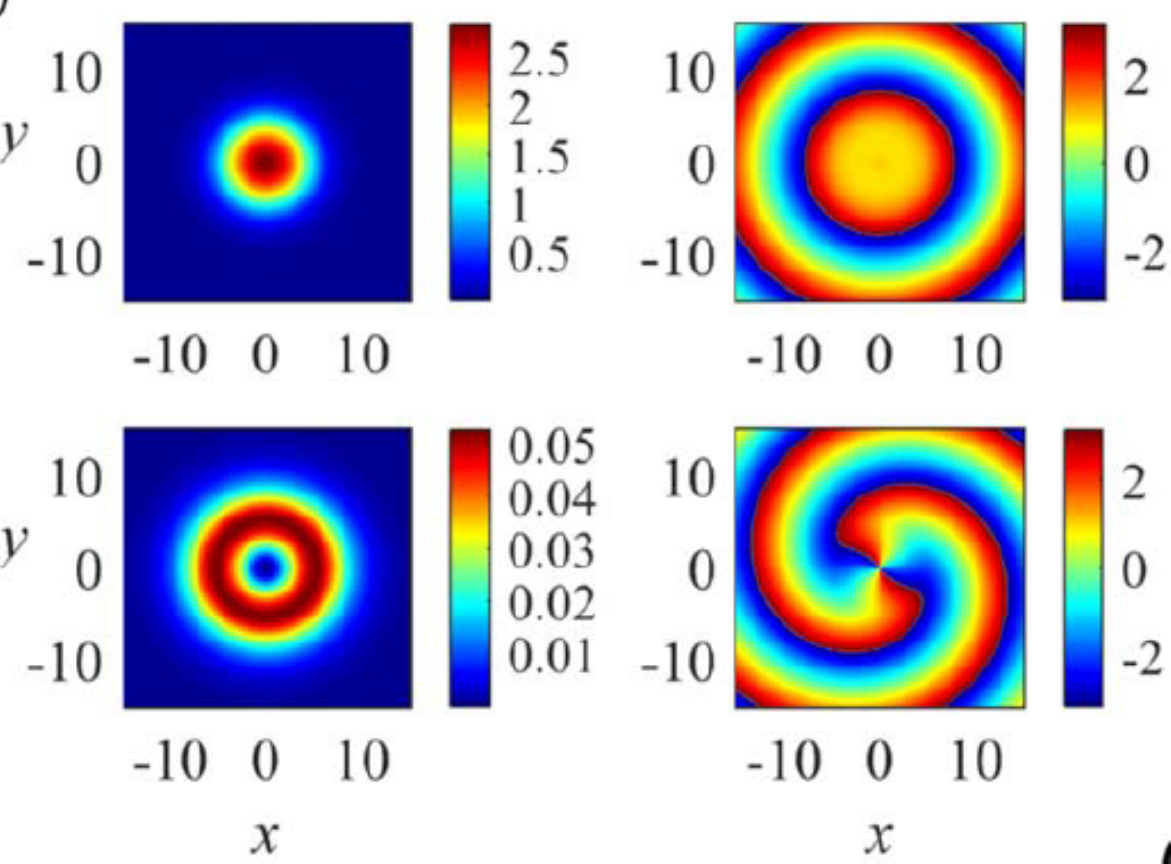}
\caption{The same as in Fig. \protect\ref{fig14.add3=fig_extra7}, but for a
stable SV (semi-vortex), with $\protect\beta =0.1$ and $g=2.10$ (source:
Ref. \protect\cite{Mayteevarunyoo-Malomed-Skryabin-2018b}).}
\label{fig14.add4=fig_extra8}
\end{figure}

\begin{figure}[tbp]
\includegraphics[scale=0.5]{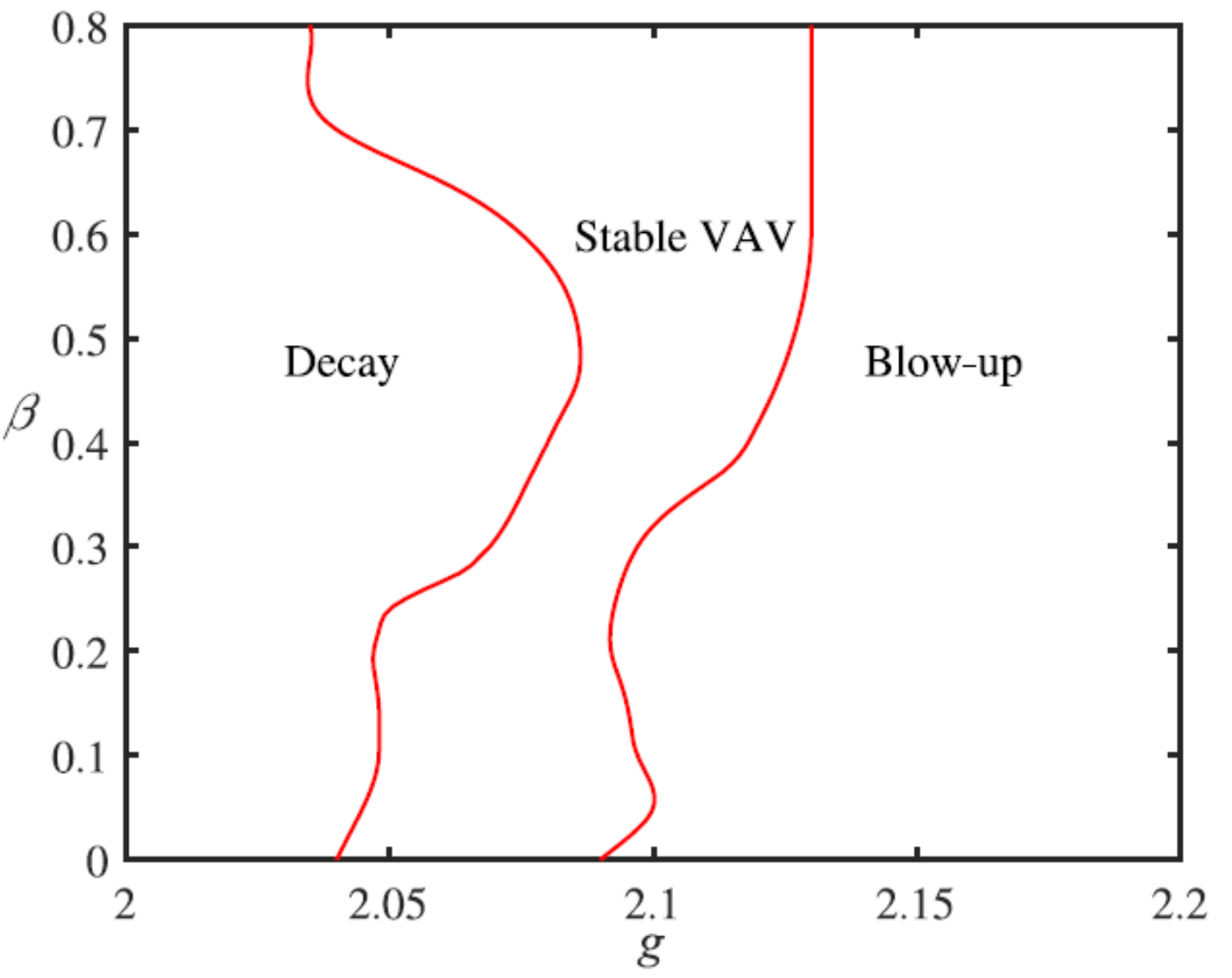}
\caption{Stability boundaries of the VAV\ solutions in the parameter plane $%
\left( g,\protect\beta \right) $ of the system of equations (48)-(50), while
other parameters are fixed according to Eq. (51) (source: Ref. \protect\cite%
{Mayteevarunyoo-Malomed-Skryabin-2018b}).}
\label{fig14.add5=fig_extra9}
\end{figure}

The stability chart of the VAVs in the plane of control parameters $\left(
g,\beta \right) $, which represent the gain and SOC strength in Eqs. (48),
(49) and (50), while the saturation and loss parameters are fixed as per Eq.
(51), is displayed in Fig. \ref{fig14.add5=fig_extra9}. The stability area
as a whole lies\ within interval (57). Naturally, the blowup and decay occur
if the gain is too large or too small, as indicated in the figure.

Note that the stability region for the VAV does not vanish at $\beta =0$ (in
the absence of SOC). In this limit, Eqs. (48) and (49) decouple into the
single-component CGL equations, in which the vortex (or antivortex) DSs are
not subject to the usual splitting instability because these equations do
not contain self-focusing terms (conservative nonlinearity). If the latter
terms are included, the VAVs become stable above a certain minimum value of
the SOC strength, $\beta >\beta _{\min }$ \cite{SMS}.

Unlike the VAVs, the stability area of SVs is very narrow, therefore this
soliton species is a less significant one in the present context. It was
demonstrated that the stability of SVs can be enhanced by adding the Zeeman
splitting to the system \cite{SMS}. In terms of Eqs. (48) and (49), it means
including terms $\mp \Omega \psi _{\pm }$ on the right-hand side of the
equations, where $\Omega $ is the strength of the Zeeman splitting. Lastly,
the system (without the Zeeman effect)\ may also sustain stable solutions in
the form of MMs (mixed modes), which combine zero-vorticity terms and ones
with vorticities $\pm 2$ in both components \cite{SMS}.

\section{Three-dimensional dissipative solitons}

\subsection{3D solitons in the free space}

A natural 3D extension of the CGL equation (2), written in terms of the
spatiotemporal right propagation in optics, includes the temporal GVD and
spectral filtering (the imaginary part of GVD), with the respective
coefficients $D\gtrless 0$ and $\gamma \geq 0$:%
\begin{equation}
\left( \frac{\partial }{\partial z}+\delta \right) U=\left( \beta +\frac{i}{2%
}\right) \left( \frac{\partial ^{2}}{\partial x^{2}}+\frac{\partial ^{2}}{%
\partial y^{2}}\right) U+\left( \gamma +\frac{i}{2}D\right) U_{tt}+\left(
\varepsilon +i\right) |U|^{2}U-\left( \mu +i\nu \right) |U|^{4}U.  \tag{60}
\end{equation}%
Recall that $D>0$ and $D<0$ in Eq. (60) correspond to the anomalous and
normal GVD, respectively, the former case being favorable for the creation
of temporal solitons. As shown below, in the presence of the spectral
filtering, $\beta >0$, Eq. (60) generates stable 3D solitons solitons for
both $D>0$ and $D<0$ (creation of bright temporal solitons in setups with
normal GVD is a problem of considerable interest in nonlinear optics \cite%
{Wu-et-al-2009,Ilday}. Further, it is assumed that the cubic and quintic
conservative terms in Eq. (60) represent self-focusing (with the respective
coefficient scaled to be $1$) and defocusing (accounted for by $\nu >0$),
respectively. Stationary solutions to Eq. (60), with propagation constant $k$
and integer vorticity (winding number) $S$, are looked for, in the
cylindrical coordinates $\left( r,\theta ,z,t\right) $, according to the
general ansatz,%
\begin{equation}
U\left( r,\theta ,z,t\right) =\exp \left( ikz+iS\theta \right) \psi
_{S}\left( r,t\right) ,  \tag{61}
\end{equation}%
where complex function $\psi _{S}$ satisfies equation%
\begin{equation}
\left( ik+\delta \right) \psi _{S}=\left( \beta +\frac{i}{2}\right) \left(
\frac{\partial ^{2}}{\partial r^{2}}+\frac{1}{r}\frac{\partial }{\partial r}-%
\frac{S}{r^{2}}\right) \psi _{S}+\left( \gamma +\frac{i}{2}D\right) \frac{%
\partial ^{2}\psi _{S}}{\partial t^{2}}+\left( \varepsilon +i\right) |\psi
_{S}|^{2}\psi _{S}-\left( \mu +i\nu \right) |\psi _{S}|^{4}\psi _{S}.
\tag{62}
\end{equation}%
Stationary states are characterized by their 3D norm, which has the meaning
of the total soliton's energy, in terms of optics:
\begin{equation}
E=\int \int \int \left\vert U\left( x,y,t\right) \right\vert ^{2}dxdydt=2\pi
\int_{0}^{\infty }rdr\int_{-\infty }^{+\infty }\left\vert \psi _{S}\left(
r,t\right) \right\vert ^{2}.  \tag{63}
\end{equation}%
The solutions with $S\geq 1$ may be considered as three-dimensional VRs,
alias \textit{vortex tori}, or \textquotedblleft donuts".

Stable fundamental ($S=0$) and vortex DS solutions to Eq. (60) were first
reported in Ref. \cite{Mihalache-et-al-2006}. Then, essential results were
added in Refs. \cite{Mihalache-et-al-2007a} and \cite{Mihalache-et-al-2007b}%
. It was found that fundamental 3D solitons, produced by Eq. (60), are
always stable, while, similar to what is known about solutions of the
two-dimensional CGL equation (2), vortex solitons with winding number $S\geq
1$ may only be stable, against spontaneous splitting, in the presence of the
diffusion term, with $\beta >0$. On the other hand, the action of the
spectral filtering is not necessary for the stability, i.e., Eq. (60) with $%
\gamma =0$ can generate stable 3D vortex solitons. For this reason, the
results are displayed below for $\gamma =0$.

\begin{figure}[tbp]
\includegraphics[scale=0.7]{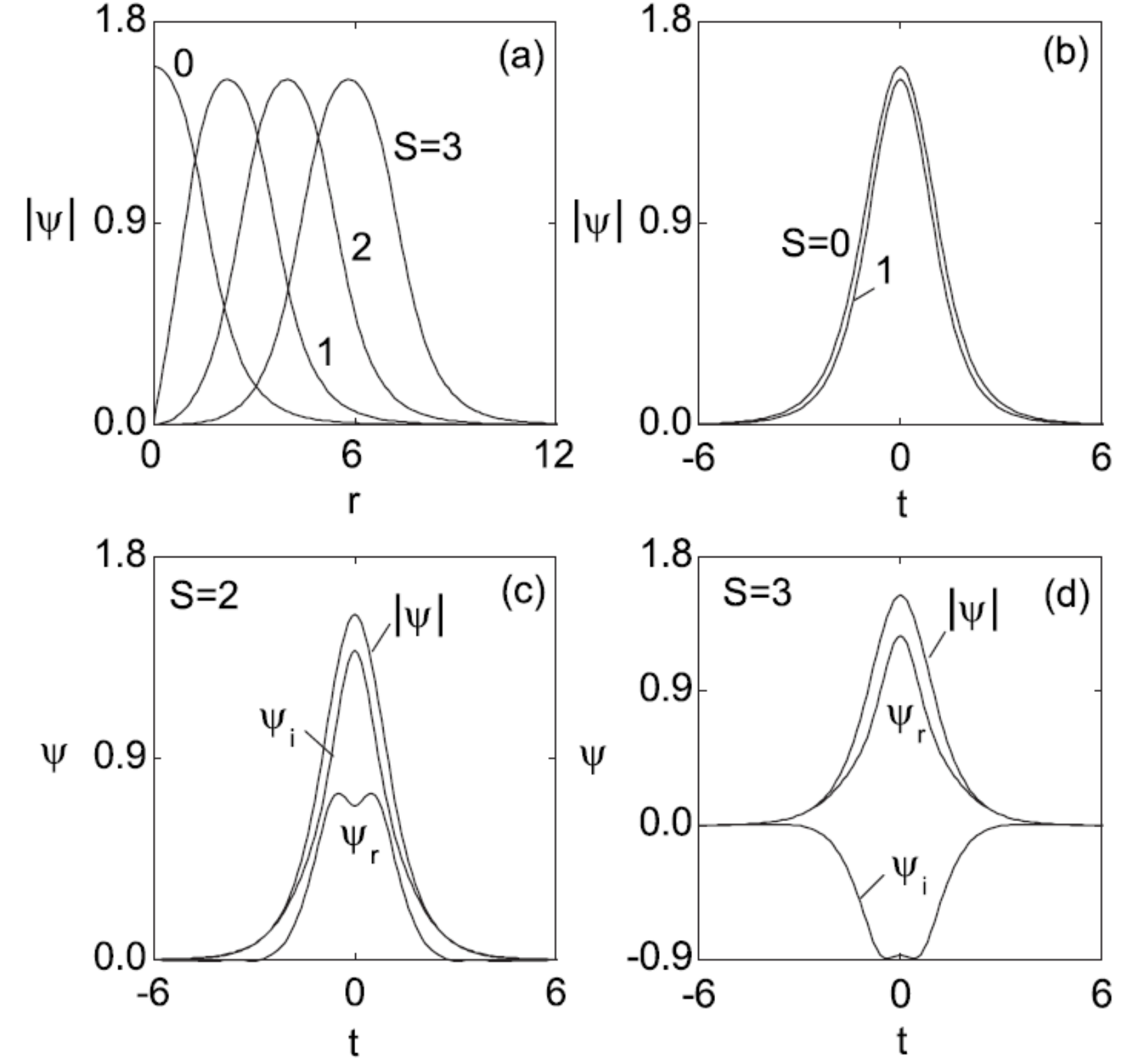}
\caption{(a) Profiles of the radial cross-section (at $t=0$) of $|\protect%
\psi _{S}(r,t)|$ in typical stable 3D dissipative solitons, with $S=0,1,2,3$%
, produced by numerical solution of Eq. (62). Panels (b)-(d): Cross sections
of the absolute value and real and imaginary parts of $\protect\psi %
_{S}(r,t) $ in the temporal direction (at $r=0$). In (b), the temporal
profiles for $S=2$ and $3$ completely overlap with that for $S=1$.
Parameters are $D=1$, $\protect\delta =0.4$, $\protect\beta =0.5$, $\protect%
\varepsilon =2.2$, $\protect\mu =1$, $\protect\nu =0.1$, and $\protect\gamma %
=0$ (source: Ref. \protect\cite{Mihalache-et-al-2007b}).}
\label{fig14.34=fig200}
\end{figure}

\begin{figure}[tbp]
\includegraphics[scale=0.65]{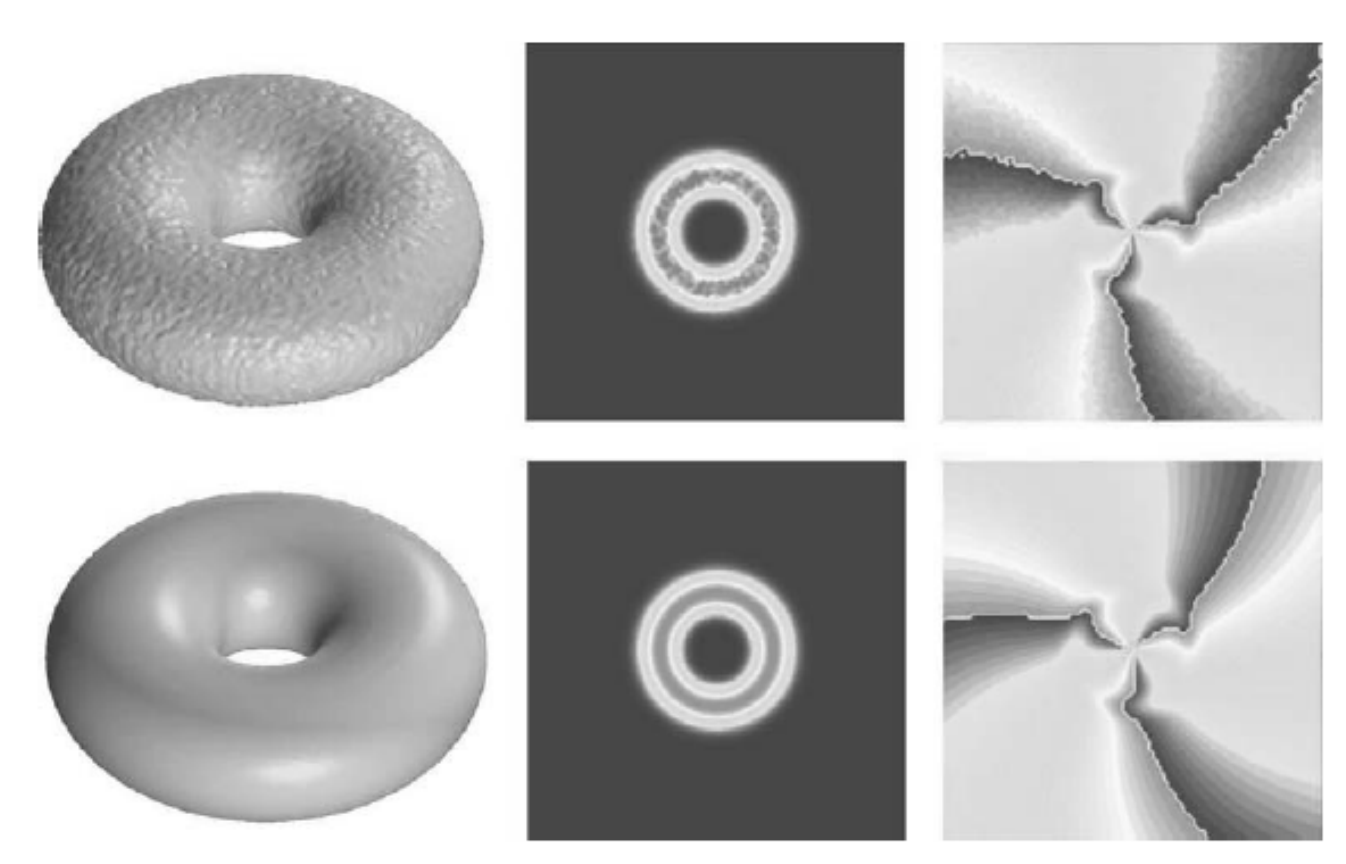}
\caption{Relaxation of an initially perturbed 3D vortex soliton with $S=3$,
produced by simulation of Eq. (60). The left and middle panels display the
intensity distribution, $\left\vert U(x,y,t)\right\vert ^{2}$, in the 3D
form, and in the temporal mid plane, $t=0$. The right panels: the phase
distribution, $\arg \left( U(x,y\right) )$, in the same plane. The top and
bottom rows show, respectively, the input and the result of the evolution at
$z=800$. Parameters are $D=1$, $\protect\delta =0.4$, $\protect\beta =0.5$, $%
\protect\varepsilon =2.3$, $\protect\mu =1$, $\protect\nu =0.1$, and $%
\protect\gamma =0$ (source: Ref. \protect\cite{Mihalache-et-al-2007b}).}
\label{fig14.35=fig201}
\end{figure}

A set of typical shapes of stable 3D solitons with $S=0,1,2,3$ are displayed
in Fig. \ref{fig14.34=fig200}, by means of their radial cross sections in
the plane of $t=0$ (temporal midplane) and temporal cross sections along the
line of $x=y=0$ (the spatial axis). A full 3D view of a stable toroidal
vortex soliton with $S=3$ is given in Fig. \ref{fig14.35=fig201}, which
demonstrates the stability of this soliton against small perturbations.

\begin{figure}[tbp]
\includegraphics[scale=0.85]{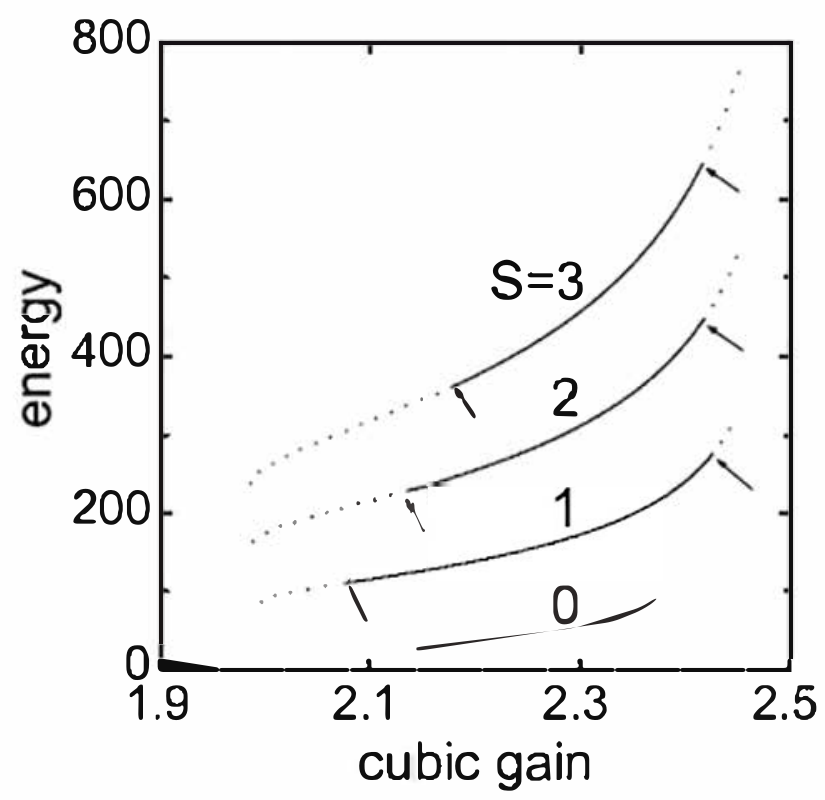}
\caption{Energy $E$ of the 3D fundamental and vortical solitons with winding
numbers $S$ vs. the cubic-gain parameter $\protect\varepsilon $. Other
parameters in Eq. (60) are $D=1$, $\protect\delta =0.4$, $\protect\beta =0.5$%
, $\protect\mu =1$, $\protect\nu =0.1$, and $\protect\gamma =0$. Dotted and
solid segments correspond to unstable and stable solutions, respectively.
Arrows indicate stability boundaries. The branch pertaining to $S=0$
(fundamental solitons) is completely stable (source: Ref. \protect\cite%
{Mihalache-et-al-2007b}).}
\label{fig14.36=fig202}
\end{figure}

Families of fundamental and vortex 3D solitons are characterized by
dependences of their energy, calculated as per Eq. (63), on control
parameters of the model. In particular, curves $E(\varepsilon )$ are
displayed, for $S=0,1,2,3$, in Fig. \ref{fig14.36=fig202}. As seen in this
figures, the family of the fundamental 3D solitons (with $S=0$) is stable in
its entire existence region, while for vortex solitons, with $S=1,2,3$, the
stability region is narrower than the existence domain. Furthermore, as
mentioned above, in the case of $\beta =0$ (no diffusion term in Eq. (60))
vortex-soliton families exist, but are completely unstable (as well as in
the 2D version of the CGL equation).

\begin{figure}[tbp]
\includegraphics[scale=0.65]{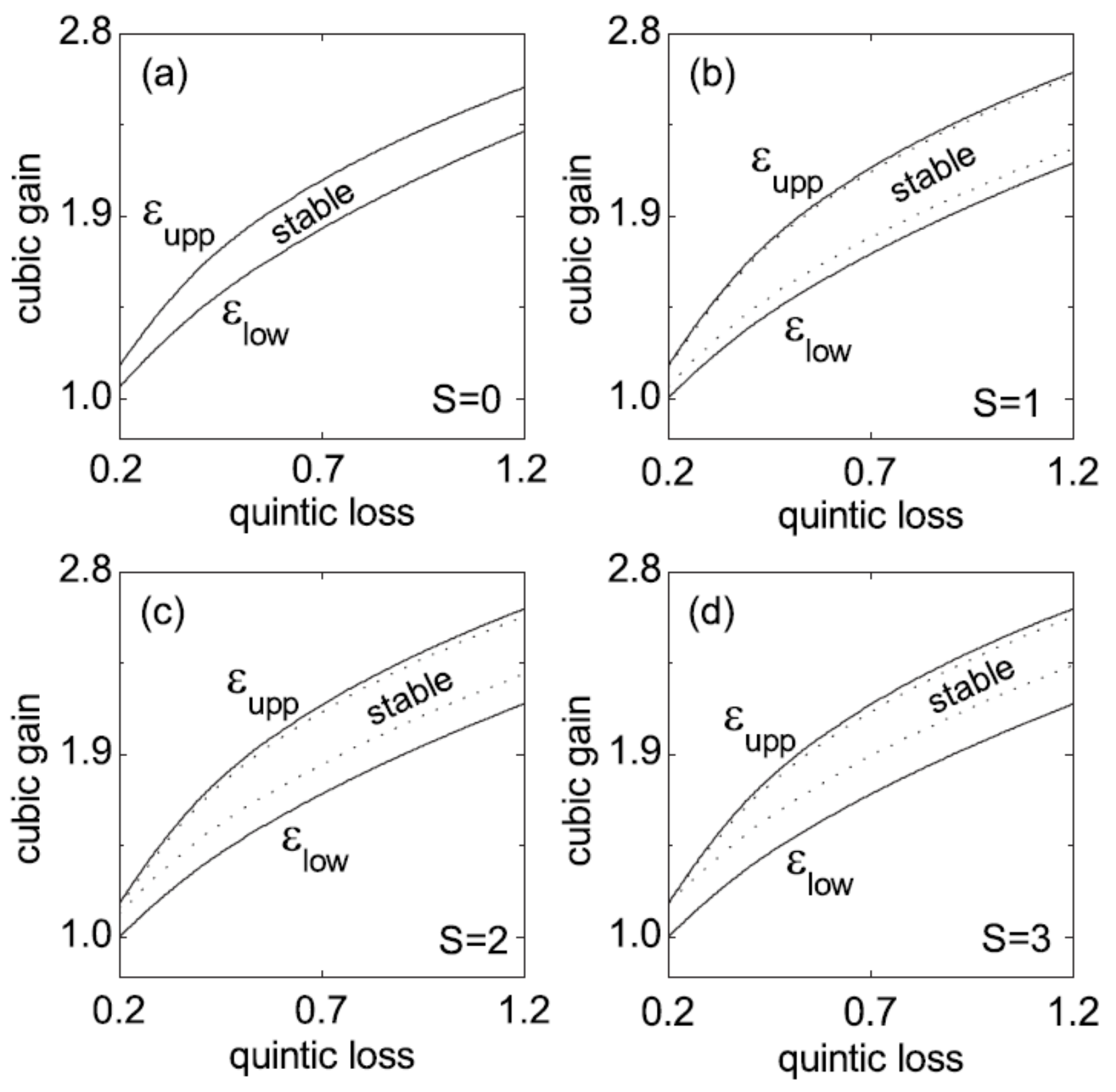}
\caption{The existence and stability domains of 3D solitons, with $S=0,1,2,3$%
, for $\protect\beta =0.5$, in the plane of the quintic-loss and cubic-gain
coefficients, $(\protect\mu ,\protect\varepsilon )$, of Eq. (60). Other
parameters are $D=1$, $\protect\delta =0.4$, $\protect\beta =0.5$, $\protect%
\gamma =0$, and $\protect\nu =0.1$. The fundamental solitons $(S=0)$ exist
and are stable between solid curves. The vortex solitons with $S=1,2,3$
exist between solid curves and are stable between dotted ones, i.e., their
stability domain is narrower than the existence range (source: Ref.
\protect\cite{Mihalache-et-al-2007b}).}
\label{fig14.37=fig203}
\end{figure}

The results for the existence and stability of the solitons are summarized
in Fig. \ref{fig14.37=fig203}, in the form of the respective charts plotted
in the plane of the nonlinear loss and gain parameters, $\left( \mu
,\varepsilon \right) $. The stability was identified by means of numerical
solution of the eigenvalue problem for small perturbations, and verified by
direct simulations of the perturbed simulations \cite%
{Mihalache-et-al-2006,Mihalache-et-al-2007a,Mihalache-et-al-2007b}) (see, in
particular, Fig. \ref{fig14.35=fig201}).

The findings presented above were obtained with $D=1$ in Eq. (42), i.e., for
the anomalous sign of the GVD term. The same CGL equation with $D<0$, i.e.,
with normal GVD, is also able to produce stable 3D fundamental and vortex
solitons. In particular, all the fundamental solitons, with $S=0$, remain
stable in this case. Because the combination of the cubic self-focusing with
normal GVD gives rise to stable CW (flat) states, the temporal cross section
of the solitons tends to be flatter in this case, as seen in Fig. \ref%
{fig14.38=fig204}.
\begin{figure}[tbp]
\includegraphics[scale=0.85]{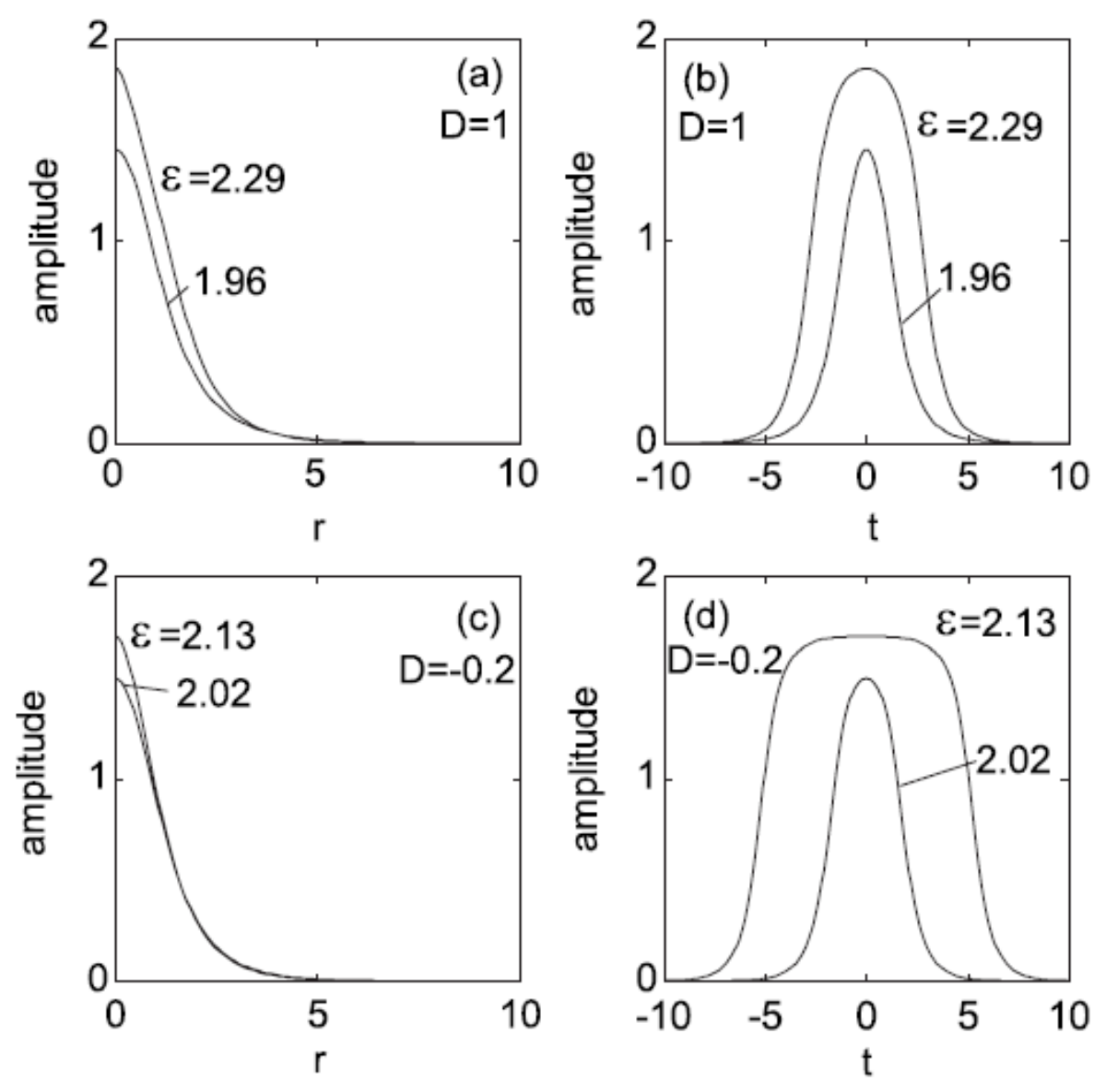}
\caption{(a,c) Profiles of the radial cross-section (at $t=0$) of $|\protect%
\psi _{S}(r,t)|$ in stable fundamental solitons, with $S=0$, obtained as the
numerical solution of Eq. (62). (b,d) Cross sections of $|\protect\psi %
_{S}(r,t)|$ in the temporal direction (at $r=0$) for the same solitons.
Values of the GVD and cubic-gain coefficients, $D$ and $\protect\varepsilon $%
, are indicated in the panels. Note that panels (c) and (d) represent a
stable fundamental soliton existing in the case of the \emph{normal GVD},
with $D=-0.2<0$. Other parameters are $\protect\delta =0.4$, $\protect\gamma %
=0.5$, $\protect\beta =0$, $\protect\mu =1$, and $\protect\nu =0.1$ (source:
Mihalache \textit{et al}. (2007a)).}
\label{fig14.38=fig204}
\end{figure}

A noteworthy finding is that the stability area for the vortex solitons with
$S=1$ produced by Eq. (60) with the normal GVD ($D<0$), is somewhat \textit{%
broader} than its counterpart for $D>0$, as seen in Fig. \ref%
{fig14.39=fig205}, which compares $E(\varepsilon )$ curves for both cases.
This fact is natural, as the normal sign of the GVD precludes the
possibility of the modulational instability in the temporal direction.
\begin{figure}[tbp]
\includegraphics[scale=0.65]{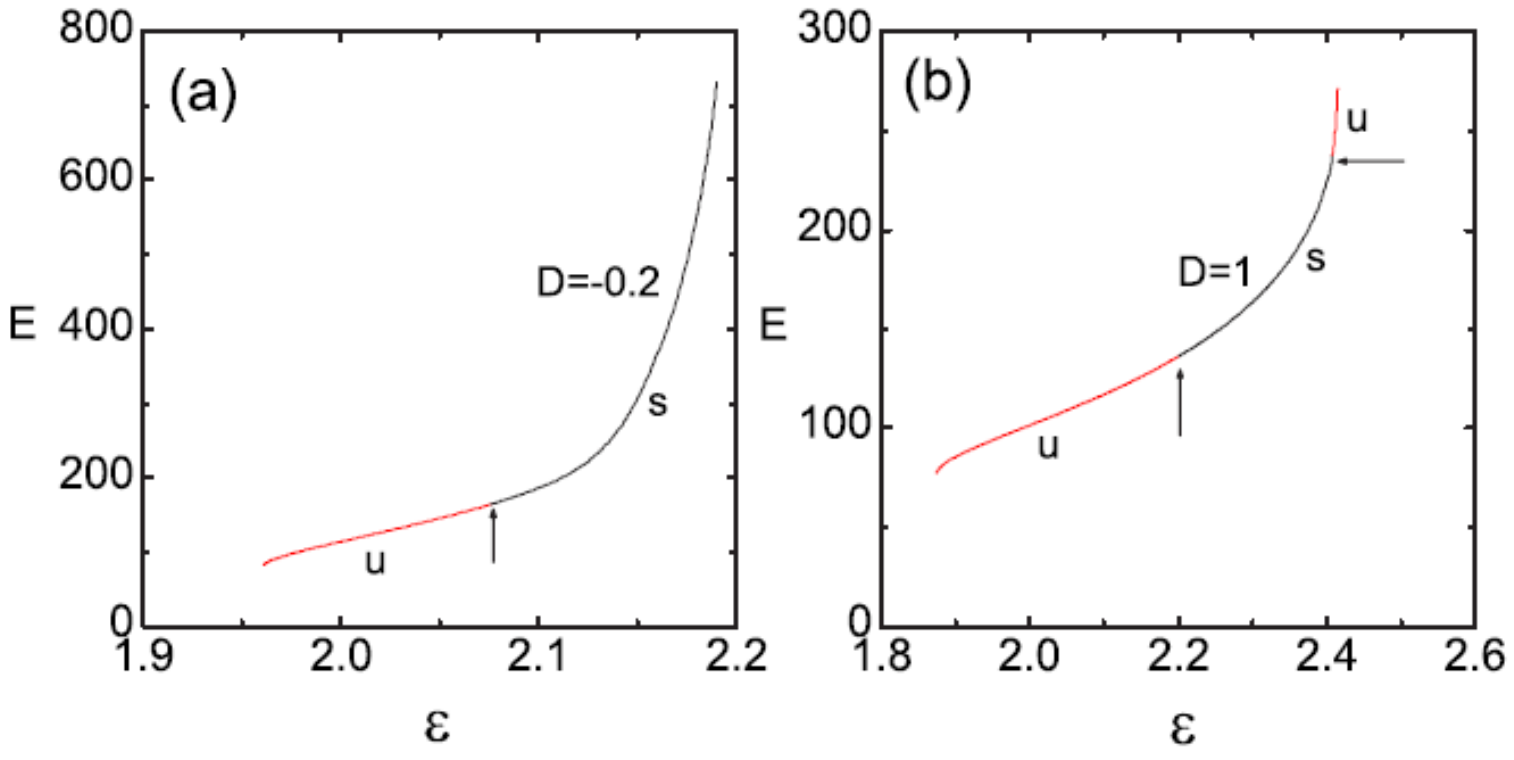}
\caption{The energy of the vortex solitons with $S=1$ vs. the nonlinear-gain
coefficient, $\protect\varepsilon $. The solutions are produced by Eq. (60)
with \emph{normal GVD}, $D=-0.2$ in (a), and anomalous GVD, $D=1$, in (b).
Other parameters are $\protect\delta =0.4$, $\protect\beta =0.1$, $\protect%
\gamma =0.5$, $\protect\mu =1$, and $\protect\nu =0.1$. Stable (black) and
unstable (red) portions of the solution branches are marked by symbols
\textquotedblleft s\textquotedblright\ and \textquotedblleft
u\textquotedblright , in addition to using the different colors in them.
Arrows indicate stability boundaries (source: Mihalache \textit{et al}.
(2007a)).}
\label{fig14.39=fig205}
\end{figure}

Essential results for 3D vortex DSs have been produced by the work with
complex models of the spatiotemporal dynamics in laser cavities. Stable
sophisticated states produced this work with those models include complexes
of bound 3D solitons which may emulate motion of a solid body \cite%
{Veretenov2008,Veretenov2010}, knotted solitons \cite{Veretenov2017}, and
ones with a tubular structure \cite{Veretenov2021}. This topic was reviewed
in Refs. \cite{Veretenov2010} and \cite{Veretenov2019}.

\subsection{Collisions between 3D vortex solitons}

As stated above, the CGL equation (60) gives rise to stable vortex solitons
only in the presence of the diffusion term with $\beta >0$, hence the
equation has no Galilean invariance in the spatial plane $\left( x,y\right) $%
. Nevertheless, the existence of stable vortex solitons in the absence of
the spectral filtering, $\gamma =0$, makes it possible to set the soliton in
free motion in the longitudinal (temporal) direction. Thus, it is possible
to simulate collisions between the 3D solitons (fundamental and vortical
ones) moving along the longitudinal coordinate. Analysis of the collisions
was developed using Eq. (60) with $\gamma =0$ and $D=1$ (the anomalous sign
of the GVD term) \cite%
{Mihalache-et-al-2008a,Mihalache-et-al-2008b,Mihalache-et-al-2009}:%
\begin{equation}
\left( \frac{\partial }{\partial z}+\delta \right) U=\left( \beta +\frac{i}{2%
}\right) \left( \frac{\partial ^{2}}{\partial x^{2}}+\frac{\partial ^{2}}{%
\partial y^{2}}\right) U+\frac{i}{2}U_{tt}+\left( \varepsilon +i\right)
|U|^{2}U-\left( \mu +i\nu \right) |U|^{4}U.  \tag{64}
\end{equation}

In the general form, simulations of collisions between 3D solitons is a
challenging problem. However, it is quite tractable in the case of
interactions between coaxial vortex solitons, which may collide moving in
opposite directions along their common pivot. In particular, simulations of
Eq. (60) for collisions between identical solitons, with equal vorticities, $%
S_{1}=S_{2}$, and a large initial temporal separation, $T$, between them,
were initiated by input%
\begin{equation}
U_{0}(r,\theta ,t)=\psi _{S}(r,t-T/2)\exp \left( iS\theta +i\chi t\right)
+\psi _{S}(r,t+T/2)\exp \left( iS\theta -i\chi t\right) ,  \tag{65}
\end{equation}%
where real $\pm \chi $ are initial kicks applied to the two solitons.
Numerical results are displayed below in Figs. \ref{fig14.40=fig206} -- \ref%
{fig14.42=fig208} for parameters%
\begin{equation}
\delta =0.4,\beta =0.5,\varepsilon =2.3,\mu =1,\nu =0.1.  \tag{66}
\end{equation}

For all values of the common vorticity, $S=0,1,2$, the simulations of Eq.\
(64) with initial conditions (65) lead to the same qualitative results \cite%
{Mihalache-et-al-2008a}: as shown in Fig. \ref{fig14.40=fig206} for $S=2$,
slowly colliding solitons merge into a single one (which demonstrates
gradually decaying intrinsic excitation), fast solitons quasi-elastically
pass through each other, while at intermediate (moderately large) velocities
the collision creates an additional soliton, with the same vorticity as
carried by the original ones. The newly created soliton stays at the
collision position.
\begin{figure}[tbp]
\includegraphics[scale=0.65]{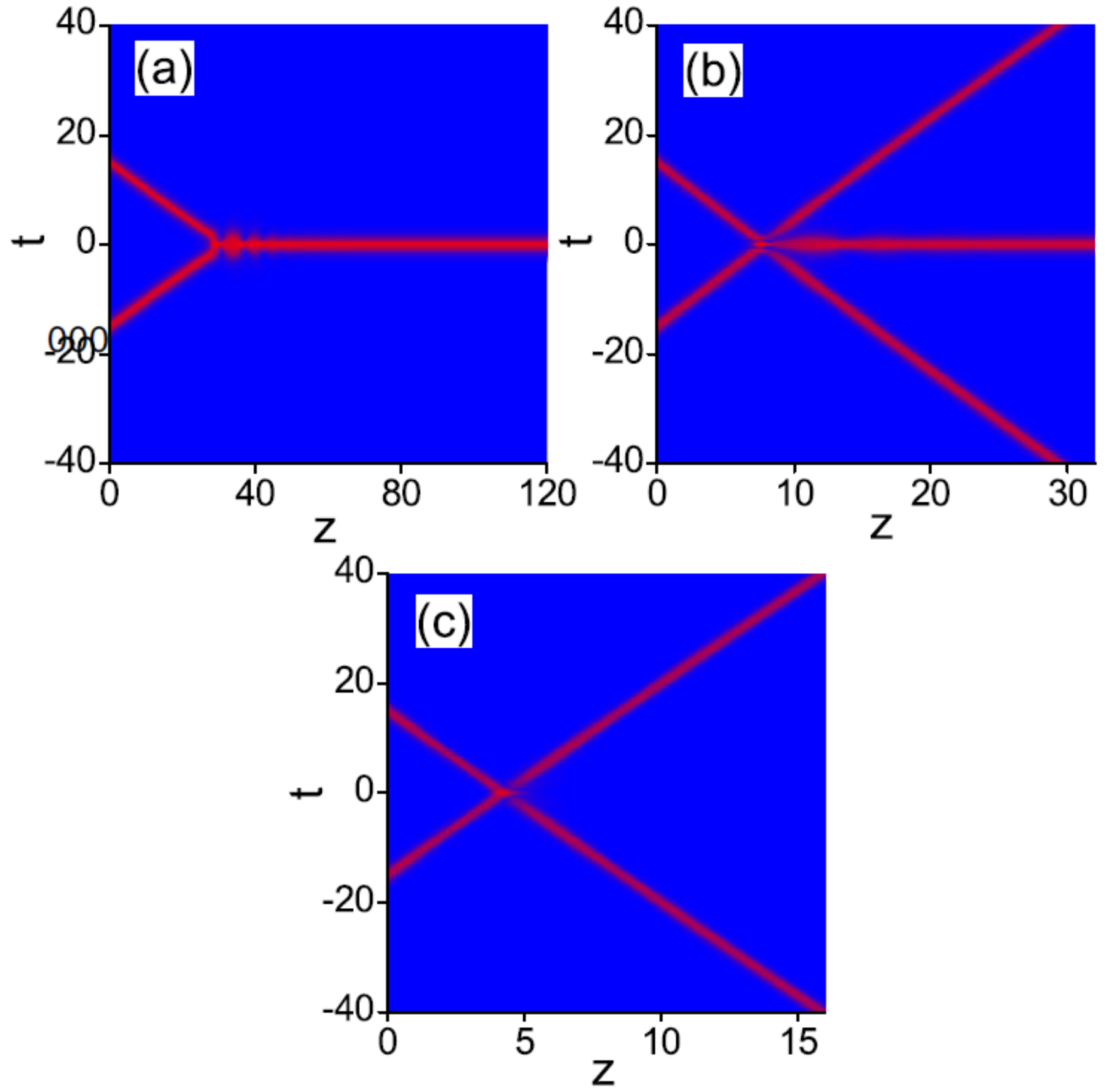}
\caption{Plots of $\left\vert U\left( x,y,t,z\right) \right\vert $ in the
plane of $x=y=0$ illustrate typical outcomes of collisions between two 3D
dissipative vortex solitons with $S=2$, produced by simulations of Eq.
(5.64) with parameters (66) and initial conditions (65). (a) Merger of the
slowly colliding solitons, which were set in motion by the kicks with $%
\protect\chi =0.5$ (see the detailed picture in Fig. \protect\ref%
{fig14.41=fig207}). (b) Creation of an additional quiescent soliton with $%
S=2 $ in the case of $\protect\chi =2$. (c) Quasi-elastic collision of fast
solitons in the case of $\protect\chi =4$, see the detailed picture in Fig.
\protect\ref{fig14.42=fig208} (source: Ref. \protect\cite%
{Mihalache-et-al-2008a}).}
\label{fig14.40=fig206}
\end{figure}

\begin{figure}[tbp]
\includegraphics[scale=0.80]{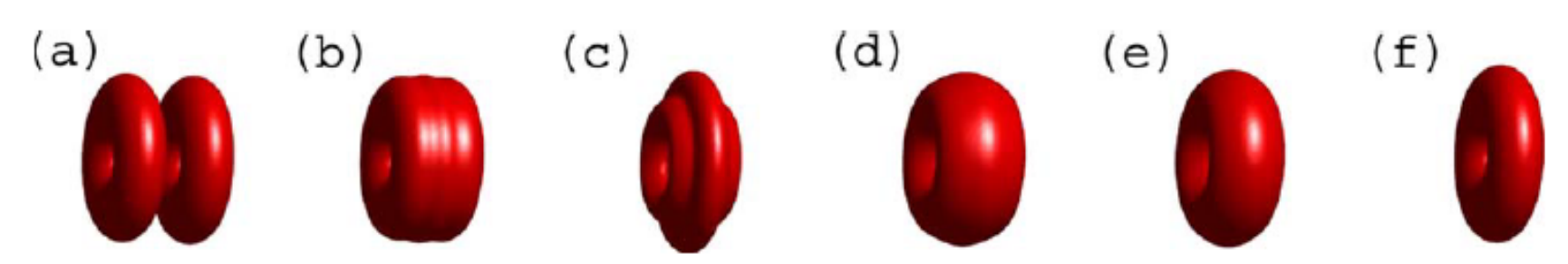}
\caption{Merger of the vortex solitons with winding numbers $S=2,$ which are
set in motion as per initial conditions (65) with $\protect\chi =0.5$.
Configurations are displayed at propagation dostances $z=25$ (a), $28$ (b), $%
30$ (c), $35$ (d), $40$ (e), and $60$ (f). The figure corresponds to panel
(a) in Fig. \protect\ref{fig14.40=fig206} (source: Ref. \protect\cite%
{Mihalache-et-al-2008a}).}
\label{fig14.41=fig207}
\end{figure}

The merger of slowly colliding solitons into a single one, and the
generation of the additional soliton by those colliding at moderately large
velocities, are illustrated in detail by Figs. \ref{fig14.41=fig207} and \ref%
{fig14.42=fig208}, respectively. These pictures adequately represent generic
outcomes of the inelastic collisions.
\begin{figure}[tbp]
\includegraphics[scale=0.80]{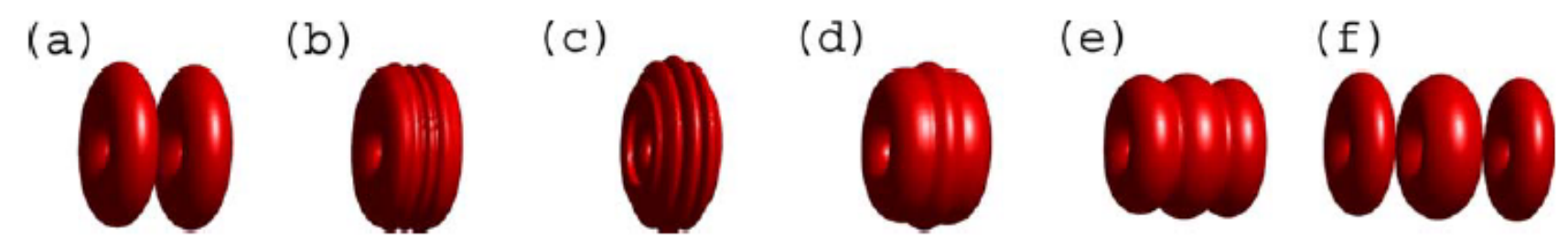}
\caption{The generation of an additional soliton with $S=2$ by the collision
of two vortex solitons (also with $S=2$), which are set in motion as per
initial conditions (65) with $\protect\chi =2$. Configurations are displayed
at propagation distances $z=3$ (a), $3.5$ (b), $4$ (c), $4.5$ (d), $5$ (e),
and $5.5$ (f).The figure corresponds to panel (b) in Fig. \protect\ref%
{fig14.40=fig206} (source: Ref. \protect\cite{Mihalache-et-al-2008a}).}
\label{fig14.42=fig208}
\end{figure}

\begin{figure}[tbp]
\includegraphics[scale=0.66]{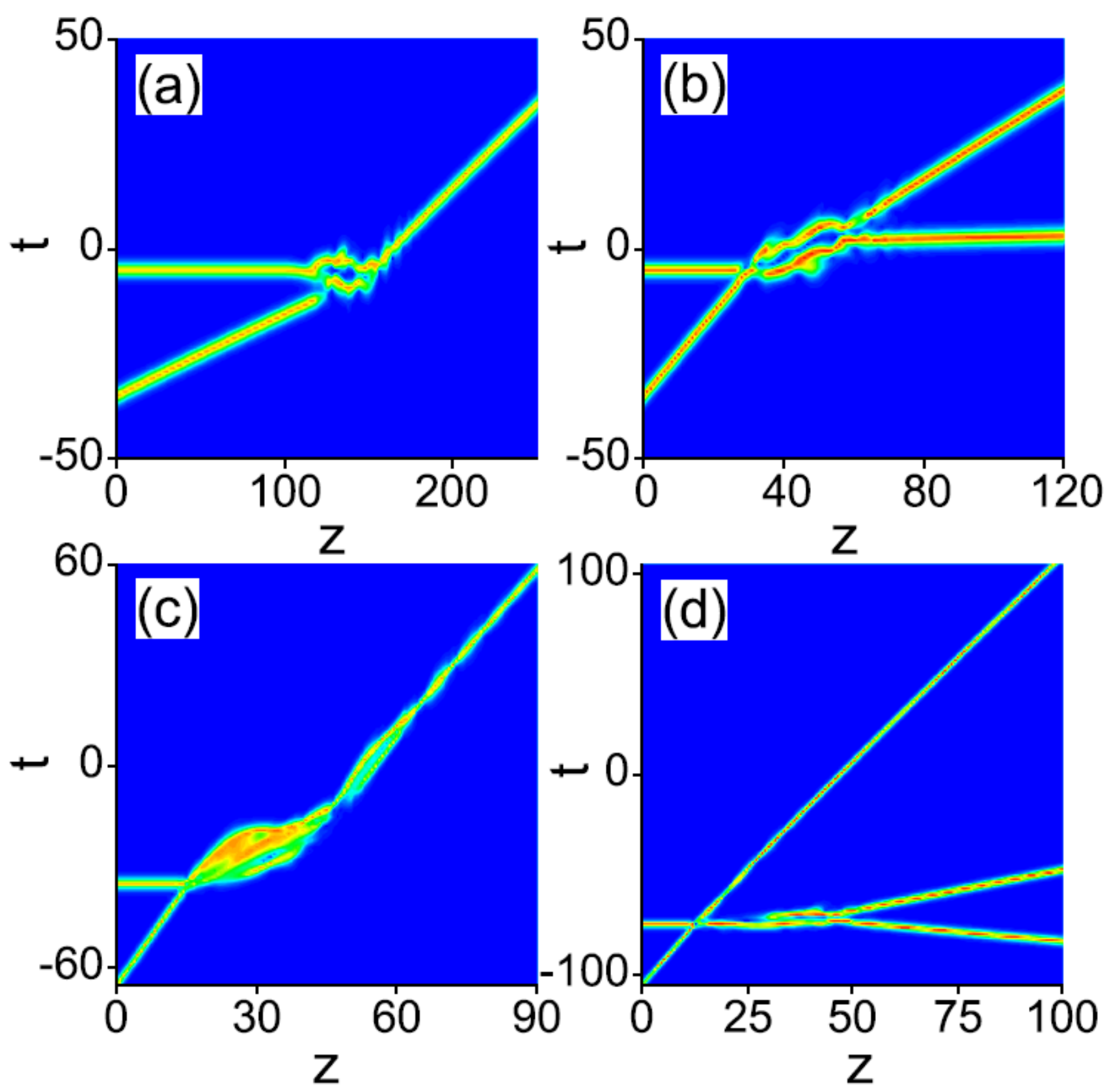}\newline
\caption{Contour plots of $\left\vert U\left( z,t\right) \right\vert $ in
the plane of $\left( x=y=0\right) $, which illustrate outcomes of collisions
of an initially kicked fundamental soliton ($S=0$) and a quiescent
(unkicked) stable one with $S=2$. The results are produced by simulations of
Eq. (64) with parameters $\protect\delta =0.4$, $\protect\beta =0.5$, $%
\protect\varepsilon =2.3$, $\protect\mu =1$, $\protect\nu =0.1$. In this
case, the unperturbed solitons are found with propagation constants $%
k(S=0)\approx 0.443,$ $k(S=1)\approx 0.500$, $k(S=2)\approx 0.504$. The
fundamental soliton is initially set in motion by kick $\protect\chi =0.2$
(a), $1$ (b), $2$ (c), and $2.5$ (d), as in Eq. (65) (source: Ref.
\protect\cite{Mihalache-et-al-2009}).}
\label{fig14.add6=fig_extra10}
\end{figure}

Collisions between a fundamental soliton and a vortex one, with $S=1$ or $2$%
, were studied, by means of simulations of Eq. (64) \cite%
{Mihalache-et-al-2009}. Typical results are displayed, for the case of the
vortex soliton with $S=2$, in Fig. \ref{fig14.add6=fig_extra10}. The slow
collision leads to merger of the solitons into a single one with $S=0$,
i.e., the vorticity is destroyed by the inelastic collision in this case,
see Fig. \ref{fig14.add7=fig_extra11}(a). A faster collision produces, in
Fig. \ref{fig14.add7=fig_extra11}(b), two separating solitons, both having $%
S=0$, i.e., the vorticity is lost in this case too. Merger of the colliding
solitons into a single one with $S=0$ again occurs at somewhat larger value
of the initial kick, as seen in Fig. \ref{fig14.add6=fig_extra10}(c).
Finally, the fast collision seems quasi-elastic, but the perturbed vortex
soliton eventually splits in two fragments, each having $S=0$. The latter
outcome is displayed in Fig. \ref{fig14.add6=fig_extra10}(d).

\begin{figure}[tbp]
\includegraphics[scale=0.92]{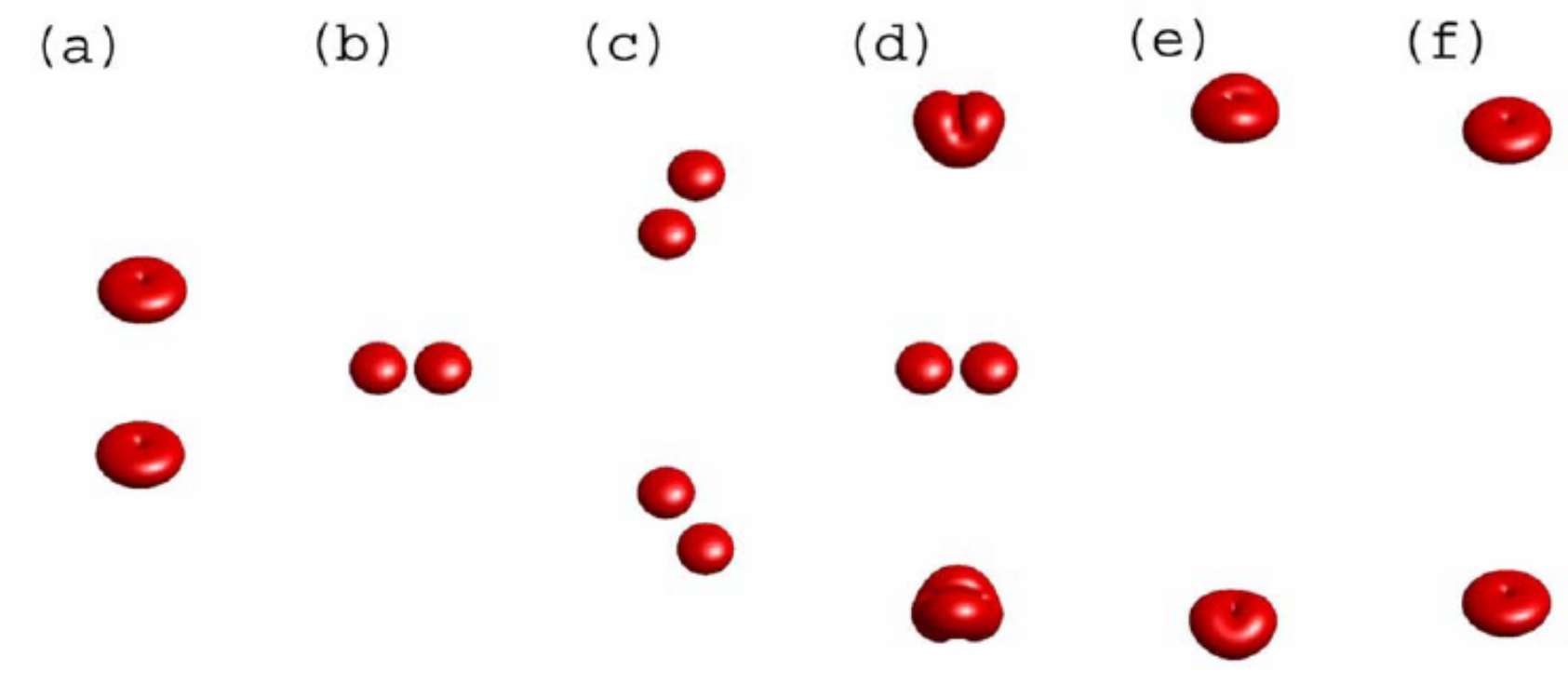}
\caption{A summary of outcomes of collisions between coaxial
counter-rotating vortex solitons with $\left( S_{1},S_{2}\right) =\left(
+1,-1\right) $, produced by input (67). Plot (a) represents the input. The
results, produced by simulations of Eq. (64), are shown in plots (c)-(f),
which correspond to increasing values of kick $\protect\chi $ in Eq. (67)
(source: Ref. \protect\cite{Mihalache-et-al-2008b}).}
\label{fig14.43=fig209}
\end{figure}

\begin{figure}[tbp]
\includegraphics[scale=0.92]{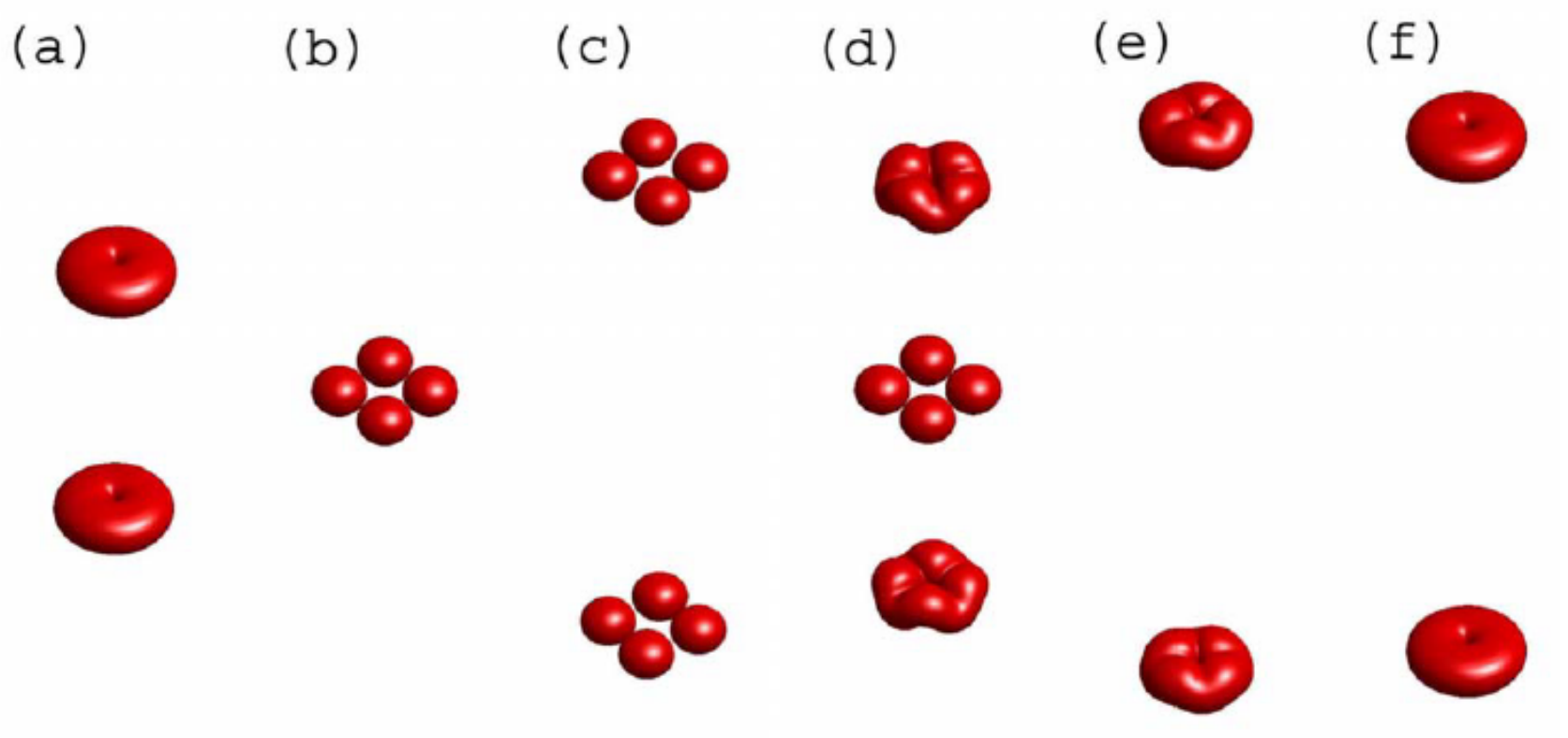}
\caption{The same as in Fig. \protect\ref{fig14.40=fig206}, but for the
input corresponding to Eqs. (67) and (69) (source: Ref. \protect\cite%
{Mihalache-et-al-2008b}).}
\label{fig14.44=fig210}
\end{figure}

Different results are produced by collisions of \textquotedblleft
counter-rotating" vortex solitons, which are taken as pairs of identical
ones with opposite intrinsic vorticities $S_{1}=-S_{2}=1$ or $S_{1}=-S_{2}=2$
\cite{Mihalache-et-al-2008b}. In this case, simulations of Eq. (64) with
coefficients (66) were initiated by input
\begin{equation}
U_{0}(r,\theta ,t)=\psi _{S}(r,t-T/2)\exp \left( iS\theta +i\chi t\right)
+\psi _{-S}(r,t+T/2)\exp \left( -iS\theta -i\chi t\right) ,  \tag{67}
\end{equation}%
cf. Eq. (65). The simulations demonstrates that, in both cases of input (67)
with the vorticity pairs
\begin{equation}
\left( S_{1},S_{2}\right) =\left( +1,-1\right)  \tag{68}
\end{equation}%
and%
\begin{equation}
\left( S_{1},S_{2}\right) =\left( +2,-2\right) ,  \tag{69}
\end{equation}%
the collision may lead, essentially, to five different outcomes, depending
on the magnitude of kick $\chi $ in the input. These outcomes are
summarized, for inputs corresponding to Eqs. (67) and (68) or (69), in Figs. %
\ref{fig14.43=fig209} and \ref{fig14.44=fig210}. In these figures, the
inputs are represented by plots (a).

First, if $\chi $ takes small values, the slow collisions are strongly
inelastic, leading to the merger of the colliding vortex and antivortex
solitons into a single compound state, which is a non-rotating dipole
cluster in Fig. \ref{fig14.43=fig209}(b), and a non-rotating quadrupole
cluster in Fig. \ref{fig14.44=fig210}(b), which are built of two or four
fundamental solitons, respectively. These clusters, as well as those
observed in plots (c) and (d) of both figures, very slowly expand in the
course of the subsequent evolution, i.e., they do not represent true bound
states.

At larger values of $\chi $, the colliding solitons pass through each other,
but they do not keep the intact structure. Instead, as seen in Figs. \ref%
{fig14.43=fig209}(c) and \ref{fig14.44=fig210}(c), each vortex is converted
into a cluster, consisting, respectively, of two or four fundamental
solitons. In both cases corresponding to Eqs. (68) and (69), the emerging
clusters exhibit slow counter-rotation, which is subject to deceleration
under the action of the viscous force induced by the term $\sim \beta $ in
Eq. (64).

Next, still larger values of $\chi $ lead to an inelastic collisions
illustrated by plots (d) in Figs. \ref{fig14.43=fig209} and \ref%
{fig14.44=fig210}. In this case, there emerge a pair of slowly
counter-rotating two-humped (in Fig. \ref{fig14.43=fig209}(d)) or
four-humped (in \ref{fig14.44=fig210}(d)) localized states. In addition, the
collision creates a single nonrotating dipole cluster (in Fig. \ref%
{fig14.43=fig209}(d)), or a quadrupole one (in \ref{fig14.44=fig210}(d)). In
the course of the subsequent evolution, both two- and four-humped states
slowly split into pairs of fundamental solitons.

Further increase of the initial kick $\chi $ in Eq. (67) leads, in Figs. \ref%
{fig14.43=fig209}(e) and \ref{fig14.44=fig210}(e), to the formation of the
pairs of the counter-rotating two- or four-humped states, without the
central cluster (unlike plots (d)). Lastly, at largest values of $\chi $,
the collision is, naturally. quasi-elastic, featuring the vortex and
antivortex solitons passing through each other. The latter outcome is shown
in Figs. \ref{fig14.43=fig209}(f) and \ref{fig14.44=fig210}(f).

\subsection{Quasi-stable 3D vortex solitons supported by 2D potential
lattices or a trapping potential}

While the 3D CGL equation (60) cannot maintain stable vortex solitons in the
absence of the diffusion term ($\beta =0$), vorticity-carrying complexes
with a four-peak structure can be made nearly stable, in the case of $\beta
=0$, with the help of the 2D potential lattice \cite{Mihalache-et-al-2010a}.
This is a significant finding because, as said above, the diffusion term
does not appear in CGL models of laser systems, while the transverse
spatially periodic structure is present if the system is based on the
photonic crystal.

\begin{figure}[tbp]
\includegraphics[scale=0.60]{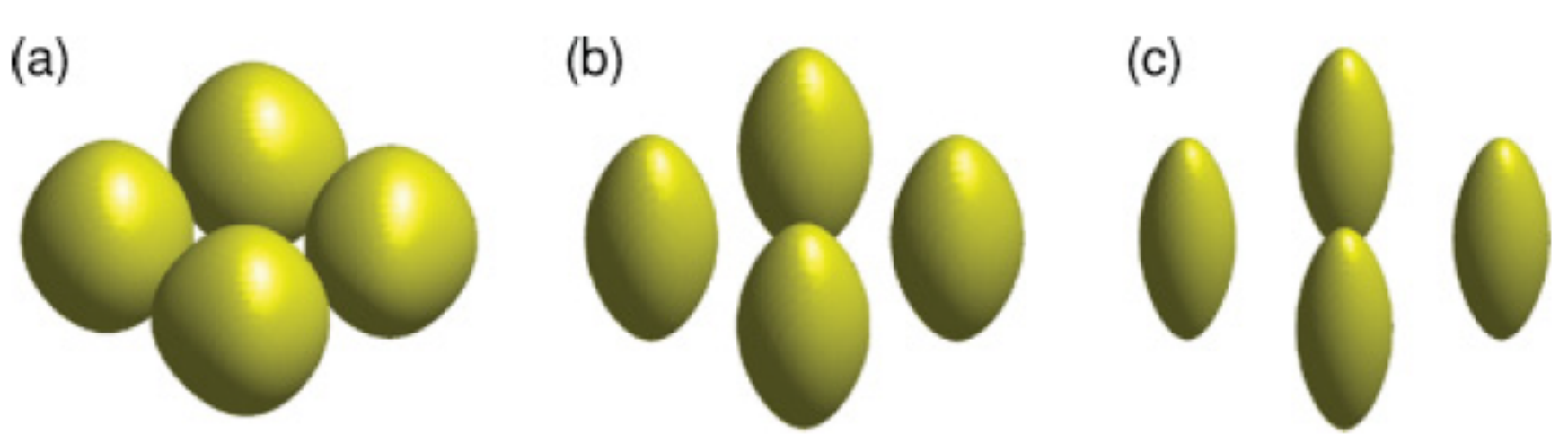}\newline
\includegraphics[scale=0.70]{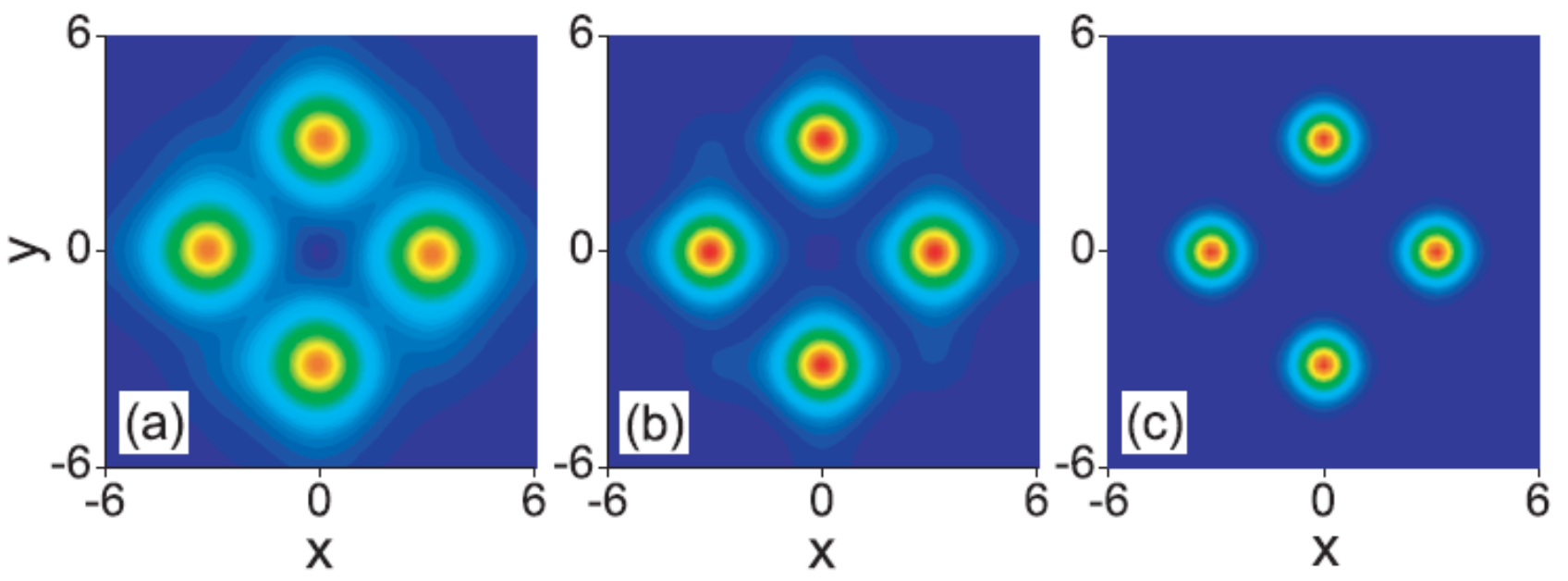}\newline
\caption{The top and bottom plots are the 3D images and 2D cross sections in
the plane of $\left( x,y\right) $ of the four-peak complexes with winding
number $S=1$, produced as solutions of Eq. (70) with parameters $\protect%
\delta =0.4$, $\protect\mu =1$. and $\protect\nu =0.1$. The cubic-gain
coefficient and strength of the lattice potential are $\protect\varepsilon %
=1.9$, $V_{0}=0.25$ in (a), $\protect\varepsilon =1.7$, $V_{0}=1$, and $%
\protect\varepsilon =1.8$, $V_{0}=4$ (source: Ref. \protect\cite%
{Mihalache-et-al-2010a}).}
\label{fig14.44=fig211}
\end{figure}

The corresponding model is similar to CGL equation (64) with $\beta =0$, to
which the lattice potential with scaled spatial period $L=\pi $ and depth $%
4V_{0}>0$ is added:%
\begin{equation}
\left( \frac{\partial }{\partial z}+\delta \right) U=\frac{i}{2}\left( \frac{%
\partial ^{2}}{\partial x^{2}}+\frac{\partial ^{2}}{\partial y^{2}}+\frac{%
\partial ^{2}}{\partial t^{2}}\right) U+\left( \varepsilon +i\right)
|U|^{2}U-\left( \mu +i\nu \right) |U|^{4}U+iV_{0}\left[ \cos (2x+\cos (2y)%
\right] U.  \tag{70}
\end{equation}%
Numerical solution of Eq. (70) give rise to the OS-centered (rhombic)
four-peak complexes with vorticity $S=1$, see typical examples in Fig. \ref%
{fig14.44=fig211}. Such solutions are stable against perturbations in the $%
\left( x,y\right) $ plane, but slightly unstable against spontaneous
separation of the constituents in the free longitudinal (temporal)
direction. Nevertheless, for $V_{0}=4$, the separation instability of the
vortex complex displayed in panels (c) of Fig. \ref{fig14.44=fig211} is
extremely weak, making it a practically stable state.

\begin{figure}[tbp]
\includegraphics[scale=1.50]{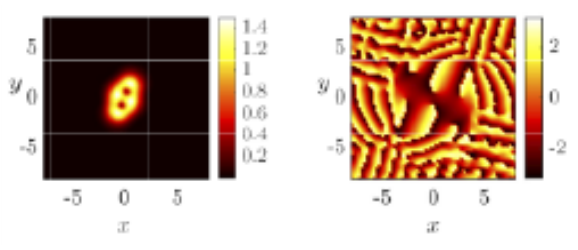}\newline
\caption{The left and right panels display the snapshots of amplitude and
phase, $\left\vert U(x,y)\right\vert $ and $\arg \left( U(x,y)\right) $, in
the temporal midplane, $t=0$, of the rotating pair of vortex solitons with $%
S=1$, produced by splitting of an unstable one with $S=2$. The results are
produced by simulations of Eq. (71) with $\protect\delta =0.1$, $\protect%
\varepsilon =0.6$, and $\protect\mu =0.4$ (source: Ref. \protect\cite%
{MMS-2019}).}
\label{fig14.add7=fig_extra11}
\end{figure}

In the absence of the diffusivity, i.e., $\beta =0$ in Eq. (60), 3D
dissipative spatiotemporal solitons with embedded vorticity may be
stabilized by the HO trapping potential, applied in the transverse plane $%
\left( x,y\right) $. The respective modification of Eq. (70) is \cite%
{MMS-2019}%
\begin{equation}
\left( \frac{\partial }{\partial z}+\delta \right) U=\frac{i}{2}\left( \frac{%
\partial ^{2}}{\partial x^{2}}+\frac{\partial ^{2}}{\partial y^{2}}+\frac{%
\partial ^{2}}{\partial t^{2}}\right) U+\left( \varepsilon +i\right)
|U|^{2}U-\mu |U|^{4}U-i\left( x^{2}+y^{2}\right) U.  \tag{71}
\end{equation}%
Numerical solution of this equation has produced results which are similar
to those outlined above in the connection to Eq. (60), where the
stabilization of vortex DSs is provided by the diffusion term. Namely, a
part of the family of 3D localized solutions with vorticity $S=1$, generated
by Eq. (71), are stable, A difference from the model based on Eq. (60) is
that all solitons with $S\geq 2$ are unstable against splitting (recall that
Eq. (60) produces stable vortex solitons with $S=2$ and $3$). Because of the
presence of the trapping potential in Eq. (71), a soliton with $S=2$ splits
into a pair of vortex solitons with $S=1$, which form a rotating bound
state, as shown in Fig. \ref{fig14.add7=fig_extra11}.

\section{Conclusion}

As said in the Introduction, this review does not aim to cover all essential
aspects of the work on dissipative multidimensional solitons. In particular,
in addition to collisions between DSs, which are considered above in some
form, it may be interesting to consider stationary bound states of spatially
separated 2D and 3D solitons, alias \textquotedblleft soliton molecules".
Excited states of such \textquotedblleft molecules", in the form of
oscillating bound states, may be quite interesting objects too. A
theoretical tool for the prediction of potentially stable bound states of
multidimensional DSs is offered by the analysis of an effective potential of
interaction between far separated solitons \cite{potential}. Experimentally,
such bound states were created only in effectively 1D setups, \textit{viz}.,
as coupled pairs of temporal solitons in fiber lasers \cite%
{Tang,Grelu,Optica}.

A new branch of studies of light propagation in complex media represents
topological photonics, which aims to create sophisticated optical patterns
carrying robust intrinsic structures, which are characterized by the
respective topological numbers. In particular, much interest has been
recently drawn to emulation of topological insulators in photonics \cite%
{Rechtsman}. In most cases, such patterns were studied in the framework of
linear optics. Solitons were addressed too, in the context of nonlinear
topological photonics, although chiefly in the 1D form, such as edge
solitons \cite{Leykam}. First results for 2D topological photonic solitons
were reported recently \cite{Bandres,Smirnova}. In the context of the
present review, a challenging possibility may be to consider 2D topological
solitons in artificially built nonlinear media, where losses are inevitable,
hence they should be compensated by gain in an appropriate form.

Another area of photonics which is relevant in the present context is the
study of self-trapped soliton-like states in exciton-polariton condensates
filling semiconductor microcavities. Most works on this topic addressed 1D
settings, but the consideration of 2D localized exciton-polariton states is
relevant too \cite{Deng-Haug-Yamamoto}. The exciton-polariton condensates
are essentially dissipative nonlinear media, in which losses are balanced by
an external pump. In particular, vortex states were created in such
condensates experimentally \cite{Lagoudakis,Roumpos}. Spatially-periodic
(lattice) potentials, which provide a major technique for the creation and
stabilization of various soliton-like states (as discussed, in various
contexts, in this review), are also available for the experimental work with
the exciton-polariton condensates \cite{Kim-et-al}. In particular, the
lattice potential was used for the creation of 2D gap solitons in the
exciton-polariton system, making use of the repulsive nonlinearity induced
by dipole-dipole interactions between polarized excitons \cite{Cerda}.

A different species of dissipative nonlinear systems in which losses are
compensated by an external pump is represented by the Lugiato-Lefever (LL)\
equation \cite{LL}. Laser cavities offer an important physical realization
of this model. Although in most works it was considered in the 1D form, the
consideration of the 2D LL equation is relevant too, which suggests a
possibility to look for 2D soliton-like states in this context \cite%
{Staliunas,Panajotov,Milian-et-al}. Additional possibilities for the
creation of localized states are offered by the 2D LL equation with a
tightly localized (focused) pump term \cite{Cardoso}.

Finally, it is relevant to mention that theoretical analysis was also
developed for various types of 2D solitons in $\mathcal{PT}$-symmetric
systems, with the cubic or CQ nonlinearity and a complex potential, composed
of spatially even real and odd imaginary parts \cite%
{Musslimani,Nixon,Burlak,Chen-et-al,Jianke,Luz,Li-et-al}. As mentioned
above, although the $\mathcal{PT}$-symmetric systems may be considered as a
very special case of dissipative settings, they give rise not to isolated
solitons, but to their continuous families (like in conservative models), up
to the $\mathcal{PT}$-symmetry-breaking point. In the theoretical works on
2D $\mathcal{PT}$-symmetric solitons, including ones with embedded
vorticity, stability is a crucially important issue.

\section*{Acknowledgments}

I would like to thank Profs. Marcel Clerc, Karin Alfaro-Bittner, Alejandro
Leon, Rene Rojas, and Mustapha Tlidi, the Guest Editors of the special issue
of Chaos, Solitons \& Fractals on the topic of \textit{New trends on
instabilities and nonequilibrium structures}, for their invitation to submit
a contribution for this special issue. The work was supported, in part, by
the Israel Science foundation through grant No. 1286/17.

\end{document}